\documentclass[12pt]{article}

 \usepackage{ bm}
\usepackage{eucal}
\usepackage{color}

\usepackage{tikz}
\usepackage{amsmath,bbm}
\usepackage{amssymb}
\usepackage{epic}

\usepackage{graphicx}
\usepackage{pict2e}

\usepackage{pdfsync}

  \usepackage{soul}

\title{A singular Toeplitz determinant and the discrete tacnode kernel for skew-Aztec Rectangles}

\author{Mark Adler\thanks{2000
{\em Mathematics Subject Classification}. Primary:
60G60, 60G65, 35Q53; secondary: 60G10, 35Q58. {\em Key
words and Phrases}:Lozenge tilings, non-convex polygons, kernels. \newline
 *Department of Mathematics, Brandeis University,
Waltham, Mass 02453, USA. E-mail: adler@brandeis.edu.
The support of a Simons Foundation Grants 
 \# 278931 is
 gratefully acknowledged. }~~~~~~Kurt Johansson\thanks{Department of Mathematics,
KTH Royal Institute of Technology, Stockholm, Sweden. E-mail: kurtj@kth.se. The support of the Swedish Research Council (VR) and grant KAW 2015.0270 of the Knut and Alice Wallenberg Foundation are gratefully acknowledged.} ~~~~~ Pierre
van Moerbeke\thanks{ Department of Mathematics,
Universit\'e de Louvain, 1348 Louvain-la-Neuve, Belgium
and Brandeis University, Waltham, Mass 02453, USA. E-mail: pierre.vanmoerbeke@uclouvain.be . The support of Simons Foundation Grants 
 \# 280945 is
gratefully acknowledged. \newline
 }
}

\date{}

\newcommand{\MAT}[1]{\left(\begin{array}{*#1c}}
\newcommand{\mat}{\end{array}\right)}
\newcommand{\qed}{\leavevmode\unskip\nobreak\penalty200\hskip2pt\null
\nobreak\hfill\rule{1.1ex}{1.1ex}
\medbreak }

\newcommand{\hh}{\widehat h}
\newcommand{\I}{{\rm i}}
\newcommand{\AR}{{\cal A}}

\newcommand{\DR}{{\cal D}}
\newcommand{\ER}{{\mathcal E}}
\newcommand{\FR}{{\cal F}}
\newcommand{\GR}{{\cal G}}
\newcommand{\HR}{{\cal H}}

\newcommand{\JR}{{\cal J}}
\newcommand{\KR}{{\cal K}}
\newcommand{\LR}{{\cal L}}
\newcommand{\RR}{{\cal R}}
\newcommand{\MR}{{\cal M}}
\newcommand{\NR}{{\cal N}}

\newcommand{\PR}{{\cal P}}

\newcommand{\SR}{{\cal S}}

\newcommand{\BC}{{\mathbb C}}

\newcommand{\BH}{{\mathbb H}}

\newcommand{\BZ}{{\mathbb Z}}

\newcommand{\iy}{\infty}
\newcommand{\pl}{\partial}
\newcommand{\al}{\alpha}

 \newcommand{\Om}{\Omega}

\newcommand{\Id}{\mathbbm{1}}

\newcommand{\0}{\Ga_0}

\newcommand{\rk}{\mathfrak r}

\newenvironment{proof}{\medskip\noindent{\it Proof:\/} }{\qed}

\newcommand{\vp}{\varphi}
\newcommand{\la}{\langle}
\newcommand{\ra}{\rangle}
\newcommand{\ga}{\gamma}
\newcommand{\ka}{\kappa}
\newcommand{\Ga}{\Gamma}
\newcommand{\dt}{\delta}
\newcommand{\Dt}{\Delta}
 \newcommand{\vr}{\varepsilon}
\newcommand{\sg}{\sigma}
\newcommand{\ze}{\zeta}
\newcommand{\BR}{{\mathbb R}}
\newcommand{\lb}{\lambda}

\newcommand{\Lb}{\Lambda}

\newcommand{\dis}{\displaystyle}

\newcommand{\gr}{{\frak r}}

\newcommand{\bl}{\begin{aligned}}
\newcommand{\el}{\end{aligned}}

\newcommand{\BK}{{\mathbb K}}

\def\be#1\ee{\begin{equation}#1\end{equation}}
\def\bea#1\eea{\begin{eqnarray}#1\end{eqnarray}}
\def\bean#1\eean{\begin{eqnarray*}#1\end{eqnarray*}}

 \newtheorem{definition}{Definition}[section]
 \newtheorem{theorem}[definition]{Theorem}
 \newtheorem{lemma}[definition]{Lemma}
 
 \newtheorem{proposition}[definition]{Proposition}

\catcode `!=11

\newdimen\squaresize
\newdimen\thickness
\newdimen\Thickness
\newdimen\ll! \newdimen \uu! \newdimen\dd! \newdimen \rr! \newdimen
\temp!

\def\sq!#1#2#3#4#5{%
\ll!=#1 \uu!=#2 \dd!=#3 \rr!=#4
\setbox0=\hbox{%
 \temp!=\squaresize\advance\temp! by .5\uu!
 \rlap{\kern -.5\ll!
 \vbox{\hrule height \temp! width#1 depth .5\dd!}}%
 \temp!=\squaresize\advance\temp! by -.5\uu!
 \rlap{\raise\temp!
 \vbox{\hrule height #2 width \squaresize}}%
 \rlap{\raise -.5\dd!
 \vbox{\hrule height #3 width \squaresize}}%
 \temp!=\squaresize\advance\temp! by .5\uu!
 \rlap{\kern \squaresize \kern-.5\rr!
 \vbox{\hrule height \temp! width#4 depth .5\dd!}}%
 \rlap{\kern .5\squaresize\raise .5\squaresize
 \vbox to 0pt{\vss\hbox to 0pt{\hss $#5$\hss}\vss}}%
}
 \ht0=0pt \dp0=0pt \box0
}

\def\vsq!#1#2#3#4#5\endvsq!{\vbox to \squaresize{\hrule
width\squaresize height 0pt%
\vss\sq!{#1}{#2}{#3}{#4}{#5}}}

\newdimen \LL! \newdimen \UU! \newdimen \DD! \newdimen \RR!

\def\vvsq!{\futurelet\next\vvvsq!}
\def\vvvsq!{\relax
  \ifx     \next l\LL!=\Thickness \let\continue=\skipnexttoken!
  \else\ifx\next u\UU!=\Thickness \let\continue=\skipnexttoken!
  \else\ifx\next d\DD!=\Thickness \let\continue=\skipnexttoken!
  \else\ifx\next r\RR!=\Thickness \let\continue=\skipnexttoken!
  \else\def\continue{\vsq!\LL!\UU!\DD!\RR!}%
  \fi\fi\fi\fi
  \continue}

\def\skipnexttoken!#1{\vvsq!}

\def\place#1#2#3{\vbox to 0pt{\vss
\rlap{\kern#1\squaresize
  \raise#2\squaresize\hbox{$#3$}}
\vss}}

\def\Young#1{\LL!=\thickness \UU!=\thickness \DD! = \thickness \RR! =
\thickness \vbox{\smallskip\offinterlineskip
\halign{&\vvsq! ##
\endvsq!\cr #1}}}

\def\blank{\omit\hskip\squaresize}
\catcode `!=12

\begin{document}

\sloppy
 
 \maketitle
 
\vspace*{-1cm}

\begin{abstract}
Random tilings of geometrical shapes with dominos or lozenges have been a rich source of universal statistical distributions. This paper deals with domino tilings of checker board rectangular shapes such that the top two and bottom two adjacent squares have the same orientation and the two most left and two most right ones as well. 
 It forces these so-called ``{\em skew-Aztec rectangles}" to have cuts on either side. For large sizes of the domain and upon an appropriate scaling of the location of the cuts, one finds {\em split tacnodes} between liquid regions with two distinct adjacent frozen phases descending into the tacnode. Zooming about such split tacnodes,  filaments appear between the liquid patches evolving in a bricklike sea of dimers of another type. This work shows that the random fluctuations in a neighborhood of the split tacnode are governed asymptotically by the {\em discrete tacnode kernel}, providing strong evidence that this kernel is a universal discrete-continuous limiting kernel occurring naturally whenever we have double interlacing pattern. The analysis involves the inversion of a singular Toeplitz matrix which leads to considerable difficulties.
\end{abstract}

 \tableofcontents

 \section{Introduction}

Random tiling models have been much investigated recently. In these models we tile a domain with tiles of a certain form possibly with different weights at different positions, typically we tile with lozenges or dominos. In many cases these models can also be thought of as dimer models on certain finite bipartite graphs. We are typically interested in limits when the size of the tiling region grows to infinity, and in the limit it is possible to look at the tiling patterns at different scales. To these tiling or dimer models we can also associate a height function, an integer valued function on the faces of the dimer model graph, which gives an interpretation of the dimer model as a random surface. At the global scale this height function converges to a deterministic limiting height function solving a variational problem, \cite{CKP}, and the region typically splits into domains with different phases. Phases refer to the different types of local limits, limiting Gibbs measures, that can be obtained. It has been shown in \cite{KOS} that in a large class of dimer models these fall into three types,
frozen, liquid (or rough) and gas (or smooth).  In this paper we are only concerned with a model that has no gas phase.
It appears, \cite{AstDuse}, \cite{KO}, that for a large class of dimer models in a domain, the boundary between the solid and liquid phases (arctic curves) is, asymptotically for the size going to infinity, a smooth curve possibly with singularities and some special points. It is shown in \cite{AstDuse} that in this class of tiling models the only singularities for the arctic curve are a cusp singularity, a tacnode or a tangency point of the arctic curve with the boundary of the domain (turning points). We have not verified that the model analyzed in this paper satisfies the conditions
in \cite{AstDuse}. Visually speaking, this means: the liquid region consists of patches in the domain, which may touch the boundary of the domain (turning points), which may intersect each other (leading to cusps along the arctic curve) or which may merely touch (tacnodes); touching can occur in two different ways: a soft and a split tacnode. 
At a {\em soft tacnode} we have the same frozen phase (same type of tiles) at both sides of the tacnode, whereas at a {\em split tacnode}, we have different frozen phases on the two sides. For a nice survey on random tilings, see \cite{Go}.
 
It is of great interest to study the random fluctuations around these limit shapes. Inside the liquid patches the height function is conjectured to fluctuate according to a Gaussian Free Field. This has been shown in a number of cases see e.g. \cite{Ke,BF,D,BuGo,Buf,BLR}.
We can also consider the fluctuations of the dimers at the interface between the liquid and frozen phases, or at the at points where the liquid region touches the boundary or another liquid region. These type of limits fall naturally into two types depending on whether one rescales in just one direction, {\em a discrete-continuous limit}, or in both directions, {\em a continuous-continuous limit}, see \cite{Jo16} for a discussion. 

There are three basic {\em continuous-continuous limits}. At generic points of the boundary of a liquid region the dimers fluctuate according to the {\em Airy kernel process}, \cite{OR1,Jo05c,Petrov2,Duse2}, at a cusp they behave as the {\em Pearcey process}, \cite{OR2,ACvM}, and at a soft tacnode the fluctuations are described by the {\em (continuous) tacnode kernel}, \cite{AJvM0}. The kernel for the soft tacnode process, was given in the context of non-intersecting random walks \cite{AFvM}, nonintersecting Brownian motions \cite{Jo13} and in the context of overlapping Aztec diamonds in \cite{AJvM0}. 
When an Airy process interacts with a boundary, a different process appears, which is called a hard tacnode process; see Ferrari-Vet\H o  \cite{FerVet}.

There are also three basic {\em discrete-continuous limits}. At turning points the natural limit process is given by the {\em GUE-minor process} (or GUE-corner process), \cite{JN,OR}, at split cusp points with two different frozen phases inside the cusp, we get the {\em Cusp-Airy process}, \cite{DJM}. Zooming about a split tacnode between the liquid regions one observes non-intersecting paths of dimers of definite types; they form {\em filaments}  between liquid patches evolving in a bricklike sea of dimers of another type \cite{ACJvM},\cite{AJvM1},\cite{AJvM2}. Here the limiting kernel is the {\em discrete tacnode kernel}. All these three cases are discrete in one direction and continuous in the other. Based on the classification of the regularity of the height function it is conjectured
in \cite{AstDuse} that the six interface processes discussed above are all that can occur in a large class
of dimer models.

The following questions arise about the asymptotic statistical fluctuations of the paths, when the size tends to infinity: 
\newline {\em (i) Does the discrete tacnode kernel indeed deserve the ``universality" label?}
  \newline {\em (ii) Is this a ``master kernel", from which all the existing kernels mentioned above can be derived by appropriate scaling limits?}

As to question {\em (i)}, the discrete tacnode kernel was found in the context of lozenge tilings of non-convex domains \cite{AJvM1,AJvM2}. The GUE-minor kernel describes
a one-sided pattern of interlacing particles, whereas the discrete tacnode kernel describes a double interlacing pattern, see also \cite{AvM1}.
The analysis was based on Skew-Young Tableaux of shape  $\lambda\backslash \mu$, filled with numbers $1,\dots,N$ for large $N$ when the size of  both, $\lambda$ and $\mu$,  grow appropriately with $N$; these tableaux fluctuate  according the discrete tacnode kernel \cite{AJvM1}. A special instance of the discrete tacnode kernel already  appeared in the context of two overlapping Aztec diamonds \cite{ACJvM} called the double Aztec diamond. In the present study it is shown that the same kernel appears in domino tilings of skew-Aztec rectangles, a model generalizing two overlapping 
 Aztec diamonds, see Theorem \ref{mainTh}. This is strong evidence that the discrete tacnode kernel is a universal limiting kernel that naturally occurs whenever we have double interlacing pattern of the type encountered here. The approach in this paper is completely different from that in \cite{AJvM1,AJvM2}. Here, we base the analysis on the approach in section 2.2.3 in \cite{Jo16}, where the difficulty is to invert a finite Toeplitz matrix. In the model studied here the symbol of this Toeplitz matrix has a zero at zero or infinity which leads to considerable technical difficulties.

The question {\em (ii)} is largely open. In a subsequent publication, we will show that an appropriate double scaling limit of the discrete tacnode kernel gives the Cusp-Airy kernel, see Theorem \ref{cuspAiryThm} below.

\section{Skew-Aztec Rectangles: geometry and main results}

 A skew-Aztec rectangle is a domain $\DR$ of width $n>0$ and length $ m+M\geq 1$ with two cuts (nonconvexities) symmetric about the middle of the rectangle along the sides $\pl \DR_R$ and $\pl \DR_L$, where $\pl \DR_U,\pl \DR_R,\pl \DR_D,\pl \DR_L$  denote the Up, Right, Down and Left sides of the rectangle; see Fig. 1. The number $m$ denotes the number of white squares and $M$ the number of blue squares on the $\pl \DR_R$ and the $\pl \DR_L$-side. Assume $m\geq 0$ and $M\geq 1$.  Notice that cuts are unavoidable, if the top two and bottom two adjacent squares of the rectangle have the same orientation (blue-white) and the two most left and two most right as well ($\mbox{\tiny blue}\atop\mbox{\tiny white}$) (as in Fig.1), unlike a regular Aztec diamond, where they have opposite orientations and thus no cuts.  
 
 We need two different system of coordinates on the skew-Aztec rectangle (with different origins): $(\xi,\eta)$ and $(s,u) $  related by 
 \be(\xi,\eta)\to (s,u)=\left( 
   \eta\! +\!1 ,~\tfrac 12 (\eta\!-\!\xi \!+
  \!1)\right), \mbox{with} ~-1\leq\xi\leq 2(m\!+\!M),~ -2\leq \eta\leq 2n+1\label{changevar}\ee
The middle of the most left boundary of the diamond serves as origin of the $(\xi, \eta)$ system, while the origin of the $(s,u)$-system is at  $(\xi, \eta)=(0,-1)$.  The middle of the blue squares are at $(\xi,\eta)\in 2\BZ\times (2\BZ+1)$, with $0\leq \xi\leq  2(m+M-1) $ and $-1\leq \eta\leq 2n$. In $(s,u)$ coordinates, they are at $(s,u)\in 2\BZ\times \BZ $, with $0\leq s\leq 2n$ and $-(m+M-1)\leq u\leq n $. One defines 

\newpage

\vspace*{-1.5cm}
\noindent
 two crucial integers  
\be\Dt:=n-m\in \BZ, \label{Dt}\ee and (the second formula in (\ref{sg}) follows at once from (\ref{Dt})) 
 \be \label{sg}
 \sg:=n-(m+M-n)+1=n-M+\Dt+1\in \BZ.
  \ee
$\Dt$ measures how much the Aztec rectangle differs from two overlapping Aztec diamonds (usual ones) and $\sg$ measures the ``amount of overlap", if one were to ignore the cuts and view the Aztec rectangle as two overlapping diamonds. Thus $\sg$ is the number of even points $\xi$ between the coordinates (and including) $\xi=2n$ and $\xi=2(m+M-n)$. Notice that $\sg$ can be negative, when two separated diamonds are connected with white and blue squares. In \cite{AJvM0}, we considered two usual  Aztec diamonds overlapping each other; then it is easily seen that $m=n$ and so $\Dt=0$. 

  Also consider the strip $\{\rho\}$ formed by the $\rho +1$ lines $\xi \in 2\BZ$ through the black squares  between (and including) the lines $\xi=2(M-1)$ and $\xi=2m$ defined by the cuts, with 
    \be \label{rho}
  \rho:=|m-(M-1)|.
  \ee
Fig. 1 gives an illustration for $m\geq M-1$ and Fig. 2 for $m<M-1$. 

Consider a covering of the skew-Aztec rectangle by means of dominos, horizontal ones with the blue square to the left or to the right $(H_L,~H_R)$ and vertical ones with the blue square on the upper-side or on the down-side $( V_U, ~V_D)$; see Fig. 5. Let these dominos be decorated as follows (as in Fig. 5): {\em (i)} with red lines and red dots, {\em (ii)} with blue lines and blue dots and for {\em (iii)} with green lines and green dots. Each domino carries an appropriately chosen height function, also indicated in Fig. 5, such that the red, blue and red lines are level lines of height $h+\tfrac 12$, for the corresponding height function. 
  
   It is easily seen that, given any domino covering by any of these sets {\em (i)}, {\em (ii)} (resp. {\em (iii)}), this produces a set of $n+m$ red/blue (nonintersecting) level lines (resp. $M$ green level lines extending all the way from the left edge of the blue (resp. white) boundary squares along $\pl \DR_U$ and $\pl \DR_L$ to the right edge of the white (resp. blue) boundary squares along $\pl \DR_R$ and $\pl\DR_D$. It is easily seen that the heights along the boundaries is specified, the same for {\em (i)} and {\em (ii)} and different for {\em (iii)}, but also fixed. For each of the models, the red, blue and green dots are intersection points (slightly moved to the middle of the blue square) of the level lines with the lines $\{\xi \in 2\BZ\}$ for {\em (i)} and $\{\eta \in \BZ_{\tiny odd}\}$ for {\em (ii)} and {\em (iii)}. Notice the duality of the blue paths and the green path (Fig. 5 {\em (ii)} and {\em (iii)}).  This duality interchanges the tiles $H_R $ and $H_L$ and the tiles $V_U$ and $V_D$; it amounts to a rotation by $180^{o}$.

\newpage 

\vspace*{-5cm}
\hspace*{-7cm}\setlength{\unitlength}{0.015in}\begin{picture}(-100,170)

\put(400,10){\makebox(0,0){\includegraphics[width=100mm,height=125mm]{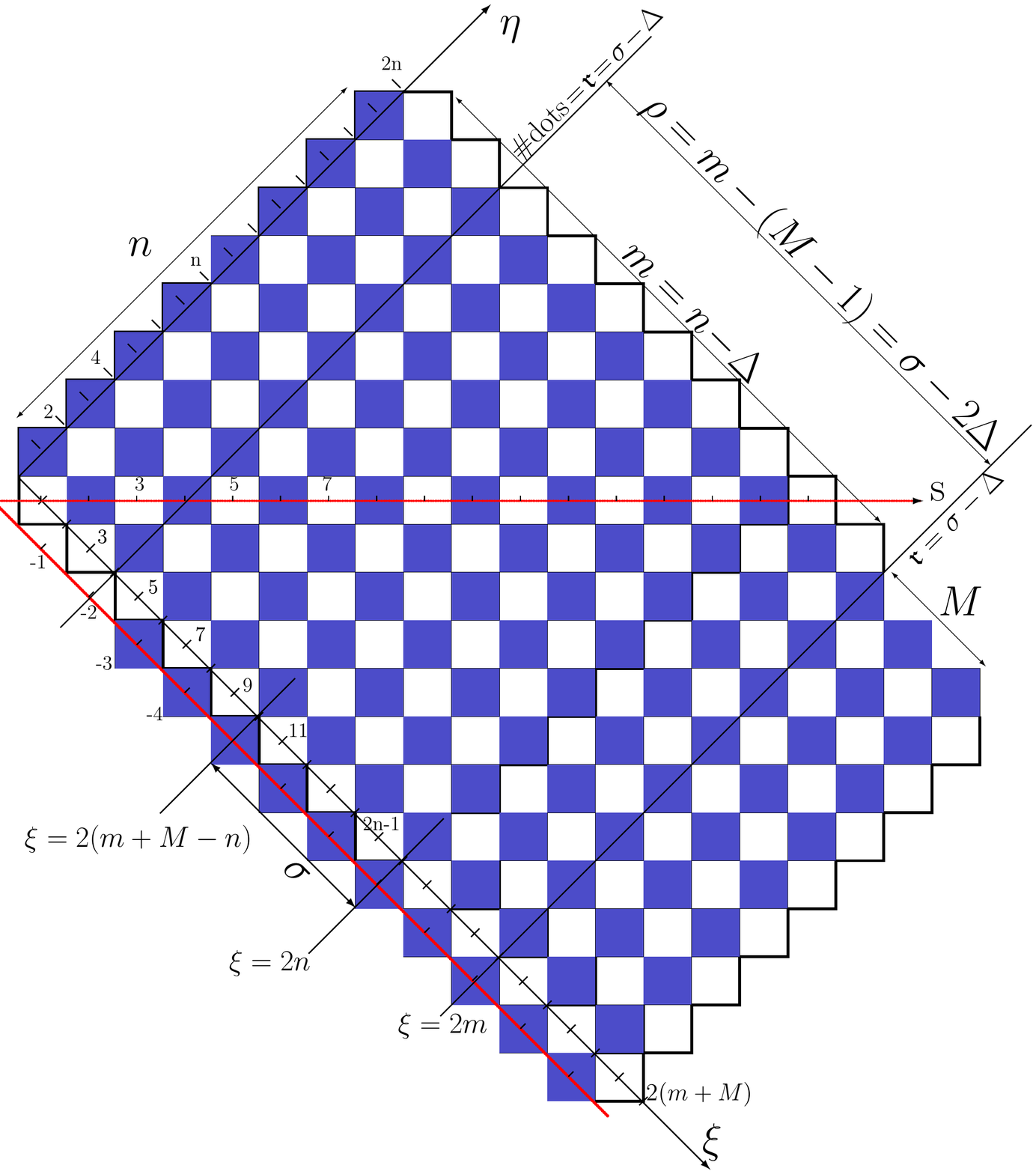}}}

 \multiput(270,30)(-1,1){50}{\makebox(0,0){\tikz\draw[red,fill=red] (0,0) circle (0.1ex);}}
 \put(222,78){\vector(-1,1){4}}
 \put(212,88){\makebox(0,0){\Large\mbox{u}}}
  \put(235,34){\makebox(0,0){\footnotesize{ $(s,u)=(0,0)$}}}

\end{picture}

\setlength{\unitlength}{0.015in}\begin{picture}(0,170)

 \put(100,-70){\makebox(0,0){\includegraphics[width=110mm,height=140mm]{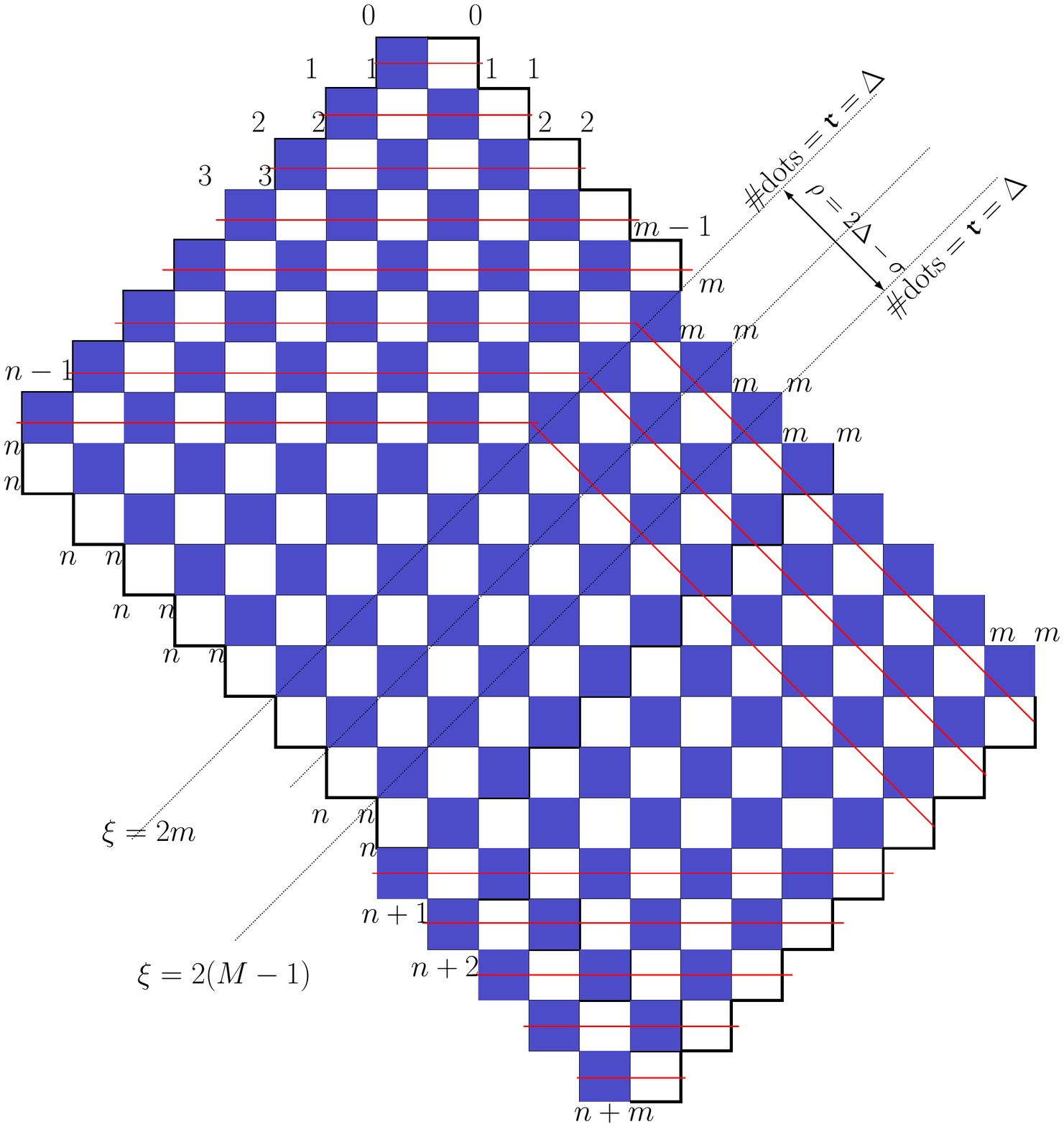}}}
 
 \put(300,80){$\pl \DR_D$}
 \put(60,140){$\pl \DR_L$}
  \put(60,300){$\pl \DR_U$}
   \put(320,300){$\pl \DR_R$}

    \put(70,170){\tiny\mbox{$\xi=2(M-1)$}} 
    
     \put(81,202.5){\tiny\mbox{$\bullet$}}

\end{picture}

\vspace*{8.0cm}

\hspace*{-1cm}Fig. 1: 
skew-Aztec rectangle with $m\geq M-1$,~$\Dt<0$ and coordinates $(\xi,\eta)$ and $(s,u)$. Here  $n=8,~m=10,~M=3,~\Dt=-2,~\rho=8,~\sg=4,~\rk=\sg-\Dt=6$.

\vspace*{.1cm}

\hspace*{-1cm}Fig. 2:  skew-Aztec rectangle with $m< M-1$ and $\Dt>0$, with height function  along the boundary, with nonintersecting red paths (level lines). Here $n=8,~m=5,~M=8,~\Dt=3,~\rho=2, ,~\sg=4,~\rk=\Dt=3~.$

\newpage

\vspace*{-4cm}
\hspace*{ 2cm}
\setlength{\unitlength}{0.015in}\begin{picture}(0,170)

 \put(190,20){\makebox(0,0){\includegraphics[width=120mm,height=155mm]{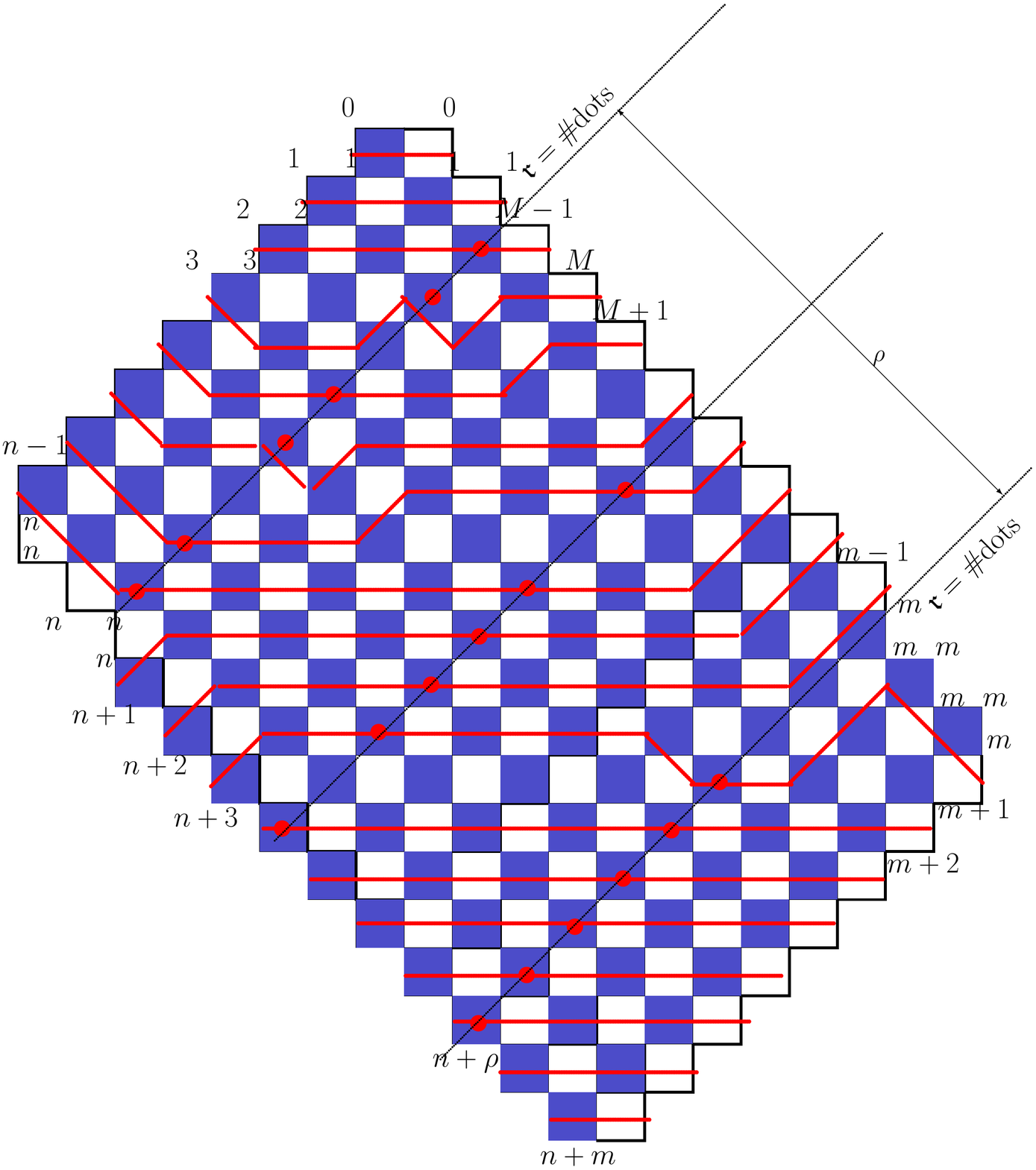}}}

\end{picture}

\vspace{2cm}\hspace*{-2cm}
\setlength{\unitlength}{0.015in}\begin{picture}(0,170)

\put(150,40){\makebox(0,0){\includegraphics[width=120mm,height=155mm]{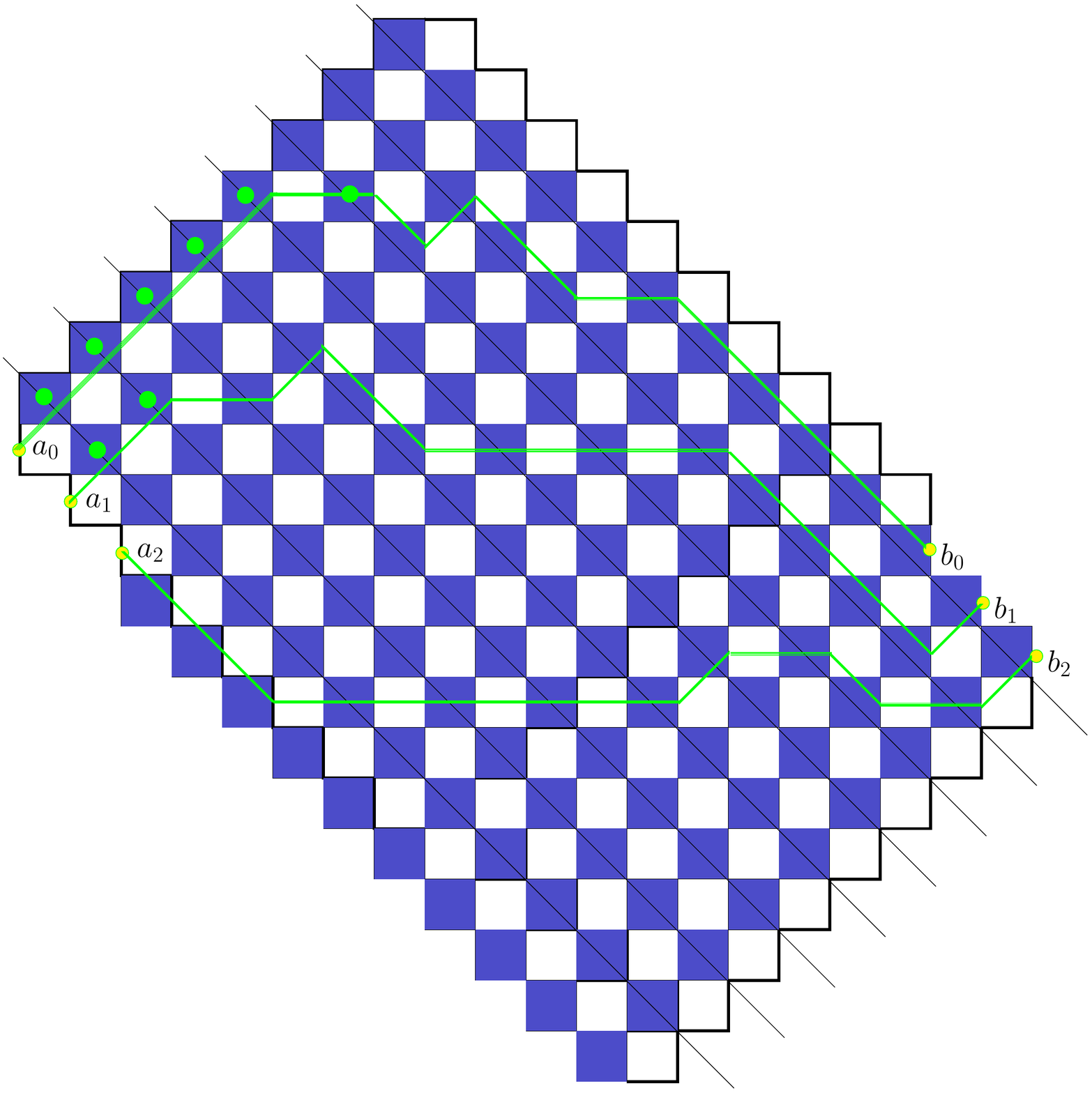}}}

 \end{picture}
 
\vspace*{4.5cm}

  \hspace*{-1cm}Fig.~3: Same Aztec rectangle as in Fig.1 with strip $\{\rho\}$ and with height function along the boundary, with  nonintersecting red paths and the corresponding red point process $\PR_{\tiny{red}}$.%

 \hspace*{-1cm}Fig. 4: Same Aztec rectangle as in Fig. 1, with nonintersecting green paths, starting and ending at {\em contiguous points} and the corresponding green point process $\PR_{\tiny{green}}$. The red and green paths in Figs. 3\&4 are dual of each other.
\newpage

\newpage

\vspace*{-3cm}

$~~~~~~~~~~~H_L~~~~~~~~~~~~~~~~~~ V_U~~~~~~~~~~~H_R~~~~~~~~~~~~~~~~V_D~~~~~~ 
$

$~~~~~~~~ Green ~~~~~~~~~~~~~Red~~~~~~~~~~~Yellow~~~~~~~ ~~~~Blue~~~~~
$
 
   \vspace*{-1.5cm}
   \setlength{\unitlength}{0.015in}\begin{picture}(0,0)
   
   \put(323,-25){\makebox(0,0){$ \mbox{Point process}$}}
   
   \put(0,-70){\makebox(0,0){$(i)~\PR_{{\mathcal R}ed} $}}
   
    \put(330,-70){\makebox(0,0){$ \begin{array}{llll}\mbox{red path} \\\cap   \{ \xi\in 2\BZ \}\end{array}$}}

\put(160,-70){\makebox(0,0){\includegraphics[width=132mm,height=170mm]{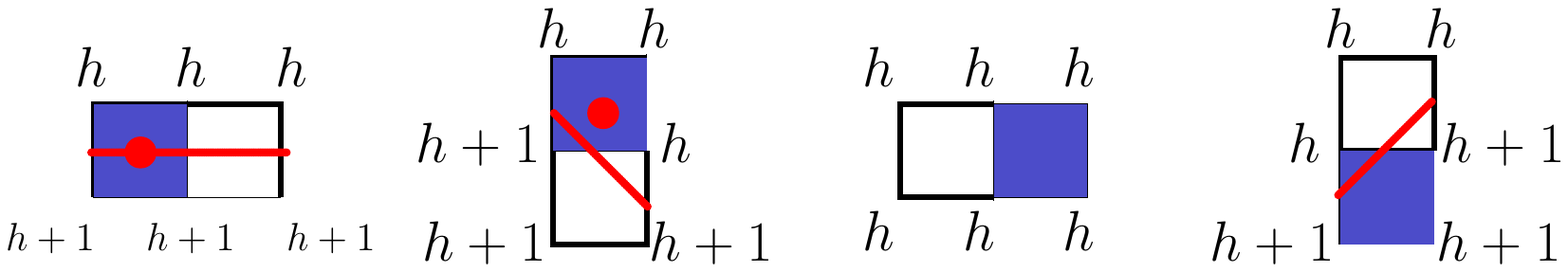}}}

\put(7,-95){\makebox(0,0){$\Uparrow \mbox{\tiny Kasteleyn}$ }}

\put(0,-120){\makebox(0,0){$(ii)~\PR_{{\mathcal B}lue} $}}

\put(160,-120){\makebox(0,0){\includegraphics[width=132mm,height=170mm]{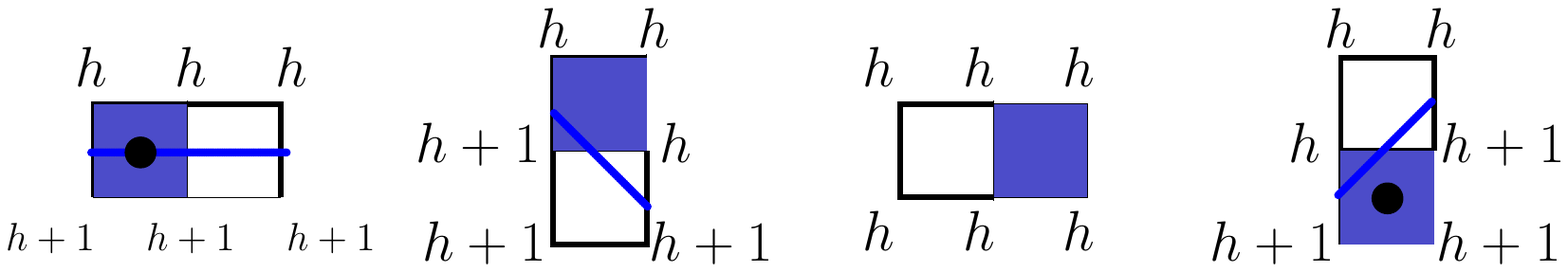}}}

 \put(330,-120){\makebox(0,0){$ \begin{array}{lllll}\mbox{blue path} \\\cap   \{ \eta\in  \BZ_{ \mbox{\tiny odd}} \}\end{array}$}}
 \put(7,-145){\makebox(0,0){$\Uparrow \mbox{\tiny Duality}$ }}
 
 \put(0,-170){\makebox(0,0){$(iii)~\PR_{{\mathcal G}reen} $}}

 \put(160,-170){\makebox(0,0){\includegraphics[width=132mm,height=170mm]{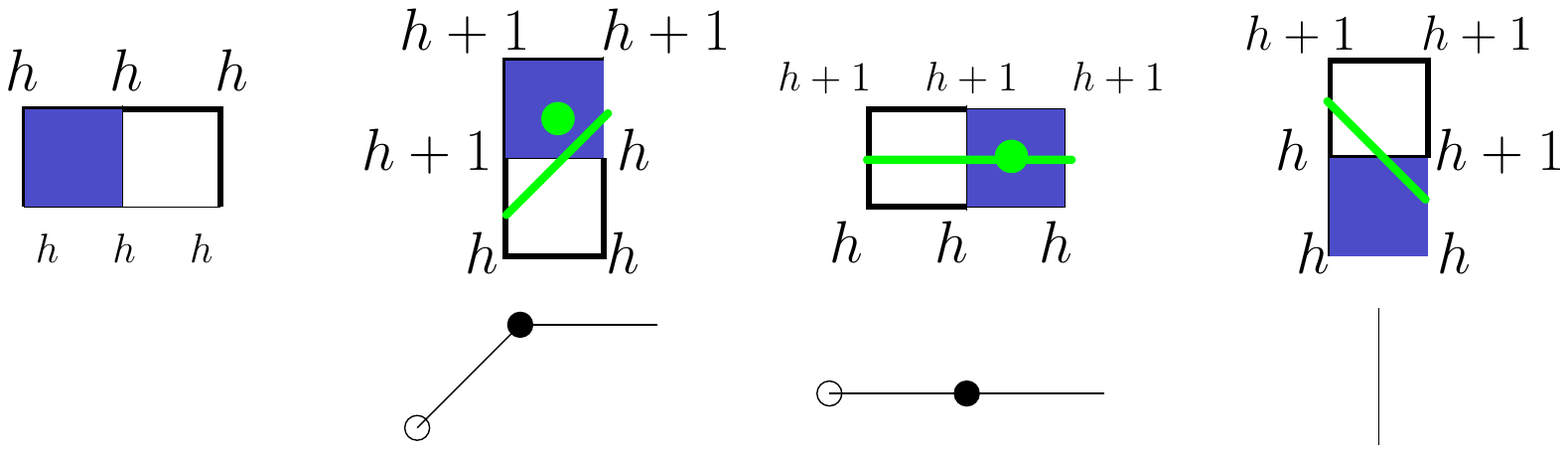}}}

   \put(330,-170){\makebox(0,0){$ \begin{array}{lllll}\mbox{green path} \\\cap   \{ \eta\in  \BZ_{ \mbox{\tiny odd}} \}\end{array}$}}
    \put(0,-187){\makebox(0,0){$\Updownarrow 
     $ }}
  
  \put(0,-205){\makebox(0,0){$(iv)~\PR_{{\mathcal G}reen} $}}
  
   \put(330,-205){\makebox(0,0){$ \begin{array}{lllll}\mbox{\footnotesize from green paths, Fig. 4 } \\ \mbox{\footnotesize to paths in Fig. 7}\end{array}$}}

\end{picture}

 \vspace{8.4cm}
 Fig. 5: Dominos with 4 different orientations: two horizontal ones with blue square to the left/right ($H_L,H_R$) and two vertical ones with blue square up/down ($V_U,V_D$). Three different point processes $\PR_{{\mathcal R}ed}$, $\PR_{{\mathcal B}lue}$, $\PR_{{\mathcal G}reen}$, associated  with the intersection of the level lines (for the given height functions) with the lines $\{\xi\in 2 \BZ\}$ and $\{\eta\in  2\BZ+1\}$.
 
  \vspace*{.6cm}
  
   This also shows for each of the cases the one-to-one correspondence between the decorated red, blue, green tilings of $\DR$   and the red, blue, green level lines; see e.g., \cite{Jo05c,AJvM0}.
    
  Notice from Fig.4, that the $M$ green paths leave from and end up at contiguous points, whereas the $n+m$ red/blue paths leave from a non-contiguous set. In \cite{ACJvM}, we called the green paths inliers, and the red/blue paths outliers.

  Putting the customary probability on the tilings by giving the weight $0<a\leq 1$ on vertical dominos and weight $1$ on horizontal dominos, 
one obtains three point processes $\PR_{{\mathcal R}ed}$, $\PR_{{\mathcal B}lue}$, $\PR_{{\mathcal G}reen}$ of red, blue and green dots. This purpose of this paper will be to show that $\PR_{{\mathcal R}ed}$ is a determinantal point process, of which an appropriate  asymptotic limit will be given by the discrete tacnode kernel ${\mathbb L}^{\mbox{\tiny dTac}}_{\rk,\rho,\beta} $ as in (\ref{Ldtac}).

   Notice that Aztec rectangles may not always be tilable; they will be in the cases below. Also we shall express  $\rho, ~\rk$ in terms of the basic geometrical data $n, ~m,~M$.

  \newpage
   \newpage
  
\vspace*{-2cm}\hspace*{-3cm}
\setlength{\unitlength}{0.015in}\begin{picture}(0,170)

\put(120,40){\rotatebox{-45}{\makebox(0,0){\includegraphics[width=158mm,height=79mm]{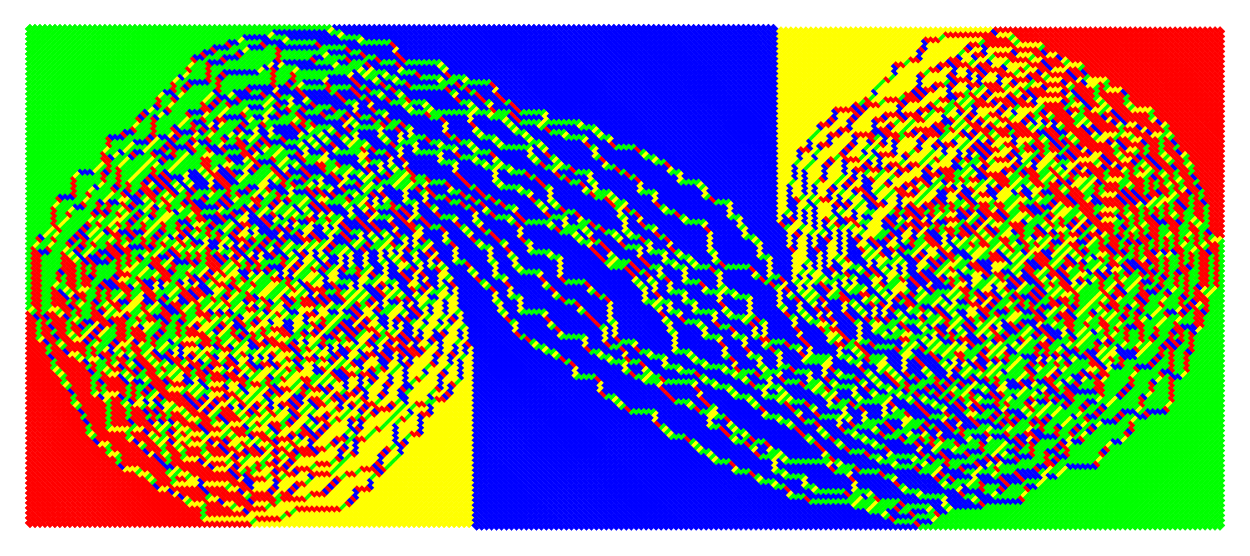}}}}

\put(-14.2,-22){\line(1,1){170}}
\put(57,-94){\line(1,1){170}}

  
  \put(130,190){\makebox(0,0){(i)}}
   \put(430,190){\makebox(0,0){(ii)}}
   \put(-10,-100){\makebox(0,0){(iii)}}
    \put(285,-100){\makebox(0,0){(iv)}}
  
   \put(390,60){\rotatebox{-45}{\makebox(0,0){\includegraphics[width=158mm,height=79mm]{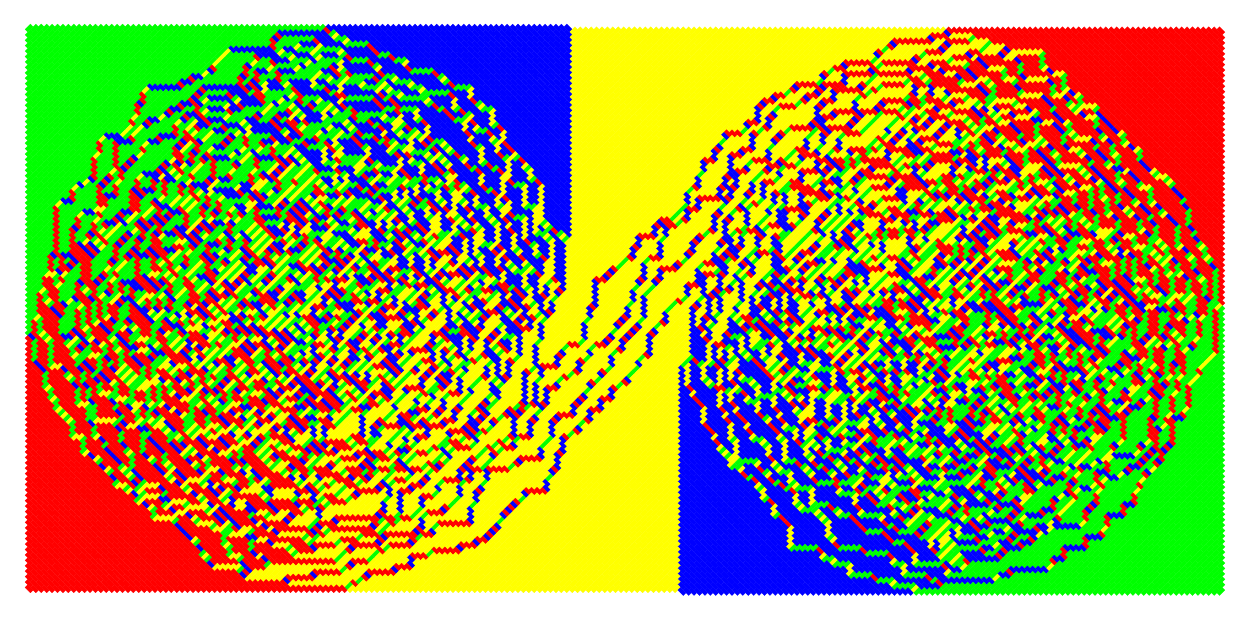}}}}
 
 \put(304,0){\line(1,1){150}}
\put(329,-26){\line(1,1){150}}

 \put(130,-260){\rotatebox{-45}{\makebox(0,0){\includegraphics[width=158mm,height=79mm]{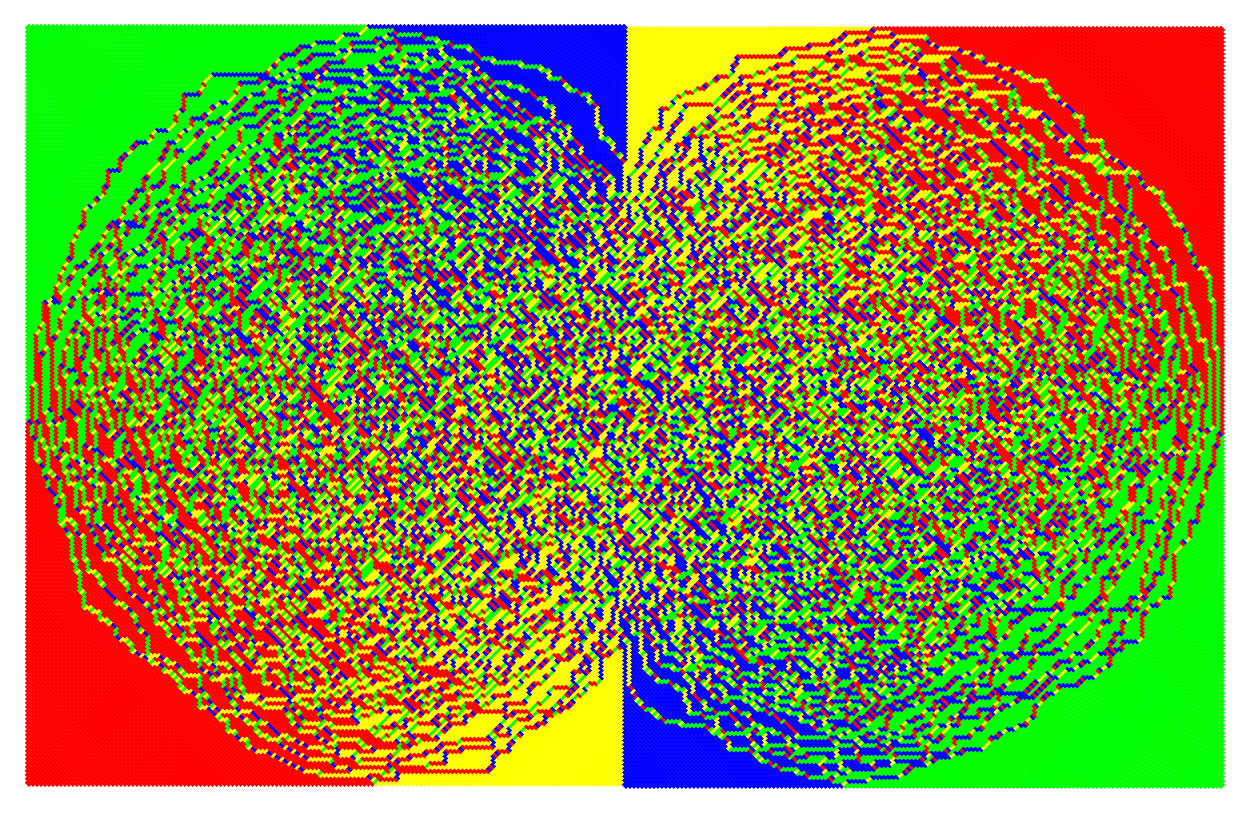}}}}

  \put(420,-260){\rotatebox{-45}{\makebox(0,0){\includegraphics[width=158mm,height=79mm]{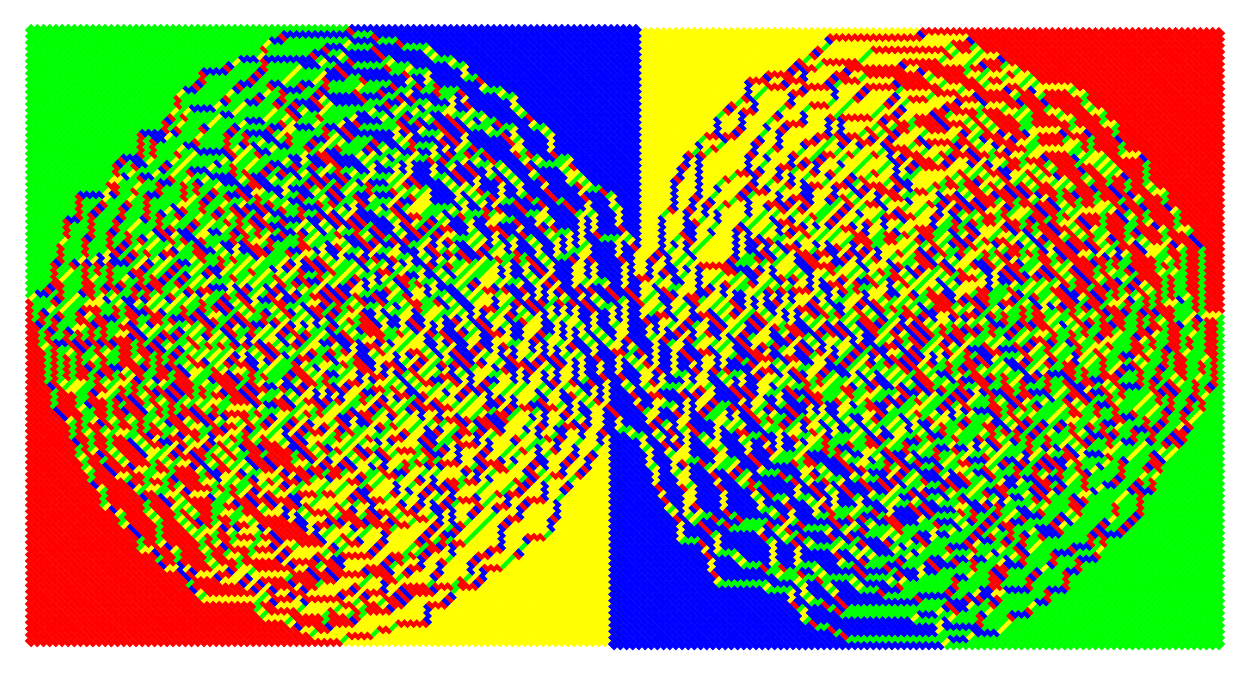}}}}
       
 \end{picture}
 
 \newpage
 
 \newpage
 
 Fig. 6. Simulations: {\em (courtesy of Sunil Chhita)}

$\begin{array}{cccccccccccc}
\mbox{Fig.}&\mbox{type}& n&m&M&\rho&\rk&\Delta=n-m&\vline&\mbox{filaments}
\\ \hline 
 \\(i)& m\geq M-1&100 &  150 &   90 & 61&  11&\Dt<0&\vline    &\mbox {all but blue}
 \\(ii)&m<M-1& 104 &  100 &   121 & 20&  4&\Dt>0&\vline   &\mbox {all but yellow}
\\ (iii)& m\geq M-1& 190&150&150&1&41&\Dt>0&\vline   &\mbox {all but blue}
\\(iv)& m\geq M-1&100&99&95&5&6&\Dt>0 & \vline   &\mbox {all but blue}
& \end{array}$

     \bigbreak

 \begin{theorem}\label{tilable}
The skew-Aztec rectangle $\DR$ is tilable iff 
\be\label{case1,2}\bl \mbox{ Case 1:} &~~~  1\leq M~ \leq ~\min (m  ,n +1 )   
\\ 
\mbox{or  Case 2:}&~~~0\leq ~m~\leq ~\min (M-1,n ).
\el\ee
The associated point process $\PR_{\mathcal R ed}$ consists of $\rk$ (red) dots on each of the $\rho+1$ lines $\xi \in 2\BZ$ within the strip $ \{\rho\}$ of width $\rho$, with the number of dots per line increasing one-by-one on either side of the strip, up to $n$, where 
 \be \label{rho,r} 
 \bl \rho&= |m-(M-1)| \\&= |\sg-2\Dt |\el
~~~ \mbox{and}~~~
 \bl \rk
  &=\max (n-(M-1), ~n-m)\\&=\max(\sg-\Dt,\Dt).\el 
   \ee
To be more specific, we have
 \newline Case 1. $ \rho=m-(M-1)=\sg\!-\!2\Dt\geq 1$ ~and ~ $\rk=n\!-\!(M\!-\!1)=\sg\!-\!\Dt\geq 0$ 
  \newline  Case 2.  $\rho=(M-1)-m=2\Dt- \sg \geq  0$ and ~ $\rk=n-m=\Dt\geq 0$.
 \vspace*{-.4cm} \be\label{rho,r'}\ee 

\end{theorem}
 
\noindent \underline  {\em Remark 1}: 
 Throughout the paper we deal with case 1 of Theorem \ref{tilable} and $\Dt\leq 0$; for this case $\rho$ and $\rk$ are given by (\ref{rho,r'}), with $\rho\geq\rk\geq 1$. 
Setting throughout $\ka:=-\Dt\geq 0$, we have $\rk=\ka+\sg$, $\rho=\ka+\rk$.
 
 \medbreak
 One then introduces the following {\em discrete-continuous} scaling into $\BK_{ n,\rk,\rho}^{^{\textrm{\tiny red}}}
 (\xi_1,\eta_1;\xi_2,\eta_2)
 $, letting $n\to \infty$ and 
  keeping $\rk=n-M+1$ and $\rho=m-M+1$ fixed:
 \be \label{scaling'}
\mbox{scaling}:~  \xi_i := 2m -2\tau_i,~
\eta_i :=  n-1+  y_i\sqrt{n}  ,~a=1+\frac{\beta}{\sqrt{n}} ,~\frac {\Dt \eta}2=\frac{dy}{2}\sqrt{n}
\ee

  For $ \tau_i\in \BZ$ and $y_i\in \BR $, the {\em discrete tacnode kernel} depending on the geometric parameters $\rho,\rk$ and the scaling parameter $\beta$, is as follows:
 \be \label{Ldtac}\bl
 {\mathbb L}^{\mbox{\tiny dTac}}_{\rk,\rho,\beta} & ( \tau_1, y_1 ;\tau_2, y_2)  
  =  - 
{\mathbb H}^{\tau_1-\tau_2} (  y_1-y_2)   
\\& +\oint_{\Ga_0}\frac{du} {(2\pi\I)^2}\oint_{\uparrow L_{0+}}  \frac{ dv}{v-u}\frac{u^{\rho-\tau_1}}{v^{\rho-\tau_2}}
\frac{e^{-\frac{u^2}2  +(\beta+y_1 )  u  }}
{e^{-\frac{v^2}2  +(\beta+y_2 )  v  }}
      \frac{ \Theta_r( u, v )} { \Theta_r(0,0)}  
\\& + \oint_{\Ga_0}\frac{du} {(2\pi\I)^2}\oint_{\uparrow L_{0+}}\frac{ dv}{v-u} \frac{u^{  \tau_ 2}}{v^{  \tau_1}}
\frac{e^{-\frac{u^2}2  +(\beta-y_2 )  u  }}
{e^{-\frac{v^2}2  +(\beta-y_1 )  v  }}
      \frac{ \Theta_r( u, v )} { \Theta_r(0,0)}  \\& 
      + \oint_{\uparrow L_{0+}  }    \frac{du} {(2\pi\I)^2} \oint_{\uparrow L_{0+}}dv \frac{u^{ -\tau_1}}{v^{\rho-\tau_2}}
\frac{e^{ \frac{u^2}2  -(\beta-y_1 )  u  }}
{e^{-\frac{v^2}2  +(\beta+y_2 )  v  }}
    \frac{ \Theta^+_{r-1}(  u, v )} {  \Theta_r(0,0)} 
 \\
 & -  \oint_{\Ga_0}\!\frac{du} {(2\pi\I)^2} \!\oint_{\Ga_0}\!dv 
 \frac{u^{ \rho-\tau_1}}{v^{ -\tau_2}}
\frac{e^{ -\frac{u^2}2  +(\beta+y_1 )  u  }}
{e^{ \frac{v^2}2  -(\beta-y_2 )  v  }}
    \frac{ \Theta^-_{r+1}( u,v  )} {  \Theta_r(0,0)}   
\el\ee
where  
 $\Ga_0=$ small circle about $0$  and 
$\uparrow L_{0+} =$  upgoing vertical line in $\BC$ to the right of $\Ga_0$. The Heaviside function and the functions $\Theta_\rk$ appearing in (\ref{Ldtac}) are as follows:
 \be \bl\BH^{m}(z)  :=\frac{z^{m-1}}{(m-1)!}\Id _{z\geq 0}\Id_{m\geq 1},~~~~(\mbox{Heaviside function}).
 \label{Heaviside}\el\ee
 \be\bl\Theta_\rk( u,v)&:=\frac1{\rk!}\left[     
\prod_1^\rk \oint_{\uparrow L_{0+}}\frac{e^{     w_\al^2-2\beta   w_\al}}{    w_\al  ^{\rho }}
 ~\left(\frac{v\!-\!w_\al}{u\!-\!w_\al}\right) \frac{dw_\al}{2\pi \I}\right]\Dt_\rk^2(w_1,\dots,w_r)
 \\  \Theta^{\pm}_{\rk\mp1}(  u,v)     &:=\frac{1}{(\rk\mp1)!}\left[     
\prod_1^{ \rk\mp 1  }\oint_{\uparrow L_{0+}}\frac{e^{     w_\al^2-2\beta  w_\al}}{    w_\al  ^{\rho }}
 ~\left( ({u\!-\!w_\al} )\ ({v\!-\!w_\al})\right)^{\pm 1} \frac{dw_\al}{2\pi \I}\right]
 \\
 &~~ ~~\qquad~ ~\qquad \qquad~~~~~~~~~~~~~~~~~~~~~~~~~~\Dt_{\rk\mp1}^2(w_1,\dots,w_{r\mp1}).
 \label{Theta} 
\el\ee

The next Theorem shows the {\em universality} of the tacnode kernel: it is identical to the asymptotics obtained in the very different context of lozenge tilings of hexagons with nonconvexities and skew Young diagrams; see \cite{AJvM1,AJvM2}.  From \cite{AvM1,AvM2},  the distribution of the red dots along the oblique lines $\xi\in 2\BZ$ is known, including their joint distributions and, in the case $\Dt=0$, it is related to coupled random matrices for a certain coupling \cite{AvM1}.

 \begin{theorem}\label{mainTh}
 Given the scaling (\ref{scaling'}) in the situation of remark 1 (just after formula (\ref{rho,r'})) the kernel $ {\mathbb K}_{ n,\rk,\rho}^{red} $ tends asymptotically  for $n  \to\iy$ to ${\mathbb L}^{\mbox{\tiny dTac}}$ as given in (\ref{Ldtac}),  where $\tau_i=m- \frac{\xi_i}{2}\in \BZ$ and $y_i\in \BR $, 
  $$ \begin{aligned}
   \lim_{t\to 0} \frac{2 (-a)^{\frac{\eta_1-\eta_2}{2}}}{1+a^2}(\tfrac{a}{\sqrt{n}})^{\frac{\xi_2-\xi_1}{2}}{\mathbb K}_{ n,\rk,\rho}^{red} &(\xi_1,\eta_1 ; \xi_2,\eta_2) 
\frac{\Dt \eta_2}{2} \Bigr|_{\footnotesize\mbox{scaling}} \!\!\! =
   {\mathbb L}^{\mbox{\tiny dTac}}_{\rk,\rho,\beta}  ( \tau_1, y_1 ;\tau_2, y_2) dy_2 
    \end{aligned}
      $$
     
\end{theorem}

The case $\ka=-\Dt=0$ has been completely studied  in \cite{AJvM0}. We conjecture that the same limiting statement (Theorem \ref{mainTh}) holds for case 1 and $\Dt>0$; then $\rho<\rk$. For case 2 of Theorem \ref{tilable}, we expect the limit to be the same kernel ${\mathbb L}^{\mbox{\tiny dTac}}_{\rk,\rho,\beta} $, but instead with $\rho= (M-1)-m$ and $\rk=n-m$.

\medbreak

{\em Outline of the paper:} 
One first computes the inlier kernel $\BK_{ n,\rk,\rho}^{^{\textrm{\tiny green}}}
 (s_1,u_1;s_2,u_2)
 $ in the coordinates $(s,u)$ of Fig. 1. The fact that $\Dt\neq 0$ is responsible for the presence of a Toeplitz determinant for a singular symbol, a special instance of a Fisher-Hartwig singularity. This can be remedied, in the case $\Dt<0$ by blowing up  
  the singularity $\ze^{-\Dt}\to \prod_1^{-\Dt}(1-\frac{\ze}{\lb_i})$, followed by integration with regard to a measure $d\mu(\lb) $. The next step is deduce the outlier kernel $\BK_{ n,\rk,\rho}^{^{\textrm{\tiny blue}}}
 (s_1,u_1;s_2,u_2)
 $ by duality and to deduce from it the $\BK_{ n,\rk,\rho}^{^{\textrm{\tiny red}}}
 (\xi_1,\eta_1;\xi_2,\eta_2)
 $. This is done by computing the inverse of the Kasteleyn adjacency matrix from $\BK_{ n,\rk,\rho}^{^{\textrm{\tiny blue}}}$ in the $(\xi,\eta)$-coordinates of Fig 1; see also (\ref{changevar}).  The $\BK_{ n,\rk,\rho}^{^{\textrm{\tiny red}}}$ will then be given in (\ref{Lkernel}) and in (\ref{Lsca} after inserting the new scale. In section 5, we reduce what seems an infinite-dimensional problem to a finite-dimensional one. This enables us to give in section 6 an integral representation of the kernel and to pull through the $d\mu(\lb)$-integration mentioned earlier. In the end one is led to a  quite simple kernel (section 8); contours have to be interchanged appropriately to be able to take a limit (section 9). Section 9 deals with the final limit.
  
\medbreak
In \cite{AJvM1} we conjectured that ${\mathbb L}^{\mbox{\tiny dTac}}$ is also a {\em master kernel}; i.e., all the kernels and corresponding statistics obtained at critical points of the arctic boundary in tiling models are appropriate scaling limits of the discrete tacnode kernel ${\mathbb L}^{\mbox{\tiny dTac}}$.
 Such a prominent example is the cusp-Airy kernel ${\mathbb L}_{}^{cusp-Airy}$, which appears near a split-cusp of the arctic boundary, where the frozen region splits into two regions; see \cite{DJM}. We shall need 
 the Airy-like function, for $\tau\in \BZ$ and given the contour ${{\nwarrow}\atop { \nearrow}}=\iy e^{4\pi \I /3}\mbox{ to } \iy e^{2\pi \I /3}$,  
$$A_\tau(u)=\oint_{{\nwarrow}\atop { \nearrow}} \frac{z^\tau dz}{2\pi \I}e^{-\frac{z^3}{3}+uz  } .
$$

We state the following Theorem, which will be shown in a subsequent publication \cite{AJvMCusp}. It deals with the limit of the discrete tacnode kernel to the cusp-Airy kernel for $\beta=0$ and $\rho=0$ (i.e., the strip $\{\rho\}$ reduces to one line $\xi=2m$ or, what is the same, $\tau=0$). It describes the statistical behavior of the red dots at a distance $ \sqrt{2\rk}$ along the lines $\tau\in \BZ$, when $\rk\to \iy$. The tip of the cusp belongs to the line $\tau=0$  in that regime. Fig. 6(iii) gives an example of a split cusp for $\rho=1$ instead. So, zooming about the split cusp, one finds:

\begin{theorem}\label{cuspAiryThm}
Setting\footnote{Together with the change $y_i\to -\frac{y _i}{\sqrt{2}}$, one rescales all integration variables $u,v,w$ as $u\to \sqrt{2}u,~v\to\sqrt{2}v,~w \to \sqrt{2}w$.}$$\widetilde{\mathbb L}^{\mbox{\tiny dTac}}(\tau_1,y_1;\tau_2,y_2):=(\sqrt{2})^{\tau_1-\tau_2-1}{\mathbb L}^{\mbox{\tiny dTac}}_{ \rk,\rho,\beta}(\tau_1,-\frac{y _1}{\sqrt{2}};\tau_2,-\frac{y _2}{\sqrt{2}})\Bigr|_{\rho=0,\beta=0},$$   the following limit holds:
$$\bl 
\lim_{\rk\to \infty}(-\rk^{\frac{\tau_1-\tau_2}{6}})
 &\widetilde{\mathbb L}^{\mbox{\tiny dTac}}  ( \tau_1, y_1 ;\tau_2, y_2) \Bigr|_{y_i= 2\sqrt{\rk}+\frac{\xi_i}{r^{1/6}}} =
  {\mathbb L}_{}^{cusp-Airy} (\tau_1,\xi_1;\tau_2,\xi_2)
 \el $$
 where 
  $$  {\mathbb L}_{}^{cusp-Airy}:=  - 
{\mathbb H}^{\tau_1-\tau_2}(  \xi_1-  \xi_2) 
+(-1)^{\tau_2}\int_0^\infty
A_{-\tau_1}(\xi_1+\lb)A_{ \tau_2}(\xi_2+\lb)d\lb
$$
for $\tau_1,~\tau_2\in \BZ$, except for $\tau_1\geq 0$ and $\tau_2< 0$, for which the expression is more complicated; see \cite{DJM}.


\end{theorem}

   
   \section{Tilability and the numbers $\rho$, $\rk$}

  \noindent {\em Proof of Theorem \ref{tilable}:} Consider consecutive green paths $\pi_i$ (as in Fig.4)  starting from $a_i$ and ending at $b_i$. Here we pick the  $i=0$th as the closest to the sides $\pl \DR_U$ and $\pl \DR_R$ (see Fig.1), for $i=1$ the closest to the path $\pi_0$, etc..., up to $i=M-1$. To visualize this, we  make the $(s,u)$-coordinates rectangular and define $(x,y)=(s-\tfrac 12,u)$, thus transforming these green level lines $\pi_i$ into the ones of Fig. 6, where we used the path transformation given in Fig.5(iii).

 In the $(x,y)$-coordinates, $  a_k=(-1,-k)$ and $ b_k=(2n,\Dt-k)$ with $0\leq k\leq M-1 $, together with  the corresponding paths $  \pi_{k-1 } $ as in Fig.7. For tilability we first need
  \be \label{cond1}n\geq \Dt=n-m, \mbox{and so} ~~ m\geq 0,\ee 
   since otherwise the point $  b_0$ could not be reached by $ \pi_{0} $.



  \vspace{.4cm}
  


\newpage

\vspace*{-1.5cm}

\setlength{\unitlength}{0.015in}\begin{picture}(0,0)

\put(155,  -70){\makebox(0,0){\includegraphics[width=130mm,height=150mm]{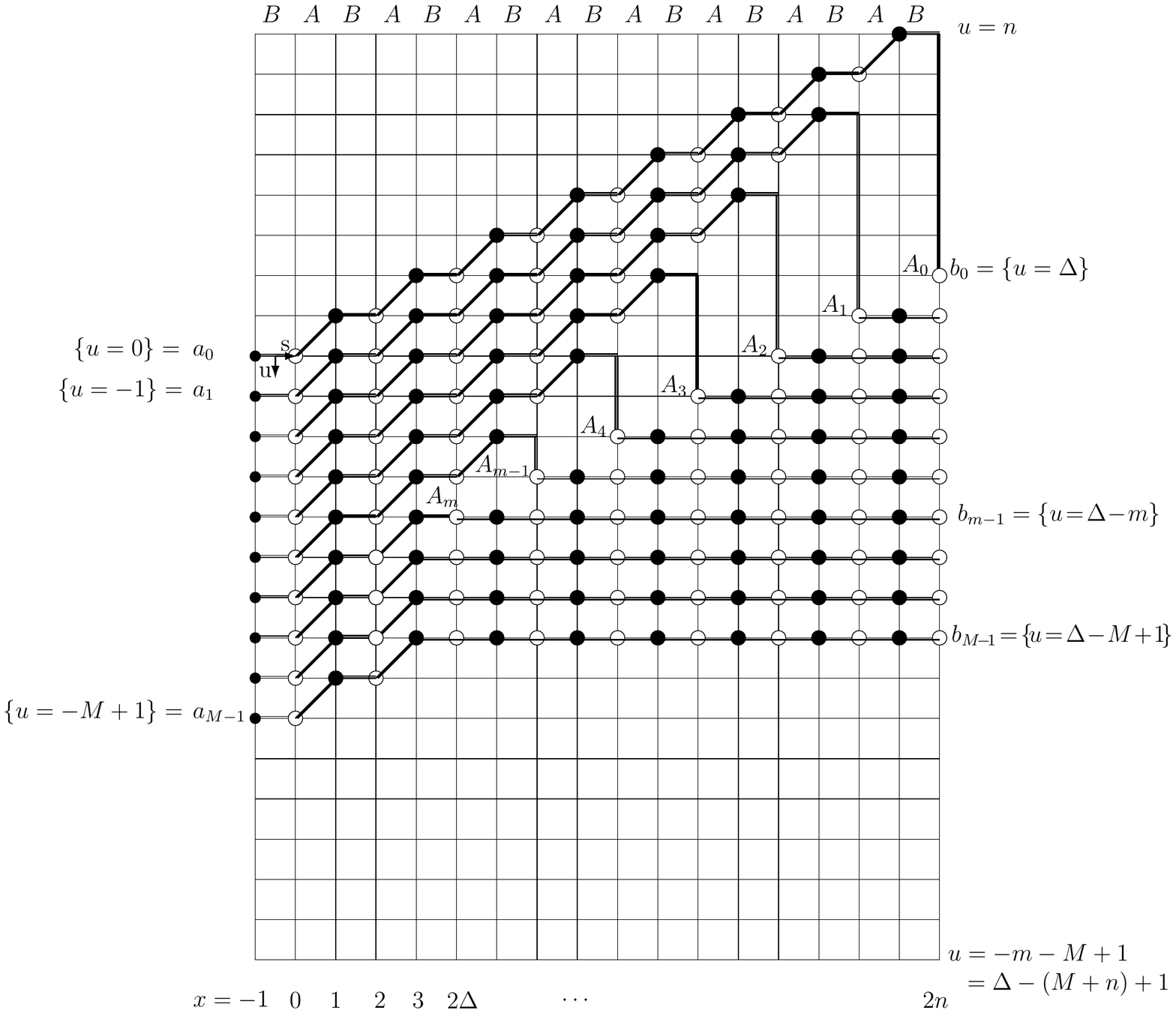}}}

\end{picture}

\vspace*{8cm}
Fig. 7: Special set of nonintersecting green paths connecting $a_i$ and $b_i$ for $0\leq i\leq M-1$ in $(x,y)=(s-\tfrac 12 , u)$ coordinates (inlier paths as in Fig.4 and $(s,u)$-coordinates as in Fig.1) for $n=8$, $m=6$,~$M=10$ and so $\Dt=2$. 

\vspace*{.8cm}

  The consecutive nonintersecting paths $\pi_i$ are maximal up to $(2n,n)$ and down to $(2n,\Dt)$, and up to $(2n-2,n-2)$ and down to $A_1$ to reach $(2n,\Dt-1)$ and so so forth for the next one. The dropdown stretches, moving from the right to left, have length $n-\Dt=m,~m-1,~m-2,\dots$. We now distinguish two cases:
 \bigbreak
    \noindent {\bf Case 1}: $m\geq M\geq 1$. Then the lowest points $A_k$ of the vertical stretches have coordinates $A_k=A_0-(2k,k)=(2n,\Dt)-(2k,k)$ for $0\leq k\leq M-1$. Then, since $\Dt=n-m$, we have 
    $$A_{M-1}=A_0-\bigl(2(M-1),M-1\bigr)=\bigl(
    2(n-M+1),n-m-M+1\bigr).$$ 
    If $A_{M-1}$ would belong to the grid, that would prove the existence of $M$ nonintersecting paths. So, belonging to the grid implies a condition on both coordinates:  $u=n-m-M+1\geq -m-M+1=\{\mbox{the $u$-coordinate of the lowest line of the grid}\}$, which is automatic (since $n\geq 1$), and $x=n-M+1\geq 0$, which is a condition.   
      \newline {\bf Case 2}: $0\leq m<M$. 
     Then, again moving from the right to the left (assuming $m>0$), the vertical down-stretches of the $M$ paths $\pi_k$ will have length $m-k>0$ for $0\leq k\leq m-1$, with the down-stretch of path $\pi_m$ having 0-length for $k=m$. The coordinates of the lowest points of these stretches have coordinate $A_k=A_0-(2k,k)$ for $0\leq k\leq m$ and so
     $$A_m=A_0-(2m,m)=\bigl(2(n-m), n-2m\bigr),$$
     which also must belong to the grid; i.e., $n-2m\geq -m-M+1$, which this time is automatic from the Case 2-condition $M-m>0$ and also we must have $n-m=\Dt\geq 0$, which together with the case 1-condition  amounts to condition (\ref{case1,2}). In the extreme case $m=0$, all the paths will move in parallel to the top $M$ locations along the line $x=2n$ and so will have no vertical stretches.
     The necessity of these conditions follows from the fact that if the rectangle $\DR$ is tilable, then in either case $\rk\geq 0$, as will be discussed below.

     To prove (\ref{rho,r}), we reconsider the two cases:
 \newline {\bf Case 1}:  Here the strip has width    
  $$\rho =m-(M-1)=\sg-2\Dt \geq 1.
 $$The second identity follows from substituting $n$ and $M$ from (\ref{Dt}) and (\ref{sg}). 
  The number of red dots along $\xi\in 2\BZ$ equals the number of intersection points of the lines $\xi\in 2\BZ$ with the red paths and also equals the difference of heights along the lines  $\xi\in 2\BZ$, measured between $\pl \DR_L$ and $\pl \DR_R$; see Fig. 2 and Fig. 5(i). The following table gives three distinct regions: the lines, above, within and below the strip $\{\rho\}$. 
 
 $$\hspace*{-2cm}\begin{array}{ccccccc}
\mbox{lines $\xi\in 2\BZ$} &\vline&\mbox{heights along ${\pl D_L }$}
  &\vline&\mbox{heights along ${\pl D_R }$}&\vline&\mbox{height difference}
 \\ \hline  
  0\leq \xi\leq 2(M\!-\!1)&\vline&h=n&\vline&
  0\leq h\leq m-\rho=M-1&\vline&
  h:n\to n\!-\!M\!+\!1
  \\ \hline
  2(M\!-\!1)\leq \xi\leq 2m &\vline&
  n\leq h\leq n+\rho&
  \vline&
  M\!-\!1\leq h\leq m\!=\!M\!-\!1\!+\!\rho&\vline&
 h= n-M+1
 \\ \hline
 2m\leq \xi\leq 2(m\!+\!M)&\vline&
 n+\rho\leq h\leq n+m&\vline&h=m
 &\vline&h:n\!-\!M\!+\!1\to n
  \end{array}
  $$
 This table tells us that the number of red dots is minimal and equals $\rk=n-M+1\geq 0$ within $\{\rho\}$ (i.e., along the lines $2(M-1)\leq \xi\leq 2m$) and increases by $1$, each time one moves away from the strip to reach $n$ along $\pl \DR_U$ and $\pl \DR_D$. The second identity for $\rk$ in (\ref{rho,r}) follows from substituting $M$ from (\ref{sg}) in $n-M+1$, and so we also have, 
 \be\label{sg'} \rk=-\Dt+\sg~~~ \mbox{and }~~~ \rho=-\Dt+\rk  .\ee  
 \medbreak
 \noindent {\bf Case 2}:  
  Here the two lines $\xi=2m$ and $\xi=2M-2$ through the cuts determine a strip of width 
  \be\label{rho'}
 \rho = (M-1)-m=2\Dt-\sg\geq 0,
  \ee
using the same substitution as before. Here we have the following table:
 $$\hspace*{-2cm}\begin{array}{ccccccc}
\mbox{lines $\xi\in 2\BZ$} &\vline&\mbox{heights along ${\pl \DR_L }$}
  &\vline&\mbox{heights along ${\pl \DR_R }$}&\vline&\mbox{height difference}
 \\ \hline  
  0\leq \xi\leq 2m&\vline&h=n&\vline&
  0\leq h\leq m &\vline&
  h:n\to n-m
  \\ \hline
  2m\leq \xi\leq 2(M-1) &\vline&
  h=n&
  \vline&
  h=m&\vline&
 h= n-m=\Dt
 \\ \hline
 2(M-1)\leq \xi\leq 2(m+M)&\vline&
 n \leq h\leq n+m&\vline&h=m
 &\vline&h:n-m\to n
  \end{array}
  $$
 The table shows that within the strip the number of dots per line $\xi\in 2\BZ$ equals $n-m=\Dt \geq 0$, and, as before, the number increases by $1$ on either side of the strip $\{\rho\}$, to reach $n$. This ends the proof of Theorem \ref{tilable}.\qed

\section{A singular Toeplitz determinant and the kernels $\BK_{n,\rk,\rho}^{^{\textrm{\tiny green}}}$, $\BK_{n,\rk,\rho}^{^{\textrm{\tiny blue}}}$ and $\BK_{n,\rk,\rho}^{^{\textrm{\tiny red}}}$}

 Throughout this paper we define the following radii,
 \be a< \rho_0<\rho_1<\sg_1<\sg_2<R<\rho_2<\rho_3<a^{-1}, \label{radii}\ee
 together with the corresponding contours $\ga_{\rho_0}\subset\ga_{\rho_1}\subset\ga_{\sg_1}\subset\dots\subset \ga_{\rho_3}$ about $0$ having those radii. A contour slightly larger (resp. slightly smaller) than, say, 
 $\ga_{\sg_2}$ or $\ga_R$,... but not containing any additional pole (resp. but not losing any pole), is denoted by $\ga^+_{\sg_2}$, $\ga^+_R$,...(resp. $\ga^-_{\sg_2}$, $\ga^-_R$,...)
Moreover for $v \in \BC$, for $\bm\lb=(\lb_1,\dots,\lb_\ka)\in \BC^\ka$ the contours $\Ga_0,~ \Ga_{v}, ~\Ga_{\bm\lb},\dots$ denote small circles about $0$, $v$, and all the $\lb_i$,..., containing no other pole than the points mentioned.

\subsection{The kernel $\BK_{n,\rk,\rho}^{^{\textrm{\tiny green}}}$ for contiguous start- and endpoints as an integral of another kernel}
 Theorem \ref{lemma5} is the main statement of this section: it shows that the $\BK_{n,\rk,\rho}^{^{\textrm{\tiny green}}}$, as in (\ref{formlemma5}), can be expressed as an integral operator $ \widetilde\LR$ of another kernel $\BK^{\textrm{\tiny green}(\bm\lb)}$, as in (\ref{formlemma5'}), depending on parameters ${\bm\lb}=(\lb_1,\dots,\lb_\ka)$. The operator $ \widetilde\LR$,  given in (\ref{Ltilde}), amounts to integration versus a measure (\ref{muhat}). 
  
We follow the machinery from \cite{AJvM0,Jo05c}, see also  \cite{Jo16}, sections 4.2 and 4.5.
The difference in the model studied here compared to that of the previous papers is that the final points have been shifted by $\Delta$. 
 We refer to the covering with tiles as in Fig. 5(iii) and the nonintersecting paths as in Fig. 4. In rectangular coordinates, as depicted in Fig. 6, these paths  run between:
\be\bl&\mbox{Initial points}~1-i,~~1\leq i\leq M
\\&\mbox{Final points} ~1-j+\Dt,~~1\leq j\leq M,
\el \label{InFin}\ee
by means of $(1-az)$-steps (A-steps) and $(1-\tfrac az)^{-1}$-steps (B-steps). 
Therefore the  transitions involving consecutive steps are given by 
\be \bl\label{trans}\varphi_{s_1,s_2} (\zeta) &= 
{(1+a\zeta)^{[\tfrac {s_2}2]-[\tfrac {s_1}2]}  \over  (1-\frac a\zeta)^{(s_2-[\tfrac {s_2}2])-
(s_1-[\tfrac {s_1}2] )}} 
,~~~0\leq s_1<s_2\leq 2n+1
\el\ee
One deduces $\varphi_{0,2n+1}(\ze)$ and one defines related functions:
\be\label{transend}\bl  
\varphi_{2n+1}(\ze)  &:=\varphi_{0,2n+1}(\ze)=\frac{(1+a\ze)^n}{(1-\frac a\ze)^{n+1}},
~~ \rho( u) :=(1+au)^n(1-\frac au)^{n+1}
~~\el\ee
The Fourier transform of $\varphi_{s_1,s_2}$ will play an important role:
\be \label{transFourier}
p_{s_1, s _2} (u_1,u_2)  : = \Id_{s_1<s_2}\oint_{\ga_{\rho_1}} \zeta^{u_1-u_2} \vp_{s_1,s_2}(\zeta){d\zeta\over  2\pi \I \zeta} 
 =:
\Id_{s_1<s_2}\widehat\varphi_{s_1,s_2}(u_1,u_2).
  \ee
According to Johansson \cite{Jo16}, using the LGV-Theorem, the (inlier) kernel $K_{n,r,\rho}^{\textrm{green}}=:K_{ }^{\textrm{green}}$ is given by 
\be \label{Kgreen'}\bl
 &K^{\textrm{green}} (s_1,u_1;s_2,u_2)  
   \\&~~~~~=
 - \Id_{s_1<s_2}\widehat\varphi_{s_1,s_2}(u_1,u_2)
  +\sum_{i,j=1}^M
  \widehat\varphi_{s_1,2n+1}
   (u_1,1-i+\Dt)
 (A^{-1})_{ij}
  \widehat\varphi_{0,s_2}(1-j,u_2)
 \el\ee
  for $0<s_1,s_2<2n+1$ by taking into account the initial and final points (\ref{InFin}) and where $A$ is the $M\times M$ matrix
 $$
 A=(A_{ij})_{1\leq i,j\leq M} \mbox{, with }A_{ij}:=\widehat\varphi_{0,2n+1}(0,i-j+\Dt).
 $$
 Notice that $\widehat\varphi_{s_1,2n+1}
   (u_1,1-i+\Dt)$ can be written:

 $$
 \widehat\varphi_{s_1,2n+1}
   (u_1,1-i+\Dt )=\oint_{\ga_{\rho_1}} \zeta^{u_1-(1-i+\Dt) } \vp_{s_1,2n+1}(\zeta){d\zeta\over  2\pi \I \zeta} 
   =\oint_{\ga_{\rho_1}} \zeta^{u_1-( 1-i)}  \psi_{s_1,2n+1}(\zeta){d\zeta\over  2\pi \I \zeta} 
   $$
   where
   \be\label{psi}
  \psi_{s_1,s_2}(\zeta)    
 :=\left\{ \bl &\varphi_{s_1,s_2}(\zeta)\mbox{  ~~       for  }
s_2<2n+1
\\
&
\zeta^{ -\Dt } ~\varphi_{s_1,s_2}(\zeta)\mbox{   for  } s_2=2n+1\el\right. 
\ee
 For a given analytic function $f(z)$ and an integer $p\geq 1$, define the Toeplitz determinant
 $$D_p(f)=\det\left(\widehat f_{i-j}\right)_{1\leq i,j\leq p}
 ,~ \mbox{with} ~ \widehat f_k=\oint_{\ga_1}\frac{d\ze}{2\pi \I \ze}\frac{f(\ze)}{\ze^k}=
  \int_0^{2\pi}\frac{d\theta}{2\pi} f(\theta)e^{-\I k\theta}
$$

\begin{proposition}(Johansson\cite{Jo16})\label{prop:KurtKernel}
 The inlier kernel $K^{\textrm{green}}$ as in (\ref{Kgreen'}) can be expressed in terms of a ratio of Toeplitz determinants: 
 \begin{equation}\label{KernelIn}
K^{\textrm{green}} 
  (s_1,u_1;s_2,u_2) = - \Id_{s_1 < s_2}  p_{s_1;s_2} (u_1,u_2) + \widetilde K^{\textrm{green}}
   (s_1,u_1;s_2,u_2)
\end{equation}
where ($a<\rho_1<\sg_1<\sg_2<\rho_2<a^{-1}$)
\be \label{Kgreen}
\bl
&\widetilde K^{\textrm{green}} (s_1,u_1;s_2,u_2)
 \\ &=   \int_{\ga_{\rho_1}} {dz \over  2\pi \I z}
 \int_{\ga_{\rho_2}} {dw\over  2\pi \I w} {z^{u_1}\over w^{u_2}} \psi_{s_1,2n+1} (z) \psi_{0,s_2} (w) 
 {D_{M-1} \left[\psi_{0,2n+1} (\ze)\left(1-{\ze \over w}\right) \left(1-{z\over \ze}\right) \right]   \over D_{M} \left[\psi_{0,2n+1}(\ze)\right]}.
\el
\ee
 \end{proposition}
 \proof Since the expression $\varphi_{s_1,s_2}$ is a product of functions admitting a Wiener-Hopf factorization, the kernel (\ref{Kgreen'})  can be written as a double integral involving a ratio of Toeplitz determinants; for details see \cite{Jo16}.\qed
 
  We need to express the ratio of these two Toeplitz determinants in terms of integrals involving Fredholm determinants. The symbol $\psi_{s_1, 2n+1}$ of the Toeplitz matrix appearing in (\ref{psi}) has a winding number due to $\zeta^{\ka}$ (special instance of Fisher-Hartwig singularity). We remove this singularity $\ze^\ka$ by blowing it up to the following, using parameters $0\neq \lb_i \in \BC$: 
 \be\label{P}
p_{\bm\lb}(\ze):=\frac{P_{\bm\lb}(\ze)}{\prod_1^\ka \lb_i}:=\prod_1^\ka \left(   1-\frac \ze {\lb_j}\right).
\ee   
We define two additional $\lb$-dependent functions  
\be \label{defs}
\varphi^{(\lb)}_{2n+1}(\ze):=\varphi _{0, 2n+1}(\ze)p_\lb(\ze),~~\mbox{and}~~
\rho_{\lb} ( u) : =\rho  ( u)p_\lb (u).
\ee 
\be \label{gal}
g_a:=(1+a^2)^{n(n+1)}\mbox{ and  }g^{(\lb)}_a:=   p_\lb(a)^{n+1}g_a=g_a\left(\frac{ P_\lb(a) }{\prod_1^\ka \lb_i}\right)^{n+1}\ee 
 The Toeplitz determinants of $\ze^{\pm \ka}f(\ze)$ and $  
   p_\lb(\ze^{\pm 1})
 f(\ze) $ are related as follows: 
  
 \begin{lemma}\label{lemma:Toeplitz1}
The following Toeplitz determinant identity holds:
$$
D_p[\ze^{\pm \ka}f(\ze)]
=(-1)^{\kappa p}   \oint_{\Ga_0^\kappa}
\prod_1^\ka\frac{\lb_j^p d\lb_j}{2\pi \I \lb_j}
D_p \left[  p_{\bm\lb}(\ze^{\pm 1})
 f(\ze)\right].
$$
\end{lemma}

 \proof
For any integer $\kappa\geq 1$, and $\lb=(\lb_1,\dots,\lb_\kappa)$, 
\be f^{(\lb)}(\ze)=f(\ze)\prod_{j=1}^\kappa (\lb_j-\ze^{\pm 1}),~~~\mbox{with}~~~f ^{(0)}(\ze)= (-1)^\kappa \ze^{\pm \kappa}f(\ze),
 \label{phi-lb}\ee
and so we have that $\widehat f_j^{(0)}
= (-1)^\ka \widehat f_{j\mp \ka}
$. Then taking residues at $\lb_j=0$ for $1\leq j\leq \ka$ of each of the entries and using (\ref{phi-lb}), one finds for an arbitrary contour $\Ga_0$ about $0$, 
$$\bl
(-1)^{\kappa p}   \oint_{\Ga_0^\kappa} 
\prod_1^\ka\frac{ d\lb_j}{2\pi \I \lb_j}
D_p \left[ P_{\bm\lb}(\ze^{\pm 1}) f(\ze)\right]
 &=
(-1)^{\kappa p}   \oint_{\Ga_0^\kappa}
\prod_1^\ka\frac{d\lb_j}{2\pi \I \lb_j}
\det\left(\widehat f^{(\lb)}_{i-j}\right)_{1\leq i,j\leq p}
\\&=D_p \left[ \ze^{\pm \ka}f(\ze)\right],\el  
 $$
 ending the proof of Lemma \ref{lemma:Toeplitz1}. \qed


\bigbreak

Given $g_a$ as in (\ref{gal}),
 define the measure $d\widetilde   {\bm\mu}(\bm\lb )$, where one uses $M-1=n-\rk$, with $\rk\geq 0$,
\be\bl \label{muhat}
d\widetilde   {\bm\mu}(\bm\lb ) &= 
g^{(\lb)}_a\dis \prod^\ka_{j=1} {\lb_j^{M-1} d\lb_j \over 2\pi i  }  
 = g_a\dis \prod^\ka_{j=1} \frac{d\lb_j (\lb_j-a)^{n+1}}{2\pi \I \lb_j^{\rk+1}} 
\el\ee
  and also an operator $\widetilde \LR$, in terms of the kernel $K_{k,\ell}^{(\lb)} {(0,0)}$ defined in (\ref{K00}) below, \footnote{\label{3}For any integer $p$ and kernel $K_{k,\ell}$, we abbreviate $\det(I-K_{k,\ell})_{k,\ell\geq p}:=\det (I-K_{k,\ell})_{\geq p}$.}
  \be \label{Ltilde}\widetilde \LR (f)=
  \int_{\ga_R^\ka} 
 d\widetilde   {\bm\mu}(\bm\lb )\det (I-K_{k,\ell}^{(\lb)} {(0,0)})_{\geq M } f(\lb).
 \ee

\begin{lemma}\label{lemma:Toeplitz2} The following holds for $a<\sg_1<\sg_2<R$, for $|z|<\sg_1<\sg_2<|w|$,~for $|z|<a^{-1}$ and for $|z|<$ all $|\lb_j|$:  (remember footnote \ref{3})
\be\label{Toeplitz2}\bl
&\frac {D_{M-1}\left[\ze^{\ka}\vp_{0,2n+1}(\ze)(1-\frac \ze w)(1-\frac z\ze)
\right]
}
{D_{M }\left[\ze^{\ka}\vp_{0,2n+1}(\ze)
\right]
}=\frac{(-1)^\ka (1-\frac aw)^{n+1}}{(1+az)^n(1-\frac zw) }\\&
~~~~~~~~~~~~~~\times \frac 1{\widetilde \LR(1)}{\dis\oint_{\ga_R^\kappa}
 \frac{d\widetilde   {\bm\mu}(\bm\lb )}{P_{\bm\lb}(z)}
\det(\Id-K_{k,\ell}^{(\lb)}(w^{-1},z))_{\geq M-1}
},
\el
\ee
where 
\be\bl
K^{(\lb)}_{k,\ell}(w^{-1},z)&:= \oint_{\ga_{\sg_1}}\frac{(-1)^{k+\ell}du}{(2\pi \I)^2} \oint_{ \ga_{\sg_2}  }\frac{dv}{v-u} \frac{u^{\ell}}{v^{k+1}} 
 \frac{(1-\frac uw)(1-\frac zv)}{(1-\frac vw)(1-\frac zu)} \frac{\rho_{\lb }( u)}{\rho_{\lb }( v)}
\\
&= K^{(\lb)}_{k+1,\ell+1}(0,0)-(z-w)h_{k+1}^{(1)}(w^{-1},\lb)
  \frac 1z h_{\ell+1}^{(2)}(z,\lb) 
\el\label{Klambda}\ee
with, for $k,\ell\geq 0$, 
\be \label{K00}
 K^{(\lb)}_{k ,\ell }(0,0):=
    \oint_{\ga_{\sg_1}}\frac{du}{(2\pi i)^2} \oint_{ \ga_{\sg_2}}\frac{dv}{v-u} \frac{u^{\ell }}{v^{k+1}} \frac{\rho_\lb  ( u)}{\rho_\lb ( v)}
\ee
\be \bl\label{hi}
h^{(1)}_i(w^{-1};\lb)&:= -\oint_{\ga_{\sg_2}} \frac{ dv}{2\pi \I  (v-w)(-v)^{i+1}\rho_{\lb}(v)}
 \\
h^{(2)}_i(z;\lb) &:= -z\oint_{\ga_{\sg_1}} \frac{(-u)^i\rho_{\lb}(u)du}{2\pi \I  (u-z)}. 
\el\ee
We also have
 \be
 D_M\left[\vp^{(\lb)}_{0,2n+1}\right]=g_a^{(\lb)}\det\left(\Id-K^{(\lb)}_{k,\ell}(0,0)\right)_{\geq M} \mbox{ and }
 D_M\left[\psi_{0,2n+1}\right]
 =(-1)^{\ka M}\widetilde\LR(1)
 \label{Dphi}\ee

\end{lemma}
\proof  Define the function $\chi_{t,s}(u)$ and its Toeplitz determinant:
\be \chi_{t,s}(u):=e^{\sum_{j=1}^\infty (t_ju^j+s_ju^{-j})}
\mbox{   and  } \tau_p(t,s):=D_p[\chi_{t,-s}(u)].
\label{chi-tau}
\ee
Considering now the loci
\be \label{locus}
\LR :
\left\{\bl
jt_j&=-n(-a)^j
\\
-js_j&=(n+1)a^j
,\el  \right\}
\mbox{    and   }
\LR^{(\lb)}:
\left\{\bl jt_j &=-n(-a)^j- \sum_{\ell}\lb_\ell^{-j}
\\ -js_j&=(n+1)a^j
\el
\right\}.
\ee
Then for $a<|u|  <\lb_\ell$, 
\be \label{phi-rho}\bl
\chi_{t,s}(u)\Bigr|_{\LR^{ }}&=\rho^{}( u)  ,~~~~~~
  \chi_{t,-s}(u)\Bigr|_{\LR^{ }}= \varphi^{}_{2n+1}(u)
\\
\chi_{t,s}(u)\Bigr|_{\LR^{(\lb)}}&=\rho_{ \lb  }( u) ,~~
\chi_{t,-s}(u)\Bigr|_{\LR^{(\lb) }}= \varphi^{(\lb) }_{2n+1}(u)
.\el
\ee
By Borodin-Okounkov one has 
\be \label{BO}\bl \tau_p(t,s)&=Z(t,s)\det(\Id-K_{\tiny \mbox{BO}}(t,s))_{\geq p},~~\mbox{with   }
Z(t,s)  :=e^{-\sum_1^\infty jt_js_j} 
,\el
\ee
for the kernel
\be
\begin{aligned}
K_{\tiny \mbox{BO}}
({\bf t},{\bf s})_{k,\ell}&
  =\frac{1}{(2\pi i)^2} \oint_{\ga_{\sg_1}}du \oint_{\ga_{\sg_2}}\frac{dv}{v-u} \frac{u^{\ell}}{v^{k+1}} 
  \frac{\chi_{t,s}(u)}{\chi_{t,s}(v)}
%
\end{aligned}
\ee
We also have, using $|z|<|w|$ , $a<|w|$, $|z|<a^{-1}$ and $|z|<\mbox{all}~|\lb_i|$, that\footnote{\label{4} For $\al\in \BC$, we define $[\al]=(\al,\frac{\al^2}{2},\frac{\al^3}{3},\dots)\in \BC^\iy$.}
\be\label{Z-shift}\bl   
  \frac{Z(t-[w^{-1}],s+[z])}{Z(t,s)}\Bigr|_{\LR^{(\lb)}}
 &=e^{\sum_{j=1}^\infty (s_jw^{-j}-t_jz^j+\frac1{j}(z/w)^j)}\Bigr|_{\LR^{(\lb)}}=\frac{ (1-\frac aw)^{n+1}p^{-1}_\lb(z)
  }{(1+az)^n(1-\frac zw) }
\\
 Z(t,s)\Bigr|_{\LR^{(\lb)}}& =e^{-\sum_1^\infty jt_js_j}\Bigr|_{\LR^{(\lb)}}
  =g^{(\lb)}_a 
%
 .\el\ee
 Moreover, as long as $a<|z| < |u| < |v|<|w|$ and $ |v| < \mbox{all}~|\lb_j|$, we have thus for $a<\sg_1<\sg_2<R$, with $|\lb_j|$ in the annulus between $\sg_2$ and $R$:
%
\be\footnotesize\bl 
(-1)^{k+\ell}K^{(\lb)}_{k,\ell}&(w^{-1},z)
\\&:=K_{\tiny \mbox{BO}} ( t-[w^{-1}],s+[z])_{k,\ell}\Bigr|_{\LR^{(\lb)}}
\\&=
     \oint_{\ga_{\sg_1}}\frac{du}{(2\pi \I)^2}\oint_{ \ga_{\sg_2}}\frac{dv}{v-u} \frac{u^{\ell}}{v^{k+1}} 
 \frac{(1-\frac uw)(1-\frac zv)}{(1-\frac vw)(1-\frac zu)} \frac{\chi_{t,s}(u)}{\chi_{t,s}(v)}\Bigr|_{\LR^{(\lb)}}
\\
&=
     \oint_{\ga_{\sg_1}}\frac{du}{(2\pi i)^2} \oint_{ \ga_{\sg_2}}\frac{dv}{v-u} \frac{u^{\ell}}{v^{k+1}} 
 \frac{(1-\frac uw)(1-\frac zv)}{(1-\frac vw)(1-\frac zu)} \frac{\rho^{(\lb) } ( u)}{\rho^{(\lb) } ( v)} 
 \\&=:(-1)^{k+\ell}\left(K^{(\lb)}_{k+1,\ell+1}(0,0)-(z-w)h_{k+1}^{(1)}(w^{-1},\lb)
  \frac 1z h_{\ell+1}^{(2)}(z,\lb)\right)\el \label{newK}\ee 
 and
\be\bl (-1)^{k+\ell}K^{(\lb)}_{k,\ell}(0,0)
 :=K_{\tiny \mbox{BO}} ( t ,s )_{k,\ell}\Bigr|_{\LR^{(\lb)}}
 =
     \oint_{\ga_{\sg_1}}\frac{du}{(2\pi i)^2} \oint_{ \ga_{\sg_2}}\frac{dv}{v-u} \frac{u^{\ell}}{v^{k+1}} 
%
  \frac{\rho_{ \lb } ( u)}{\rho_{ \lb  } ( v)}  ,\el\ee
  using in the last equality of (\ref{newK}), formula (\ref{hi}) and the identity
 $$\frac{u^\ell}{v^{k+1}}\frac{1}{v-u}\frac{(1-\frac uw)(1-\frac zv)}{(1-\frac vw)(1-\frac zu)}
=\frac{u^{\ell+1}}{v^{k+2}}\left(  \frac{1}{v-u}+\frac{z-w}{(u-z)(v-w)}    \right).$$
 %
 Applying a shift to formula (\ref{chi-tau}), we have on the one hand the formula
%
\be \label{tau1}\bl
  \tau_p (t-[w^{-1}],s+[z])\Bigr|_{\LR^{(\lb)}}&=D_p[(1-\tfrac {\ze}w)(1-\tfrac z\ze)\chi_{t,-s}(\ze)] \Bigr|_{\LR^{(\lb)}}\\
  &=
  D_p\left [\varphi^{(\lb) }_{2n+1} (\ze)(1-\frac \ze{w})(1-\frac z{\ze})
 \right]
 \el
 \ee
 \be\bl\label{tau1'}  
  \tau_p (t ,s )\Bigr|_{\LR^{(\lb)}} &=D_p[ \chi_{t,-s}(\ze)] \Bigr|_{\LR^{(\lb)}} 
   =
  D_p\left [\varphi^{(\lb) }_{2n+1} (\ze) 
 \right]
\el
\ee
and on the other hand, using the BO formula (\ref{BO}), formula (\ref{newK}) and (\ref{Z-shift}), we have for all $|\lb_j|\!>\!a$, $|\lb_j|>|z|$, $|z|<|w|$, $a\!<\! |w|$, $|z|<a^{-1}$, (see footnote \ref{3})
\be\label{tau2}\bl \tau_p&(t-[w^{-1}],s+[z])\Bigr|_{\LR^{(\lb)}} 
\\
&=Z(t-[w^{-1}],s+[z])\det(\Id-K_{\tiny \mbox{BO}}(t-[w^{-1}],s+[z]))_{\geq p}\Bigr|_{\LR^{(\lb)}}
\\
&  =Z(t,s)\Bigr|_{\LR^{(\lb)}}\frac{ (1-\frac aw)^{n+1}p^{-1}_{\bm\lb}(z)
 }{(1+az)^n(1-\frac zw) }
\det(\Id-K_{k,\ell}^{(\lb)}(w^{-1},z))_{\geq p} ~
\\&=g_a^{(\lb)}\frac{ (1-\frac aw)^{n+1}p^{-1}_{\bm\lb}(z)
 }{(1+az)^n(1-\frac zw) }
  \det(\Id-K_{k,\ell}^{(\lb)}(w^{-1},z))_{\geq p}
.\el \ee

\be\label{tau2'}\bl  \tau_p &(t ,s )\Bigr|_{\LR^{(\lb)}} 
 =Z(t,s)\det(\Id-K_{\tiny \mbox{BO}}(t ,s ))_{\geq p}\Bigr|_{\LR^{(\lb)}}
  = g_a^\lb  
\det(\Id-K_{k,\ell}^{(\lb)}(0,0))_{\geq p} ~
\el\ee
%
 Then applying Lemma \ref{lemma:Toeplitz1} and equating (\ref{tau1}) and (\ref{tau2}) and using the expression $g_a^{(\lb)}$ as in (\ref{defs}), we obtain the following, upon setting $p=M-1$  
 \be\bl\label{Toeplitz3}
 D_{M-1}&\left[\ze^{\ka}\vp_{ 2n+1}(\ze)(1-\frac \ze w)(1-\frac z\ze)
\right]
\\&=(-1)^{\ka (M-1)}\oint_{\ga_R^\ka}
\left(\prod^\ka_{j=1} {\lb_j^{M-1} d\lb_j \over 2\pi i \lb_j}  \right)
D_{M-1}\left[ \vp^{(\lb)}_{ 2n+1}(\ze)(1-\frac \ze w)(1-\frac z\ze)
\right]
\\&=\frac{ (-1)^{\ka (M-1)}(1-\frac aw)^{n+1}}{(1+az)^n(1-\frac zw)}
\oint_{\ga_R^\ka}\left(\prod^\ka_{j=1} {\lb_j^{M-1} d\lb_j \over 2\pi i  }  \right)
\frac{g_a^{(\lb)}}{P_{\bm\lb}(z)}\det(\Id-K^{(\lb)}_{k,\ell}(w^{-1},z))_{\geq M-1}
\el \ee
 Moreover, equating the right hand sides of (\ref{tau1'}) and (\ref{tau2'}) and setting $p=M$, we find the first formula of (\ref{Dphi}).  Using  Lemma \ref{lemma:Toeplitz1} on the expression below, using the first formula of (\ref{Dphi}) and the definition (\ref{muhat}) of the measure $d\widetilde \mu$, we have for $p=M$,   
 \be\bl  \label{Toeplitz3'}
D_{M} \left[\psi_{0, 2n+1}(\ze) 
\right] 
&= D_{M} \left[\ze^{\ka}\vp_{0, 2n+1}(\ze) 
\right]
\\
& =(-1)^{\ka M}\oint_{\ga_R^\ka}
\left(\prod^\ka_{j=1} {\lb_j^{M} d\lb_j \over 2\pi i \lb_j}  \right)
D_{M}\left[ \vp^{(\lb)}_{ 2n+1}(\ze) 
\right]
\\&=(-1)^{\ka M}\oint_{\ga_R^\ka}
\left(\prod^\ka_{j=1} {\lb_j^{M-1} d\lb_j \over 2\pi i  }  \right)
g_a^\lb\det(\Id-K^{(\lb)}_{k,\ell}(0,0))_{\geq M}
\\&=(-1)^{\ka M}\oint_{\ga_R^\ka}
d\widetilde   {\bm\mu}(\bm\lb )\det(\Id-K^{(\lb)}_{k,\ell}(0,0))_{\geq M}=(-1)^{\ka M}\widetilde \LR(1),
\el\ee
proving the second formula of (\ref{Dphi}).

 So, computing the ratio of (\ref{Toeplitz3}) and (\ref{Toeplitz3'}) leads at once to expression  (\ref{Toeplitz2}), with the kernel $K^{(\lb)}_{k,\ell}(w^{-1},z)$ as in   (\ref{Klambda}) and using the measure $d\widetilde \mu$ as defined in (\ref{muhat}), 
  thus ending the proof of Lemma \ref{lemma:Toeplitz2}. \qed

 We shall need the following lemma:
 
 \begin{lemma}\label{lemmaRes}For $M\geq 1$, we have for $R<\rho $: 
\be
\bl
(-1)^{\ka} \oint_{\ga_{R}^\kappa}\prod_1^{\ka} \frac{\lb_j^{M-1}d\lb_j}{2\pi \I }
 D_M&\left[\varphi_{0,2n+1}^{(\lb)} \right]
\oint_{\ga_\rho} \frac{z^{u_1-u_2+\ka }dz}{2\pi \I z P_{\bm\lb} (z) }
\\
&=\Id_{\{u_1=u_2\}}\oint_{\ga_{R}^\kappa} \prod_1^{\ka} \frac{\lb_j^{M }d\lb_j}{2\pi \I\lb_j}D_M\left[\varphi_{0,2n+1}^{(\lb)} \right],
\el\label{Res}\ee
which is the same as,
\be\label{Res'}\bl
(-1)^{\ka}\frac{\dis\widetilde\LR_{ } \left(\oint_{\ga_\rho} \frac{z^{u_1-u_2-\Dt}}{2\pi \I z P_{\bm\lb}(z)
 }\right)}{\widetilde\LR_{ } (1)} 
=\Id_{\{u_1=u_2\}} .
\el\ee
\end{lemma}

\proof Expanding the product in $(-z)^\ka P_{\bm\lb}(z)^{-1}= \prod_1^\ka (1-\frac{\lb_i}{z})^{-1}$ in powers of $z$, using homogeneous symmetric functions $h_r(\lb)$ of degree $r$ and $h_0$=1, we have, for $|\lb_i|<\rho$ and 
\be\bl
(-1)^\ka  \oint_{\ga_\rho }
\frac{dz}{2\pi \I z} \frac{z^{u_1-u_2+\ka}}{P_{\bm\lb} (z)}
 &=
\oint_{\ga_\rho } \frac{z^{u_1-u_2}dz}{2\pi \I z}\prod_1^\ka (1-\frac {\lb_i}z)^{-1}
\\
&=\oint_{\ga_\rho } \frac{z^{u_1-u_2}dz}{2\pi \I z}
\sum_{r\geq 0}\frac{h_r(\lb)}{z^r}
\\
&=\Id_{u_1\geq u_2}h_{u_1-u_2}(\lb),
\el\label{Res1}\ee
 Inserting this expression (\ref{Res1}) in the left hand side of (\ref{Res}) and using the explicit expression for the Toeplitz matrix, one finds, assuming $u_1-u_2>0$, and distributing the $\prod_1^{\ka} \lb_j$ over the $M$ columns of the Toeplitz matrix (using (\ref{P})),
$$\bl
&\oint_{\ga_{R}^\kappa}  \Bigl(\prod_1^{\ka}   \frac{\lb_j^{M }d\lb_j}{2\pi \I }\Bigr)
 \frac{h_{u_1-u_2}(\lb)}{ \prod_1^{ \ka } \lb_j}\det \left(\oint\frac{ \varphi_{2n+1}(\ze)d\ze}{2\pi \I\ze^{i-j+1}}\frac{P_{\bm\lb}(\zeta)}{\prod_1^\ka \lb_i}
   \right)_{1\leq i,j\leq M}
\\
&
 =
\oint_{\ga_{R}^\kappa} \left(\prod_1^{\ka}\frac{d\lb_j}{2\pi \I }\right)
 \frac{h_{u_1-u_2}(\lb_1,\dots,\lb_\ka)}
 { \prod_1^{\ka} \lb_j}
 \det \left(\oint\frac{\varphi_{2n+1}(\ze)d\ze}{2\pi \I\ze^{i-j+1}}P_{\bm\lb}(\ze) \right)_{1\leq i,j\leq M} 
 =0\el$$ 
The latter equals $0$, because each term in $(\prod_1^\ka \lb^{-1}_j )h_{u_1-u_2}(\lb_1,\dots,\lb_\ka)$ has at least one $\lb_\al$ appearing only in the numerator (or not at all) and not at all in the denominator. 
Moreover a typical term in the expansion of the determinant will be an $M$-fold integral in $\ze_1,\dots,\ze_M$ containing $\prod_{k=1}^M\prod_{i=1}^\ka (\lb_i-\zeta_k)$, which is holomorphic in the $\lb_i$'s. Therefore the $\lb_\al$-integration of that term vanishes and so does the integral of the Toeplitz determinant. Therefore  the only contribution of the first integral on the LHS of (\ref{Res}) will come from $u_1=u_2$ in the indicator function $\Id_{\{u_1\geq u_2\}}$, establishing formula (\ref{Res}) of Lemma \ref{lemmaRes}.\qed



\bigbreak
 
 In order to compute the kernel $\BK^{\textrm{green}}
  (s_1,u_1;s_2,u_2)$ as in (\ref{KernelIn}), we need to introduce the following functions: 
 \be \label{FG}
\bl
G^{(s)}_u(z) &:=\frac{z^{u-1}\psi_{s,2n+1}(z)}{(1+az)^n P_\lb(z)},~~F^{(s)}_u(w)=\frac{(1-\frac aw)^{n+1}\psi_{0,s}(w)}{w^{u}},
\el\ee
and the following transforms, remembering $|z|<\sg_1<\sg_2<|w|$ from the inequalities in Lemma \ref{lemma:Toeplitz2},
 \be\bl
 a^{(s)}_u(k) &:= \oint_{\ga_{\rho_2}}  F^{(s)}_u (w) h^{(1)}_k (w^{-1},\lb) {dw\over 2\pi \I  }   
   ~\mbox{and}~ b^{(s)}_u(\ell)  
    :=   \oint_{\ga_{\rho_1}} G^{(s)}_u (z) h_\ell^{(2)} (z,\lb) {dz\over 2\pi \I z}.
\el \label{FGab}
\ee
Considering the radii $\sg_2<R<\rho_2 <\rho_3<a^{-1}$ and the further contour $\Ga_{\bm\lb}$ (see (\ref{radii}) and below), define for $\dt=0,1$,
\be
\bl
  S^{(\lb)}_\dt (s_1  , u_1 ; s_2,u_2)  
&:=  
\oint_{\ga_{\rho_2}} {dw \over 2\pi \I  }  
\oint_{\ga_{\rho_3}-\Ga_{\bm\lb}} \frac{dz}{2\pi \I} 
  {F^{( s_2 )}_{u_2} (w) G^{( s_1 )}_{u_1} (z)  \over  z-w }   \left(\frac{1+a^2}{ w-a}\right)^\dt
%
\el
\label{S}\ee
$$\bl
  a^{(s)}_{u,\dt}(k):=
  \oint_{\ga_{\rho_2}} {dw\over 2\pi \I } F^{(s)}_u (w) h_k^{(1)} (w^{-1},\lb)  \left(\frac{1+a^2}{ w-a}\right)^\dt
  \mbox{  with }a^{(s)}_{u,0}(k)=a^{(s)}_{u }(k).
  \el $$

\begin{proposition}\label{lemma5} Picking the contours $\ga_{\rho_1},\dots$ 
 and the contour $\Ga_{\bm \lb}$ as in (\ref{radii}), the kernel $\BK^{\textrm{green}}$ reads, for $s_i, u_i\in \BZ$:
\be\label{formlemma5}
\bl
  \BK^{\textrm{green}}
   ( s_1 ,u_1 ;  s_2  ,u_2) &=  \frac{\widetilde\LR}{\widetilde\LR(1)} \left(\BK^{\textrm{green}(\lb)}( s_1 ,u_1 ;  s_2  ,u_2)\right),   
\el\ee
where
\be\label{formlemma5'}\bl   \BK^{\textrm{green}(\lb)}=&-  \Id_{s_1 < s_2}  p_{ s_1 , s_2 }(u_1,u_2)
 \\
&+ (-1)^{\ka}   \left\{ \int_{\ga_{\rho_1}} \frac{dz}{2\pi \I} \int_{\ga_{\rho_2}} {dw \over 2\pi \I  } {F^{( s_2 )}_{u_2} (w) G^{( s_1 )}_{u_1} (z) \over w-z}  -1 \right.
\\
&\left.   ~~~~~~~~~~~~~~~ + {\det (I-K^{(\lb)} {(0,0)} - a^{( s_2 )}_{u_2} \otimes b^{( s_1 )}_{u_1})_{\geq M}  \over  \det (I-K^{(\lb)}(0,0))_{\geq M}}
\right\}.
\el
\ee
Then setting $s_1=s_2=n$ (even), which is the half-way point on the $s$-axis within the Aztec rectangle, one finds  
\begin{equation}\label{Kernelnn}
\bl
     \BK^{\textrm{green}(\lb)}
 &(n,u_1;n,u_2) \\&=   \Id_{u_1=u_2}- 
   (-1)^\ka  \left[S_0^{(\lb)}(n,u_1;n,u_2)
 +    \left\langle \left(I-K^\lb_{(0,0)}\right)^{-1}_{\geq M} a^{(n)}_{u_2}, b^{(n)}_{u_1} \right\rangle_{\geq M}
 \right].
\el
\end{equation}

\end{proposition}


\proof
In order to prove formula (\ref{formlemma5}), 
 we put the expression (\ref{Toeplitz2}), together with (\ref{Klambda}) (be aware of the shift of the indices $k,\ell$), into the equation below, with  $a<\rho_1<\sg_1<\sg_2<\rho_2<R$. Since the $\lb$-integration encounters a pole in the $z$-variable, but none in the $w$-variable, one can take $a<\rho_1<\sg_1<\sg_2<R<\rho_2 $. We further need the following identity for a trace-class operator $A$: (see \cite{AJvM0})
$$
\det (I-A+c a(w) \otimes b(z)) = (1-c) \det (I-A) + c\det (I-A+a(w)\otimes b(z))
$$
for $A = K^{(\lb)}(0,0)$ , $c=w-z$ , $a=h^{(1)}(w^{-1}, \lb)$ , $b = {h^{(2)} (z,\lb) \over (-z)}$ and then use another identity for a trace class operator $A$, for continuous functions $F(w)$, $G(z)$ on $|w|=r_2$ and $|z|=r_1$ and for vectors  $a(w)$, $b(z)$  depending on $w$ and $z$:
$$
 \int_{|z|=r_1}     {dz  \over   2\pi \I}   \int_{|w|=r_2}   {dw  \over   2\pi \I}    F(w) G(z) \det \Bigl(I-A+a(w)\otimes b(z)\Bigr)
$$
$$
= \det \left(I-A + \left(\int_{|w|=r_2}F(w)a(w){dw  \over   2\pi \I}\right)   \otimes \left(\int_{|z|=r_1}G(z)b(z){dz  \over   2\pi \I}\right)\right)
$$
$$
+\left[ \left(\int_{|w|=r_2}F(w){dw  \over  2\pi \I}\right) 
\left(\int_{|z|=r_1}G(z){dz  \over   2\pi \I}\right) -1 \right] \det (I-A).
$$
 This gives for $\widetilde \BK^{\textrm{green}}$, as in (\ref{Kgreen}), using moreover  (\ref{Dphi}):
 \begin{equation}\footnotesize
\bl
&\widetilde \BK^{\textrm{green}}
 (s_1,u_1;s_2 ,u_2)\\  =&   \oint_{\ga_{\rho_1}} {dz \over  2\pi \I z}
 \oint_{\ga_{\rho_2}} {dw\over  2\pi \I w} {z^{u_1}\over w^{u_2}} \psi_{s_1,2n+1} (z)  \psi_{0,s_2} (w) 
 {D_{M-1} \left[ \psi_{0,2n+1} (\ze)\left(1-{\ze \over w}\right) \left(1-{z\over \ze}\right) \right]   \over D_M \left[ \psi_{0,2n+1}(\ze)\right]} 
\\
 \stackrel{ }{=}&\frac{(-1)^{\ka }}{\widetilde\LR_{ }(1)}
\oint_{\ga_R^\kappa}
d\widetilde \mu(\lb)
 \oint_{\ga_{\rho_1}}\frac{dz}{2\pi \I}\oint_{\ga_{\rho_2}}\frac{dw}{2\pi \I   }
 \frac{G_{u_1}^{(s_1)}(z)F_{u_2}^{(s_2)}(w) }{w-z}
\\
&\times \det\left[ \Id-K^{(\lb)}_{k ,\ell }(0,0)+(w-z)h_{k }^{(1)}(w^{-1},\lb)
  (-\tfrac 1z) h_{\ell }^{(2)}(z,\lb) \right]_{\geq M},
\\
=&\frac{(-1)^{\ka }}{\widetilde\LR_{ }\left(1\right)}
\oint_{\ga_R^\kappa}d\widetilde \mu(\lb) 
\\&\times \left\{\bl &\det\left[ \Id-K^{(\lb)}_{  }(0,0)\right]_{\geq M}
\left(\oint_{\ga_{\rho_1}}\frac{dz}{2\pi \I}\oint_{\ga_{\rho_2}}\frac{dw}{2\pi \I   }
 \frac{G_{u_1}^{(s_1)}(z)F_{u_2}^{(s_2)}(w) }{w-z}-1\right)%
 \\
 &+
 \det\left[ \Id\!-\!K^{(\lb)}_{ }(0,0)\!-\!  
 \oint_{\ga_{\rho_2}}  F^{(s_2)}_{u_2} (w) h^{(1)} (w^{-1},\lb) {dw\over 2\pi\I  }
\otimes \oint_{\ga_{\rho_1}}  G^{(s_1)}_{u_1} (z) h^{(2)} (z,\lb) {dz\over 2\pi \I z}
  \right]_{\geq M}.
\el \right\}\el
\end{equation}
Formula (\ref{formlemma5}) then follows from dividing the expression in curly brackets by $\det\left[ \Id-K^{(\lb)}_{  }(0,0)\right]_{\geq M}$ and using the definitions (\ref{FGab}).

Next we prove formula (\ref{Kernelnn}). 
We pick the contours as announced in the statement of Proposition \ref{lemma5}. Deforming the contour $\ga_{\rho_1}$ into $\ga_{\rho_3}$ picks up a residue when traversing $\ga_{\rho_2}$ and $\ga_{R}$, the latter containing the $\lb_i$'s. Therefore we have 
$$
\bl
 \oint_{\ga_{\rho_1}} &\frac{dz}{2\pi \I}\oint_{\ga_{\rho_2}}{dw  \over   2\pi \I  }   {F^{(n)}_{u_2}(w) G^{(n)}_{u_1}(z)  \over w-z  } 
 \\
& =\oint_{\ga_{\rho_3}-\Ga_{\bm\lb}} \frac{dz}{2\pi \I} \oint_{\ga_{\rho_2}}   {dw  \over  2\pi \I  }   {F^{(n)}_{u_2}(w) G^{(n)}_{u_1}(z)  \over  w-z  }+   \oint_{\ga_{\rho_2}} {dw  \over  2\pi \I  } F^{(n)}_{u_2} (w) G^{(n)}_{u_1} (w).
\el
$$
Taking into account the definitions (\ref{FG}) of $F$ and $G$, together with the ones (\ref{defs}) of $\psi_{0,n}$ and $\psi_{ n,2n+1}$, 
 one computes 
$$
  \oint_{\ga_{\rho_2}}  {dw  \over   2\pi \I  } F^{(n)}_{u_2} (w) G^{(n)}_{u_1} (w) =  \oint_{\ga_{\rho_2}}{dw  \over   2\pi \I w }   {w^{u_1-u_2+\ka}  \over  P_\lb ({w }) }.
$$
Remember formula (\ref{Res'}) requires $R < \rho_2$. So, hitting the latter with $\widetilde\LR_{ }$, one finds that 
for $|\lb_i|<\rho_2$,  
$$\bl
\frac{(-1)^{\ka }}{\widetilde\LR_{ } \left(1\right)}
&\widetilde\LR_{ }\left(\oint_{\ga_{\rho_2}}  {dz  \over   2\pi \I  } F^{(n)}_{u_2} (z) G^{(n)}_{u_1} (z) \right)
 =\frac{(-1)^{ \ka} }{\widetilde\LR_{ }\left(1\right)}
\widetilde\LR_{ }\left(\oint_{\ga_{\rho_2}}\frac{z^{u_1-u_2-\Dt} dz}  {  2\pi \I zP_\lb ({z })   }    \right)
 = \Id_{\{u_1=u_2\}}
\el$$
and so, using the notation (\ref{S}) for $S_0^{(\lb)}$,
\be\label{Kernelnn'}\bl
&\frac{(-1)^{\ka }}{\widetilde\LR_{ } \left(1\right)}
 \widetilde\LR_{ }\left( \oint_{\ga_{\rho_1}}  \frac{dz}{2\pi \I}\oint_{\ga_{\rho_2}}{dw  \over   2\pi \I  }  
 {F^{(n)}_{u_2}(w) G^{(n)}_{u_1}(z)  \over w-z  }\right)
\\&=
 \Id_{\{u_1=u_2\}}-\frac{(-1)^{\ka }}{\widetilde\LR_{ } \left(1\right)}
 \widetilde\LR_{ }\left(
S_0^{(\lb)}(n,u_1;n,u_2) \right).
\el\ee
One then substitutes the expression (\ref{Kernelnn'}) into (\ref{formlemma5}) for $s_1=s_2=n$ even, and then the identity (for $\det A \neq 0$)
\[
\bl
 {\det (A-b\otimes c)  \over   \det A} 
 &= \det (I-(A^{-1} b) \otimes c)
= 1- \langle A^{-1} b,c \rangle,
\el
\]
leads to formula (\ref{Kernelnn}), thus concluding the proof of Proposition \ref{lemma5}.\qed

 
 \subsection{The dual kernel $\BK_{n,\rk,\rho}^{^{\textrm{\tiny blue}}}$ of the point process along $\eta\in 2\BZ+1$} 
 
 Given the formal multiplication \be f(x) \ast_x~ g(x)= \sum_{x\in \BZ} f(x)g(x),\label{ast}\ee the following statement holds and can be found in \cite{AJvM0}:
\begin{lemma}\label{lemma4}
The extended outlier kernel  $\BK_{}^{^{\textrm{\tiny blue}}}:=\BK_{n,\rk,\rho}^{^{\textrm{\tiny blue}}}$  is given in terms of the outlier kernel $\BK_{n,\rk,\rho}^{^{\textrm{\tiny blue}}}$ at level $n$ (even) by  
\be\bl
 \BK_{}^{^{\textrm{\tiny blue}}}  &( s_1  ,u_1; s_2  ,u_2)\\
  =& -  \Id_{s_2 < s_1}  \widehat \vp_{ s_1 , s_2 } (u_1,u_2)
+ \widehat \vp_{ s_1 ,n}(u_1,\bullet)  \ast_{\bullet} ~ \BK_{}^{^{\textrm{\tiny blue}}}  (n,\bullet ; n,\circ) \ast_{\circ}~ \widehat \vp_{n, s_2 } (\circ,u_2)
\el \label{inout'}\ee
with the  kernel $\BK_{n,\rk,\rho}^{^{\textrm{\tiny blue}}}$ given by duality in terms of  $\BK_{n,\rk,\rho}^{^{\textrm{\tiny green}}}$ at level $n$:
\be
 \BK_{}^{^{\textrm{\tiny blue}}}  (n,u_1 ; n, u_2) = \Id_{u_1=u_2} -  \BK_{}^{^{\textrm{\tiny green}}} (n,u_1;n,u_2).
\label{inout}\ee
\end{lemma}

%

\begin{proposition}\label{th07}
For even \footnote{actually true for n odd as  well.} $n$ , the kernel $\BK_{}^{^{\textrm{\tiny blue}}}$ is given by the following expression, 
\be\label{Kout}
\bl
  \BK^{\textrm{\tiny blue}}
   ( s_1 ,u_1 ;  s_2  ,u_2) &=  \frac{\widetilde\LR}{\widetilde\LR(1)} \BK^{\textrm{\tiny blue}(\lb)}( s_1 ,u_1 ;  s_2  ,u_2)   
\el\ee
where
\be\label{Kout}
\bl
     \BK_{}^{^{\textrm{\tiny blue}(\lb)}}    
( s_1 & , u_1 ;  s_2  ,u_2)   
  =  -\Id_{ s_2   <  s_1 }  \widehat \vp_{s_1,s_2}(u_2-u_1)
\\
&+   (-1)^{ \ka} 
    \left[ S_0^{(\lb)}(s_1,u_1;s_2,u_2)
  \right.
  \left. 
+ \left\langle \left(I-K^\lb{(0,0)}\right)^{-1}_{\geq M} a^{(s_2)}_{u_2}, b^{(s_1)}_{u_1} \right\rangle_{\geq M}
 \right]
\el
\ee
 
\end{proposition}

\proof   Given the Fourier transform defined about a contour $\Ga_{0,a}$ about $0$ and $a$, where $z=a$ is a pole of $g(z)$,
$$\widehat g(x)=\oint_{\Ga_{0,a}}\frac{dz}{2\pi \I z}\frac{g(z)}{z^x},$$
one checks the following Fourier operations:
\be\bl \widehat g_1(x-u_1)\ast_x
\widehat g_2(u_2-x)&=
\widehat {(g_1g_2)}(u_2-u_1)
\\ \\
z^{\pm x} \ast_x \widehat F(\pm x) = z^{\pm x} \ast_x \oint {dw  \over   2\pi \I w} w^{\mp x} F(w)& = \oint {dw  \over   2\pi \I w} \sum_{u\in \BZ} \left({z  \over   w}\right)^{u} F(w) = F(z).
\label{Fourier-ast}
 \el\ee
 One first substitutes (\ref{inout}) in the right hand side of (\ref{inout'}), and the one uses formula (\ref{Kernelnn}) for $\BK^{\textrm{green}}
 (n,u_1;n,u_2) $. 
%
Next, pull the operation $\ast$ \ $\ast$ through the operator $\widetilde\LR_\lb$, the $z$ and $w$ integrations and the bracket $\langle \ ,\rangle$; from (\ref{Fourier-ast}), one has 
\[
\bl
&   z^{-v} \ast_v \widehat \vp_{n,s}(y-v) 
 = z^{-y}\vp_{n,s}(z),~~~~~
 \widehat \vp_{ s,n} (u-x) \ast_u z^{u} = z^x \vp_{ s,n} (z)
.\el 
\]
One concludes that, since  
$ \widehat \vp_{ s,n} (u_1,v) \ast_v 
\psi_{n,2n+1}(v,u_2)
=\psi_{s,2n+1}(u_1,u_2),~~
\psi_{0,n}(u_1,v)\ast_v \widehat \vp_{ n,s }(v,u_2)=\psi_{0,s}(u_1,u_2)$, taking into account (\ref{FG}), (\ref{FGab}) and (\ref{S}),
$$\footnotesize
\bl
&  \widehat \vp_{s_1,n} (u_1,v)  \ast_v 
  \left\{\!\begin{array}{lll} G^{(n)}_v (z) \\[0.2cm]  b^{(n)}_v \end{array}\!\!\right\}    =  
 \left\{\!
\begin{array}{lll}
G^{(s_1)}_{u_1} (z)
\\[0.2cm] 
b^{(s_1)}_{u_1} 
\end{array}
 \!\!\right\} 
,
 &    \left\{\begin{array}{lll} F^{(n)}_v (w)  \\[0.2cm]  a^{(n)}_v  \end{array}\right\}      \ast_v  ~  \widehat \vp_{n,s_2} {(v,u_2)}\! =\!\!
~ \left\{\!
\begin{array}{lll}
F^{(s_2)}_{u_2} (w)
\\[0.2cm] 
a^{(s_2)}_{u_2} 
\end{array}
 \!\!\right\} .
\el 
$$
establishing formula (\ref{Kout}) for 
   $  \BK_{n,\rk,\rho}^{^{\textrm{\tiny blue}}}  $, ending the proof of Proposition \ref{th07}.\qed

 \subsection{The outlier kernel $\BK_{n,\rk,\rho}^{^{\textrm{\tiny red}}}$ of the point process along $\xi\in 2\BZ $}
 
 In this section we deduce the $\BK_{ }^{^{\textrm{\tiny red}}}$ in terms of the kernel $\BK_{ }^{^{\textrm{\tiny blue}}}$.

 
\begin{proposition} \label{Kernelred}Given that $\BK_{n,\rk,\rho}^{^{\textrm{\tiny red}}}$ is expressed in $(\xi,\eta)$ coordinates, whereas $\BK_{n,\rk,\rho}^{^{\textrm{\tiny blue}}}$ is expressed in $(s,u)$-coordinates, the following holds:
\end{proposition}
   \vspace{6cm} 
   
 \newcommand\crule[3][blue!70!black!70!]{\textcolor{#1}{\rule{#2}{#3}}}
 
  \vspace*{-5cm}
  
$$  \hspace*{-9cm} \begin{picture}(0,0)   
        \put(0,20){\makebox(0,0) {      
 $  \squaresize 0.500cm \thickness .000008cm \Thickness
.060cm \Young{ 
   \blank       & ul \crule{0.5cm}{0.5cm}       
     \cr
     } $}}  
      \put(26,20){\makebox(0,0) {      
 $\squaresize 0.500cm \thickness .000008cm \Thickness
.060cm \Young{ 
   \blank       & ul \crule{0.5cm}{0.5cm}       
     \cr
     }$}}     
\put( 0,30){$B_1$}
\put( 30,30){$B_2$} 
\put( -35,20){$\BK_{n,\rk,\rho}^{^{\textrm{\tiny red}}}\Bigl(~~~~~~,~~~~~~\Bigr)=$}

\put( 70,20){$\BK_{n,\rk,\rho}^{^{\textrm{\tiny blue}}}\Bigl(~~~~~~,~~~~~~~~~~~\Bigr)-a\BK_{n,\rk,\rho}^{^{\textrm{\tiny blue}}} \Bigl(~~~~~~,~~~~~~\Bigr)$}
\put(255,20){\makebox(0,0) {      
 $  \squaresize 0.500cm \thickness .000008cm \Thickness
.060cm \Young{ 
   \blank       & ludr
     \cr
     } $}}    

\put(105,20){\makebox(0,0) {      
 $  \squaresize 0.500cm \thickness .000008cm \Thickness
.060cm \Young{ 
   \blank       & ul \crule{0.5cm}{0.5cm}       
     \cr
     } $}}    \put( 105,30){$B_2$} 
     
  \put(135,20){\makebox(0,0) {      
 $  \squaresize 0.500cm \thickness .000008cm \Thickness
.060cm \Young{ 
   \blank       & ludr
     \cr
     } $}}

  \put( 129,30){$ ~W_1$}

\put(225,20){\makebox(0,0) {      
 $  \squaresize 0.500cm \thickness .000008cm \Thickness
.060cm \Young{ 
   \blank       & ul \crule{0.5cm}{0.5cm}       
     \cr
     } $}}    \put( 225,30){$B_2$} 
  \put( 255,30){$ ~W'_1$}
      
  \end{picture}
  \begin{picture}(0,0)
       \put( 100, -20){\makebox(0,0) {      
  $\squaresize 0.500cm \thickness .000008cm \Thickness
 .060cm \Young{ 
    \blank      & uld \crule{0.5cm}{0.5cm}    &urd   \cr}
   $ }} 
   \put( 30, -26){$\mbox{with}~~W_1\in ~~~~~~~~~~~~~~~ ~~~\mbox{ and }~~~~~~   W'_1\in $}
   \put( 90, -10){$  ~B_1~~W_1 $}
   
   \put( 235, -26 ){\makebox(0,0) { $ 
  \squaresize 0.500cm \thickness .000008cm \Thickness
.060cm \Young{ 
      ul  \crule{.5cm}{.5cm}       
      \cr   ldr    \cr }
      $}} 
      \put(250,-22){$B_1$}
       \put(250,-42){$W'_1$}
 \end{picture}
 $$ \be \label{BW'}\ee%


   \vspace*{1cm}
\noindent {\it Expressed in $(s,u)$-coordinates, taking into account the change of variables\footnote{Since the middle of the blue squares are given by $(\xi,\eta)\in 2\BZ\times (2\BZ+1)$, it is advantageous to replace $s_i$ by $2s_i$ in the map (\ref{changevar}); see comments following (\ref{changevar}).} (\ref{changevar}), the blue squares $B_i$ for $i=1,2$ and the white squares $W_1$ and $W'_1$ read:
 $$ 
B_i:  (\xi_i,\eta_i)\Leftrightarrow(2s_i,u_i)=(\eta_i+1,\tfrac 12 (\eta_i-\xi_i+1))
$$
\be \label{BW}W_1=B_1+(1,0),~~~W'_1=B_1-(1,1)
,~~\mbox{in $(s,u)$-coordinates}.\ee
Coordinatewise, one finds
 \be 
\bl
\BK_{ n,\rk,\rho}^{^{\textrm{\tiny red}}}
 (\xi_1,\eta_1;\xi_2,\eta_2)&= \BK^{blue}_{ } \ \left( 2s_2,u_2 ; 2s_1\!+\!1,u_1  \right)
\!  -a  \BK^{blue}_{} \ \left(2s_2,u_2 ; 2s_1\!-\!1,u_1\!-\!1\right)
\\& = \frac{\widetilde\LR}{\widetilde\LR(1)} (-{\mathbb K}_0^{(\lb)}+(-1)^\ka {\mathbb K}_1^{(\lb)})(\xi_1,\eta_1 ; \xi_2,\eta_2)
,\el\label{Lkernel}
\ee
where
 \be \bl
  {\mathbb K}_0^{(\lb)}&=(1+a^2)  \Id_{\xi_1 < \xi_2} \int_{\Ga_{0,a}} {dz  \over  2\pi \I}  z^{(\xi_1-\xi_2)/2}   
{(1+az)^{ \frac{\eta_1-\eta_2}2-1 }  \over   (z-a)^{ \frac{\eta_1-\eta_2}2+1 }}
\\  
 {\mathbb K}_1^{(\lb)}&=  
 \left[
  \bl &S_1^\lb(2s_2,u_2 ; 2s_1\!-\!1,u_1\!-\!1) 
  \\&+ \left\langle \left(I-K^\lb {(0,0)}\right)^{-1}_{\geq M} a_{u_1-1,1}^{(2s_1-1)} , b^{(2s_2)}_{u_2} \right\rangle_{\geq M}
  \el \right],\label{Li}
  \el\ee
  in terms of the notation (\ref{S}) for $S_1^\lb$ and $a^{(s)}_{u,1}$. In formulas (\ref{Lkernel}) and (\ref{Li}) one must view the $(2s_i, u_i)$ as expressed in the $(\xi_i,\eta_i)$ variables via (\ref{BW}). 
  }
 \medbreak

\proof The proof of (\ref{BW'}) or (\ref{BW}) requires Lemma \ref{th08} below, together with Kenyon's Theorem; for details, see \cite{ACJvM}. To check the explicit expression (\ref{Lkernel}), one puts the expression (\ref{Kout}) into the first equality in (\ref{Lkernel}). The linear combination in (\ref{Lkernel}) leads to a linear combination on the $S_0^{(\lb)}$, which itself requires knowing a linear combination of the $F_u$'s defined in (\ref{FG}). So one checks the following:
%
$$
F^{(2s+1)}_{u } (w)-aF^{(2s-1)}_{u-1 } (w)
=   \frac{1+a^2}{ w-a}F^{(2s-1)}_{u-1 } (w),
$$
and so for the $S^\lb$ and $a_\dt$ as in (\ref{FGab}), one has:
$$\bl
S_0^\lb(2s_2,u_2 ; 2s_1\!+\!1,u_1 )
-& S_0^\lb( 2s_2,u_2 ; 2s_1\!-\!1,u_1-1 )=S^{(\lb)}_1 ( 2s_2,u_2 ; 2s_1\!-\!1,u_1-1 )
\el$$
$$\bl
a^{2s_1+1}_{u_1,0}  -&a a^{2s_1-1}_{u_1-1,0}=  a^{2s_1-1}_{u_1-1,1},
\el $$
which establishes formulas (\ref{Lkernel}) and (\ref{Li}) for the kernel  and which ends the proof of Proposition \ref{Kernelred}. \qed

    

\begin{lemma}\label{th08}
In the graph dual to the tiling, let the white points be $w=(w_1,w_2)$ and black points $b=(b_1,b_2)$; then we have for the inverse Kasteleyn matrix
\begin{equation}
\bl
  K_{Kast}^{-1}& \Bigl((w_1,w_2),(b_1,b_2)\Bigr)  = -(-1)^{(w_1-w_2+b_1-b_2+2)/4}
\\
&  \times\BK^{blue}_{n,r,\rho}  \left( {b_2+1   } ,  {b_2-b_1+1  \over   2} ,  {w_2 + 1  },   {w_2-w_1+1  \over   2} \right)
.\el
\end{equation}
\end{lemma}

\proof The proof proceeds along the lines of \cite{ACJvM}, with extra-care taken with regard to the location of the boundary terms in this more general situation. A crucial part of the argument is that $\BK_{n,\rk,\rho}^{^{\textrm{\tiny blue}}}=- \BK_{n,\rk,\rho}^{^{\textrm{\tiny green}}}=\mbox{expression}~(\ref{Kgreen'})$ off the diagonal. \qed

  
\section{From $\BK_{n,\rk,\rho}^{^{\textrm{\tiny red}}}$ to the rescaled kernel ${\mathbb L}_{n,\rk,\rho}$}
 \subsection{The new scale and various rescaled functions}

Remember the various integration contours $a< \rho_1<\sg_1<\sg_2<R<\rho_2<\rho_3<a^{-1} $. Setting the integration variables $v=atv',~u=atu',~w=atw',~z=atz'$, and $ \lambda = at \lambda'$, we introduce the same scaling as in (\ref{scaling'}), but, much later in section 10, with $\tau_i=x_i+m-n=x_i+\ka$ and $y_i$ replaced by $\frac{y_i}{\sqrt{2}}$,
\be\label{scaling}
n=t^{-2},~a=1+\beta t ,~  \xi_i := \frac 2{t^2}-2x_i,~~~
\eta_i :=  \frac1{t^2}+ \frac{y_i\sqrt{2}}t-1.
\ee
$\eta$ being odd along the blue tiles, the difference $\Dt \eta=2$ and so  \be 1=\frac {\Dt \eta_2}2=\frac{dy_2}{t\sqrt{2}  }.
\label{oneform}
\ee
Then the map (\ref{BW}) together with the scaling (\ref{scaling}) implies that whenever  $u_i$ and $s_i$ appear, as in (\ref{BW}), (\ref{Lkernel}) and (\ref{Li}), they must be replaced by
 \be \label{sr}\bl
   2s_i&= \eta_i+1 =  \tfrac 1{t^2}+\tfrac{y_i\sqrt{2}}{t} ,~~u_i=\tfrac 12
 (\eta_i-\xi_i+1)=\tfrac 12 (-\frac 1{t^2}+\frac{y_i\sqrt{2}}{t}+2x_i)
 \el \ee
Putting the scaling above in the expression $\rho(u)$, defined in (\ref{defs}), leads to the following $t$-dependent polynomial $h(z):=(-tz)^{n+1}\rho(atz)$ in $z$,  allowing the following decomposition (omitting the $~'$ in $z'$) and having the following limit for large $n=t^{-2}$: 
\be \label{h}\bl h(z)&:=
 \left(\frac{(1-tz )^{n+1}}{  \psi}\right)\left((1+a^2zt)^n
  \psi \right)=: h_0(z) h_1(z)= e^{-z^2+2\beta z}(1+O(t))
\el \ee
with
 \be\label{h'} \bl
 h_0(z)&=\frac{(1-tz )^{n+1}}{  \psi}=\sum_{i=0}^{n+1}\al_i z^i \mbox{,    with  } \al_0=\frac1\psi\mbox{ and }\al_\rk=1,
 \\
 \psi :=&\int_{\Ga_0}  \frac{(1-t{\lb } )^{n+1}d{\lb } }{2\pi \I {\lb } ^{\rk+1}}=\left({{n+1}\atop{\rk}}\right)(-t)^\rk\simeq \frac{ (-1/t)^\rk}{\rk!} 
 ,~  \frac1 \psi  =h_0(0) \simeq  \rk!(-t)^\rk.\el \ee
%
 %
%
%
The Fourier coefficients of $h(z)$, as in (\ref{h}) will play an important role; indeed, consider the power series about $z=0$:
 \be \label{h/h}\bl 
 h(z)&
 =\sum_{i=0}^\infty \hat h_i z^i
 \mbox{,   with  } \hh_i=\oint_{\Ga_0}
 \frac{dz~h(z)}{ 2\pi \I z^{i+1}}\mbox{   and   } h(0)=\hh_0=1
 \\
 \frac{1}{h(z)}&=\sum_{i=0}^\infty e_iz^i= \sum_{i=0}^\infty  \hat {\left({\tfrac1h}\right)}_i z^i
 \mbox{,   with  } e_i=\oint_{\Ga_0}
 \frac{dz~}{ 2\pi \I z^{i+1}h(z)},
\mbox{   and  }e_0=1.
 \el
\ee
and thus $1=(\sum_{i=0}^\infty \hh_i z^i)
(\sum_{j=0}^\infty e_j z^j)$.
The following integrals will also play an important role:
\be E_i=
 \oint_{\ga_{\sg_2}}
 \frac{dz~}{ 2\pi \I z^{i+1}h(z)}.
 \label{E}\ee
 Define for $\al\geq 1$, the matrix $\ER_\al$ and its inverse $\HR_\al$ (from(\ref{h/h})), with  $e_0=\hh_0=1$, 
 \be\label{EH}
 {\mathcal E}_\al:
  =\left(\begin{array}{cccccc}
   e_0&
   \\
   e_1&e_0   &&  O
  \\ \vdots&&\ddots
  \\
  e_{\al-1}&\dots  &   e_1&e_0
   \end{array}
   \right)\mbox{     and  } 
   \HR_{\al} := \ER_\al^{-1}=\left(\begin{array}{llllll}
\hh_0&  \\
\hh_1& \hh_0 &&O\\
\vdots \\
\hh_{\al-1}&\hh_{\al-2}&\dots& \hh_0
\end{array}\right) .
 \ee
 We shall also need  
  the permutation matrix of size $\al$,
 \be \label{permut}
 \PR_\al:=\left(\begin{array}{ccccc}
  & &&& 1
 \\O& &&1& 
 \\ \\ &1 &&&O \\ 1& &&& 
 \end{array}
  \right), \mbox{   with   }\det(\PR_\al)=(-1)^{\frac{\al(\al-1)}{2}}.
 \ee
For $1\leq i\leq \rk$, the $i$th column-vector of the matrix $\PR_{\rk} \HR_\rk$ will be denoted by
 \be\label{hhat}\bl
  {\widehat{\bf h}}_{\rk-i}=\left(\begin{array}{ccc}
   \widehat{\bf h}_{\rk-i}(0)
   \\
   \vdots
   \\
   \widehat{\bf h}_{\rk-i}(\rk-1)
   \end{array}
   \right)=\left(\begin{array}{ccc}
  {\widehat{  h}}_{\rk-i}
  \\\vdots \\
   {\widehat{  h}}_{1-i}
   \end{array}\right),~~\mbox{where  }~
  \widehat{\bf h}_{\rk-i} (\ell) :=
  \hat h_{\rk-i-\ell  }
  \mbox{  for   }{0\leq \ell\leq \rk-1} ,~~~
    \el\ee
 and so from $ \HR_{\al}   \ER_\al =I_\al$,  we have 
 \be
 ( {\widehat{\bf h}}_{\rk-1} , {\widehat{\bf h}}_{\rk-2},\dots, {\widehat{\bf h}}_{0}){\mathcal E}_\rk \PR_\rk=\Id_\rk.\label{hEP}\ee

We now define the following expressions in terms of the rescaled variables $u',v',z'$ (omitting the $'$'s and later  the indices $x_i,y_i$ in the the expressions $\FR_{x_1,y_1},~\GR_{x_2,y_2}$)
\be\bl  
h(z)&=(1+a^2tz)^n (1-tz)^{n+1}
 =e^{-z^2+2\beta z} (1+O(t))=:\widetilde h(z) (1+O(t))
\\ \\
~\FR(v )&:=\FR_{x_1,y_1}(v )  =v^{ -x_1}(1+a^2tv )^{s_1-1}
(1-tv )^{n-s_1}
\\&=v ^{-x_1}e^{-\frac 12 v ^2+(\beta+y_1\sqrt{2})v }(1+O(t))=:
{\widetilde \FR}_{ 1}(v ) (1+O(t))
\\ \\
 \GR_{ }(z )&=\GR_{x_2,y_2}(z )=z ^{-x_2-\ka}(1+a^2tz )^{s_2}
(1-tz)^{n+1-s_2}
\\
&=z ^{-x_2-\ka}e^{-\frac 12 z^2+(\beta+y_2\sqrt{2})z }(1+O(t)) =:
{\widetilde \GR}_{ 2}(z ) (1+O(t)).
 \el \label{series}\ee
%
%
 %
%
 %
The expressions (\ref{series}) enable us to define the following measures, depending on the polynomials $P_{ \bm \lb}(u)=\prod_1^\ka (\lb_i-u)$ only (as in (\ref{P})) :
$$
\bl
d\al_i(u )&:=\left\{\bl &\frac{P_{\bm \lb}(u )du }{ 2\pi \I u ^{i+1}}  \mbox{for  $0\leq i\leq \ka\!-\!1$}
\\  
& \frac{(-\FR)(u )du  }{ 2\pi \I  }\mbox{   for  $  i=\ka $}
\el\right. ,d\beta_j(v) :=\left\{\bl
&\frac{v^{\rk-j-1}dv}{2\pi \I h(v)P_{\bm \lb}(v) }
\mbox{         for  $0\leq j\leq \rk\!-\!1$}
\\
&\frac{\Phi(v)dv}{2\pi \I h(v)P_{\bm \lb}(v) }
\mbox{   for  $  j=\rk $}
\el \right. 
\el$$

 \be\bl 
d\theta_\rk(v)&:=  \frac{   dv}{  2\pi \I   \GR(v ) P_{\bm \lb}(v ) },\mbox{where}~
 \Phi(v):=\oint_{\ga_{\sg_2^+}}\frac{h(u)du}{2\pi \I \GR(u)(v-u)}
 \mbox{   , for $|v|<\sg_2^+$}. 
\el
\label{al-beta}
\ee
%
 It will be useful to define 
 \be\bl &d\widehat\al_i(u)\mbox{   with $P_{ \bm \lb}(u)$ removed for $0\leq i\leq \ka-1$}
\\ &d\widehat\beta_j(v)\mbox{   and  }
 d\widehat\theta_\rk(v)\mbox{   with $1/P_{ \bm \lb}(v)$ removed for $0\leq j\leq \rk $}.
 \el\label{al-beta-hat}\ee 

\subsection{
 The rescaled kernel ${\mathbb L}_{n,\tau,\rho}$ as a  $d {  \bm{\mu}}({\bm\lb})$-integration of another kernel ${\mathbb L}^{(\lb)}$}
  The rescaling enables us to define a new measure $d \bm\mu(\bm\lb)$ in terms of $h_0(\lb)$, as defined in (\ref{h'}), and a new operator $\LR $, compared to $d  \widetilde{\bm\mu}(\bm\lb)$ in (\ref{muhat}) and $\widetilde\LR$ in (\ref{Ltilde})
   \be \label{mu}\bl d {  \bm{\mu}}({\bm\lb}):=\prod_{j=1}^\ka d\mu(\lb_j)
  := &\prod_{j=1}^\ka\frac{h_0(\lb_j)d\lb_j }{ 2\pi \I \lb_j^{\rk+1}}
    =%
    \prod_{j=1}^\ka \left( \frac 1\psi\frac{(1-t{\lb_j } )^{n+1}d{\lb_j } }{  2\pi \I {\lb_j } ^{\rk+1}}\right)
   ,
   \el\ee
   \be\label{Ltil}\bl \LR(f)&:=\oint_{\ga_R^\ka} d \bm\mu(\bm\lb ) \det(I-{\mathcal K}^{(\lb )} { })_{\geq 0 }   f(\lb )
  ,\el\ee   
  where ${\mathcal K}^{(\lb )}:={\mathcal K}^{(\lb )} {(0,0) }$ will be defined in (\ref{RABK}) below.
  One checks that
 \be\label{h0exp}
 \oint_{ \Ga_0 }  d  \mu (\lb)=1. \ee

\begin{proposition}\label{newL}
For $-\Dt=\ka>0$, the kernel (\ref{Lkernel}) of Proposition \ref{Kernelred} reads as follows, upon inserting the scaling (\ref{scaling}) :  
\be\label{Lquadr}\bl
 \BK_{n,\rk,\rho}^{^{\textrm{\tiny red}}}  &(\xi_1,\eta_1 ; \xi_2,\eta_2) 
\frac{\Dt \eta_2}{2}
  = (-1)^{s_2-s_1 } a^{u_2-u_1}t^{x_2-x_1 }
   {\mathbb L}_{n,\tau,\rho}
(x_1,y_1;x_2,y_2)
\sqrt{2}dy_2\tfrac{ 1+a^2 }{  2}
\el\ee
where ${\mathbb L}={\mathbb L}_{n,\tau,\rho}
(x_1,y_1;x_2,y_2)$
\be  \label{Lsca}
 {\mathbb L} 
=-  {\mathbb L}_0 
+(-1)^\ka \left(   {\mathbb L}  _1
+  {\mathbb L} _2\right)
\mbox{,  with } {\mathbb L} =\frac{\LR}{\LR(1)}\left({\mathbb L}^{(\lb)}  \right)~\mbox{and}~
{\mathbb L}^{(\lb)}=\sum_{i=0}^2  {\mathbb L}^{(\lb)}_i  
 \ee
with (remember $s_1-s_2=\frac{(y_1-y_2)\sqrt{2}}t$)
 \be\label{Lixy}\bl
-  {\mathbb L}^{(\lb)}_0 (x_1,y_1;x_2,y_2) & =\Id_{x_1>x_2 }\oint_{\ga_{\rho_1}}\frac{\FR(u) du}{2\pi \I u^\ka \GR(u)}=\oint_{\ga_{\rho_1}}\tfrac{\Id_{x_1>x_2 }du}{2\pi \I u^{x_1-x_2}}\tfrac{(1+a^2tu)^{  s_1-s_2-1} }{(1-tu)^{ s_1-s_2  +1  }}
\\
  {\mathbb L}^{(\lb)}_1 (x_1,y_1;x_2,y_2) &= -\SR(\lb)
\\
  {\mathbb L}^{(\lb)}_2 (x_1,y_1;x_2,y_2) &=
  -c_{\ka+1}^\lb +\left\la (\Id-  {\mathcal K}^{(\lb)} {(0,0)})^{-1}_{\geq 0}  A^{\lb}_{x_1,y_1},    B^{\lb}_{x_2,y_2}\right\ra  ,
\el \ee
%
where 
 for $k,\ell\geq 0$,  (keeping in mind the radii: $a< \rho_1<\sg_1<\sg_2<R<R+<\rho_2<
 \rho_3<a^{-1}$), upon setting ${\mathcal K}^{(\iy)}_{k ,\ell }={\mathcal K}^{(\iy)}_{k ,\ell } (0,0)$
  \be \label{RABK}\bl
 %
%
 %
 {\mathcal K}^{(\lb)}_{k ,\ell }(0,0)&
 =\oint_{\ga_{\sg_1}}\frac{h(u)du}{(2\pi \I)^2 u^{ \rk-\ell  }}
    \oint_{\ga_{\sg_2}}\frac{ v^{\rk-k- 1}dv}{(v-u)h(v)}
  \frac{P_{\bm \lb}(u)}{P_{\bm \lb}(v)}
  =:{\mathcal K}^{(\iy)}_{k ,\ell }
 -\sum_{i=1}^{\ka } 
 b^\lb_{i   }(k)\hat h_{\rk-\ell-i  }
 \\   B^{\lb}_{x_2,y_2}(\ell)& =
\oint_{\ga_{\rho_1}}\frac{du}{(2\pi \I)^2  \GR_{ } (u)}\oint_{\ga_{\sg_1}}
    \frac{ h(v) dv}{  (  v\!-\!u) v^{\rk-\ell} }\frac{P_{\bm \lb}(v)}{P_{\bm \lb}(u)}
    =:
  B^{\iy}_{x_2,y_2}
  (\ell)+\sum_{i=1}^{\ka  } c^\lb_{i }  \hat h_{\rk-\ell-i   }
\\  \SR(\lb)
& = 
  -\oint_{\ga_{\rho_2}}d\al_\ka (u)\oint_{  \ga_{\rho_3}
  } \frac{d\theta_\rk(v)}{v-u}%
%
%
, \el\ee
 and for $1\leq i\leq \ka$ and $  j\geq 0$: (see notation (\ref{h/h}), (\ref{E}) for $e_i$ and $E_i$) 
 \be  \footnotesize
  \label{cb}\bl
c^\lb_{i}
  &=\oint_{\Ga_0}\frac{dw}{2\pi \I w^{i }}\oint_{\ga_{\rho_1}}\frac{du}{ 2\pi \I   \GR_{} (u)}\frac{\frac{P_{\bm \lb}(w)}{P_{\bm \lb}(u)}-1}{w-u}
 %
 ~\mbox{and}~ c_{\ka+1}^\lb    :=
  \oint_{\ga_{\rho_2}}d\al_\ka (u)\oint_{  \Ga_{\bm\lb} 
  }  \frac{d\theta_\rk(v)}{v-u} %
 \\b^\lb_{i   }(j) 
&=
\oint_{\ga_{\sg_2}}\frac{u^{\rk-j-1}du}{ 2\pi \I   h (u)}\oint_{\Ga_0}\frac{dw}{2\pi \I w^{i }}
\frac{\frac{P_{\bm \lb}(w)}{P_{\bm \lb}(u)}-1}{w-u} \mbox{,~~for~}j\geq 0
\\&= %
\oint_{\Ga_0}d\al_{i-1}(u)\oint_{\ga_{\sg_2}-\Ga_0^+}
\frac{d\beta_j (v)}{u-v}-e_{i+j-\rk }+E_{i+j-\rk } \mbox{,    for $0\leq  j \leq \rk-1$,}
  \\
    b^\lb_{\ka+1}(j)&: = A^{\lb}_{x_1,y_1}(j)= \oint_{\ga_{\rho_2}}\frac{\FR (u)du}{(2\pi \I)^2 }
   \oint_{\ga_{\sg_2}}\frac{  v^{\gr-j-1}dv}{(v-u)h(v) P_{\bm \lb}(v)  }\mbox{,\qquad  ~for~}j\geq 0,\\&=
  \oint_{\ga_{\rho_2}}d\al_\ka (u)\oint_{\ga_{\sg_2}-\Ga^+_0}
\frac{d\beta_j (v)}{u-v }
 \mbox{, \qquad\qquad\qquad   for $0\leq  j \leq \rk-1$.}
  \el\ee
  Notice that $ {\mathcal K}^{(\lb)}_{k ,\ell }(0,0)=0$  for $  \ell \geq \rk$.

\end{proposition}

\proof Inserting the scaling (\ref{scaling}) in the expression (\ref{Li}) for ${\mathbb K}^{(0)}$, one finds 
$$-{\mathbb K}^{(0)}\frac{\Dt \eta_2}{2}
=(-1)^{s_2-s_1}a^{u_2-u_1} t^{x_2-x_1}\oint_{\ga_{\rho_1}}\frac{\FR(w) dw}{2\pi \I w^\ka \GR(w)}\sqrt{2}dy_2\tfrac{1+a^2}{2}.
$$
Next the expressions which appear in the kernel $\BK_{n,\rk,\rho}^{^{\textrm{\tiny red}}}$ as in (\ref{Lkernel}) with the scaling (\ref{scaling}) give:
 $$\bl
\frac{1+a^2}{w-a}F_{u_1-1}^{2s_1-1}(w)&=(-1)^{n-s_1 }(1+a^2)a^{-u_1}t^{-x_1}\FR_{s_1,u_1}(w')
\\
G_{u_2}^{2s_2}(v)& =(-1)^{n+1-s_2}\frac{a^{ u_2 +\ka-1} t^{x_2+\ka}}{P_{\bm \lb}(v')\GR_{s_2,u_2}(v')},
\el $$

 From (\ref{Li}) and (\ref{S}), it follows that, using $n-M=\rk-1$, 
 \be\label{S1}
\bl
   S^{(\lb)}_1   (2s_2  , u_2 ; &2s_1-1,u_1-1)  
\\
&=:(-1)^{s_2-s_1 -1}(1+a^2)  a^{u_2-u_1+\ka }t^{x_2-x_1+\ka+1}  ( \SR_{s_2s_1}^\lb+c_{\ka+1}^\lb
 )\el \ee
 $$
 \bl
 a_{u_1-1,1}^{(2s_1-1)}(k)\Bigr|_{k=M+k'}
  =&(-1)^{s_1+k+1}(1+a^2)a^{-u_1-k }t^{n-x_1-k+1}  {\prod_1^\ka \lb'_i}
 \\ & \times 
 \oint_{\ga_{\rho_2}}\frac{\FR_{x_1,y_1}(w')dw'}{(2\pi \I)^2 }
  \oint_{\ga_{\sg_2}}\frac{  v'^{n-k}dv'}{ (v'-w')h(v')P_{\bm \lb}(v')}\Bigr|_{k=M+k'}
 \\=&(-1)^{s_1+k+1}(1+a^2)a^{-u_1-k }t^{n-x_1-k+1}
 \left({\prod_1^\ka \lb'_i} \right)  A^{\lb}_{x_1,y_1}(k')
 \el
 $$
 \be\bl
  b^{(2s_2)}_{u_2}(\ell)\Bigr|_{\ell=M+\ell'}&:=(-1)^{s_2+\ell+1}a^{ u_2+\ell+\ka}t^{-n+x_2+\ell+\ka}\tfrac1{\prod_1^\ka \lb'_i}
   \\& \times\oint_{\ga_{\rho_1}}\frac{dz'}{(2\pi \I)^2  \GR_{x_2,y_2} (z')}\oint_{\ga_{\sg_1}}
    \frac{ h(u') du'}{  (  u'-z') u'^{n+1-\ell} }\frac{P_{\bm \lb}(u')}{P_{\bm \lb}(z')} \Bigr|_{\ell=M+\ell'} 
 \\
 &:=(-1)^{s_2+\ell+1}a^{ u_2+\ell}t^{-n+x_2+\ell}  \tfrac1{\prod_1^\ka \lb'_i}
  B^{\lb}_{x_2,y_2}(\ell') 
\el\label{a1b}\ee
   where $  A^{\lb}_{x_1,y_1}(k')$ and $ B^{\lb}_{x_2,y_2}(\ell')$ are as in (\ref{RABK}) and (\ref{cb}). For future use we rename $A^\lb_{x_1,y_1}=b_{\ka+1}$ and use  the measures (\ref{al-beta}); this is only valid in the range $0\leq j\leq \rk-1$, since the $d\beta_j$ integral is pole-free at $0$. To show the second formula for $  B^{\lb}_{x_2,y_2}(\ell)$ in (\ref{RABK}), notice %
 the following expression is a polynomial of degree $\ka-1$ in $v$,
 \be \label{rat}
 \frac{ \frac{P_{\bm \lb}(v)}{P_{\bm \lb}(u)}-1 }{v-u}
=\frac{1}{v-u}
\left(\prod_1^\ka (1+\frac{v-u}{u-\lb_i})-1\right)
=\sum_1^\ka D_i(u,\lb)v^{i-1}\ee
with coefficients
\be \label{Di} D_i(u,\lb)=\oint_{\Ga_0}\frac{dw}{2\pi \I w^i}
 \frac{ \frac{P_{\bm \lb}(w)}{P_{\bm \lb}(u)}-1 } {w-u}.
\ee%
Using formula (\ref{rat}) for the $D_i(u,\lb)$ as in (\ref{Di}), we find the triple integral below, where 
 the last integral about $\ga_{\sg_1}$ can obviously be replaced by $\Ga_0$,
$$
\bl
 B^{\lb}_{x_2,y_2} (\ell)\! -\!
  B^{\iy}_{x_2,y_2}(\ell)
 &=\sum_{i=1}^\ka 
\oint_{\Ga_0}\frac{dw}{2\pi \I w^i}
\oint_{\ga_{\rho_1}}\frac{du}{ 2\pi \I   \GR_r (u)}
\frac{\frac{P_{\bm \lb}(w)}{P_{\bm \lb}(u)}-1}{w-u} \oint_{\ga_{\sg_1}\to \Ga_0}
    \frac{ h(v)  dv}{2\pi \I    v^{\rk-\ell-i+1} } 
\\&=: \sum_{i=1}^{\ka } c^\lb_{i }  \hat h_{r-\ell-i }  ~\mbox{ for}~~0\leq \ell\leq \rk-1
 \el$$
 with $
 c_{i }^\lb$ as in (\ref{cb}). In vector notation, this reads:
\be \label{Bvector}{\bf   B^{(\lb)}}
 ={\bf    B^{(\infty)}}+\sum_{i=1}^{\ka } c^\lb_{i}\widetilde{ {\bf \hat h}}_{r -i}.
 \ee
%
%
    %
 with understanding that here $\widetilde{ {\bf \hat h}}_{r -i}=(\hat h_{\rk-i},\dots,\hat h_{1-i},\hat h_{ -i},\dots)^\top\in \BC^\iy$ extends the vector ${ {\bf \hat h}}_{r -i}\in \BC^{\rk}$ to an infinite vector by adding $\hat h_{ -i}=\hat h_{-1 -i}=\dots=0$'s.
 
    Using the scaling $u=atu'$ and $v=atv'$, the kernel (\ref{K00}) for $k,\ell\geq 0$, can be expressed as follows, again using $n-M=\rk-1$,
    \be\bl
&  K^{(\lb)}_{k+M,\ell+M}(0,0) 
\\&:= \oint_{\ga_{\sg_1}}\frac{(-1)^{k+\ell}du}{(2\pi \I)^2} \oint_{ \ga_{\sg_2}  }\frac{dv}{v-u} \frac{u^{\ell+M}}{v^{k+M+1}} 
%
  \frac{\rho_{\lb }( u)}{\rho_{\lb }( v)}
%
   \\
   &=(-at)^{\ell-k }
   \oint_{\ga_{\sg_1}}\frac{du'}{(2\pi \I)^2}
    \oint_{\ga_{\sg_2}}\frac{dv'}{v'-u'}
   \frac{ v'^{\rk-k- 1}}{  u'^{ \rk-\ell  }} \frac{h(u')P_{\bm \lb}(u')}{h(v')P_{\bm \lb}(v')}
     =:(-at)^{\ell-k }{\mathcal K}^{(\lb)}_{k,\ell}(0,0)
\el \ee
The second formula for ${\mathcal K}^{(\lb)}_{k,\ell}(0,0)$ in (\ref{RABK}) follows from (\ref{rat}) and (\ref{Di}), 
 where again the $v$-integration about $\ga_{\sg_1}$ can be replaced by one about $\Ga_0$,
 $$\bl
 &{\mathcal K}^{(\lb)}_{k ,\ell } (0,0)  -{\mathcal K}^{(\iy)}_{k ,\ell }(0,0)~~~\mbox{ for $0\leq k $ and}~~0\leq \ell\leq \rk-1
\\&=-\oint_{\ga_{\sg_2}}\frac{u^{\rk-k-1}du}{ 2\pi \I   h (u)}
\oint_{\ga_{\sg_1} }
    \frac{ h(v)  dv}{2\pi \I    v^{\rk-\ell } } 
    \frac{ \frac{P_{\bm \lb}(v)}{P_{\bm \lb}(u)}-1 }{v-u}
\\&=
-\sum_{i=0}^{\ka -1}
\oint_{\ga_{\sg_2}}\frac{u^{\rk-k-1}du}{ (2\pi \I)^3   h (u)}\oint_{\Ga_0}\frac{dw}{  w^{i+1}}
\tfrac{\frac{P_{\bm \lb}(w)}{P_{\bm \lb}(u)}-1}{w-u} \oint_{\ga_{\sg_1}\to \Ga_0}
    \frac{ h(v)  dv}{    v^{\rk-\ell-i } } =-\sum_{i=1}^{\ka } 
 b^\lb_{i   }(k)\hat h_{\rk-\ell-i  }.
\el $$
This establishes the first formula (\ref{cb}) for $b^\lb_{i}(j)$.
The previous formula reads in matrix notation as follows:
\be\label{Kmatrix}\KR^{(\lb)} (0,0)=
\KR^{(\infty)} (0,0)   -  \sum^{\ka }_{i =1} \widetilde{{\bf b}}_{i  }^\lb  \otimes   \widetilde 
 {\bf \hat h}_{r -i}
.\ee
Using the measures (\ref{al-beta}), the expressions $E_i$ as in (\ref{E}), working out the residue of $w^{-i-1}(w-u)^{-1}$ about $w=0$, one finds\footnote{Notice that for $0\leq k\leq \ka-1$ and $0\leq \ell\leq \rk-1$.
 $  
 \oint_{\Ga_0}d\al_k (u)\oint_{ \Ga_0+ }
 \frac{d\beta_\ell (v)}{u-v}
 =-e_{k+\ell-\rk+1  }
 $
} for $0\leq i\leq \ka-1$ and $0\leq j\leq \rk-1$ only:
\be\label{balbet}\bl
b_{i+1}^\lb(j)&=\oint_{\Ga_0}\frac{dw}{2\pi \I w^{i+1}}
\oint_{\ga_{\sg_2}}\frac{u^{\rk-j-1}du}{ 2\pi \I   h (u)}
\frac{\frac{P_{\bm \lb}(w)}{P_{\bm \lb}(u)} }{w-u} 
  + {\oint_{\ga_{\sg_2}}\frac{u^{\rk-j-1}du}{2\pi \I u^{i+1}h(u)} }
  \\&=
 \oint_{\Ga_0}d\al_i(w )\oint_{\ga_{\sg_2} }\frac{d\beta_j (u )}{w-u }+E_{i+j-\rk+1}
  \\&=
 \oint_{\Ga_0}d\al_i(w )\oint_{\ga_{\sg_2} -\Ga_0+}\frac{d\beta_j (u )}{w-u }+ \oint_{\Ga_0}d\al_i(w )\oint_{ \Ga_0+}\frac{d\beta_j (u )}{w-u }+E_{i+j-\rk+1}
 \\&=
 \oint_{\Ga_0}d\al_i(w )\oint_{ \ga_{\sg_2} -\Ga^+_0}\frac{d\beta_j (u )}{w-u }  -e_{i+j -\rk+1  }+E_{i+j-\rk+1} 
 ,  \el\ee
 establishing the second formula (\ref{cb}) for $b^\lb_{i}(j)$ in the range $0\leq j\leq \rk-1$. %
 We now compute
  , using $M=n-\rk+1$,   remembering  ${  \KR}^{(\lb)} : ={  \KR}^{(\lb)} {(0,0)}$ and setting $\lb_i=at\lb_i'$,   
    \be \label{Lratio}
    \bl
     \frac{\widetilde\LR(\frac{ f(\lb)}{\prod_1^\ka \lb_i})}{\widetilde\LR( 1)}
&=
\frac{\displaystyle 
 \int_{\ga_R^\ka}   \prod^\ka_{j=1} \frac{(\lb_j-a)^{n+1}d\lb_j}{2\pi \I \lb_j^{\rk+2}}
 \det (I-{  \KR}^{(\lb)} { })_{\geq 0 }f( \lb )}
{\displaystyle 
 \int_{\ga_R^\ka}    \prod^\ka_{j=1} \frac{(\lb_j-a)^{n+1}d\lb_j}{2\pi \I \lb_j^{\rk+1}}
  \det (I-{  \KR}^{(\lb)} { })_{\geq 0 }}
    \\
&=\frac{1}{(at)^\ka}
\frac{\displaystyle 
 \int_{\ga_R^\ka}  \frac{ d \bm\mu(\bm\lb')}{\prod _1^\ka \lb_j'}
 \det (I-{\mathcal K}^{(\lb)} { })_{\geq 0 }f(at\lb')}
{\displaystyle 
 \int_{\ga_R^\ka}    d \bm\mu(\bm\lb')
  \det (I-{\mathcal K}^{(\lb)} { })_{\geq 0 }}
   =:
   \frac{1}{(at)^{\ka}}\frac{ \LR(\frac{f(at\lb')}{\prod_1^\ka \lb'_i})}{  \LR(1)},
    \el
    \ee 
using the definition (\ref{Ltil}) of $  \LR$.
%
 We now check, upon inserting the expressions $f=S_1^\lb$, as in (\ref{S}), and  $f=$(the inner-product (\ref{Lixy})) into (\ref{Lratio}) and using the formulas (\ref{S1}) and (\ref{a1b}),:
 %
$$\bl  (-1)^{ \ka} \frac {\widetilde\LR(\frac{S_1^{\lb})}{\prod_1^\ka \lb_i})} 
{ \widetilde\LR     (1)} 
   \tfrac{\Dt \eta}2
&=    
(-1)^{s_2-s_1 -\ka }   a^{u_2-u_1  }t^{x_2-x_1  } \frac{ \LR(- \SR_{ }(\lb) -c^\lb_{\ka+1}
 )}{ \LR( 1) }\sqrt{2}dy\tfrac{ 1+a^2 }{2}
\el $$
and
\be\label{Lquadr'}\bl
\tfrac{ (-1)^{  \ka}}{\widetilde\LR    (1)}
 & \widetilde\LR 
      \left\langle  \left(I-K^{(\lb)} {(0,0) }\right)^{-1}_{\geq M} \frac{a_{u_1-1,1}^{(2s_1-1)}}{\prod_1^\ka \lb_i} , b^{(2s_2)}_{u_2}  \right\rangle_{\geq M} \frac{\Dt\eta}{2}
 \\&\hspace*{-.5cm}=
 (-1)^{s_2\!-\!s_1\!-\!\ka } a^{u_2-u_1}t^{x_2-x_1 }
 \frac{  \LR\left(\left\la (\Id\!-\!  {\mathcal K}^{(\lb)} { })^{-1} _{\geq 0}   A^{\lb}_{x_1,y_1},   B^{\lb}_{x_2,y_2}\right\ra
 \right) }{   \LR\left(1
 \right)}
 \sqrt{2}dy  \tfrac{1+a^2}{2} 
 . \el\ee
This proves formula (\ref{Lquadr}) in Proposition \ref{newL}.
\qed


  

 \section{Reduction from an infinite to a finite-dimensional problem}



Since $\KR^{\infty}_{k\ell} =0$ for $\ell\geq \rk$, we write $\KR^{\infty}_{k\ell} $ as a block matrix, with $\KR_1$ being  square  of size $\rk$, and $\KR_2$ of size $(\infty,\rk)$; so we have 
$$\bl (\KR^{\infty}_{ k,\ell} )_{0\leq k,\ell\leq \iy}=\left(\begin{array}{cccc}
\KR_1&0\\
\KR_2&0
\end{array}\right),~ (I-\KR^{\infty}_{ } )^{-1}=
\left(\begin{array}{cccc}
(I-\KR_1)^{-1}&0\\
\KR_2(I-\KR_1)^{-1}&I
\end{array}\right),~
\el $$
and so
\be\label{K1K2}\bl
\det(I-\KR^\iy)=\det(I\!-\!\KR_1) ,~(I-\KR^{\infty\top}_{ }  )^{-1}=
\left(\begin{array}{cccc}
(I\!-\!\KR_1^{\top})^{-1}&(I\!-\!\KR_1^{\top})^{-1}\KR_2^{\top}\\
0&I
\end{array}\right).
\el\ee
Decomposing $B^\iy$ into a column vector $B_1$ of size $\rk$ and  $B_2$ of size $\iy$, we define ${\bf B}^{\infty \KR^\top} $ and 
${\bf b}_{i }^{ \lb \KR_1 }$
\be \label{B1B2'}\bl
 {\bf B}^\iy =\left(B_1\atop B_2\right),~~{\bf B}^{\infty \KR^\top} :=  (I-\KR^{\infty \top})^{-1 } {\bf B}^{\infty } ,~~{\bf b}_{i }^{ \lb \KR_1 }=(\Id-\KR_1)^{-1}{\bf b}_{i }^{ \lb}
,\el\ee
we have
\be\label{B1B2}\bl {\bf B}^{\infty \KR^\top} &:=(I-\KR^\top)^{-1}{\bf B}^{\infty}=(I-\KR^\top)^{-1}\left(B_1\atop B_2\right)
\\&= \left(\begin{array}{ccc}(I-\KR_1^{\top})^{-1}(B_1+\KR_2^\top B_2)\\ B_2\end{array}
\right)=:\left(\begin{array}{lll}
{\bf B}_1^{\infty \KR^\top}\\ {\bf B}_2^{\infty \KR^\top}\end{array}\right).
 \el\ee
Introducing the  notation $[\sum_{i\geq 0}a_iz^i]_{[0,\ell]}:=
 \sum_{i= 0}^{\ell}a_iz^i
 $ and $[\sum_{i\geq 0}a_iz^i]_{ \ell }:=
  a_\ell  .
 $ and using (\ref{B1B2}), we define the following expression for $1\leq i\leq \ka+1$ :  
\be 
\bl \label{c'}c^{\prime \lb}_{i }&:=\left\la \widetilde{\bf b}^\lb _{i },  {\bf B}^{\iy \KR^\top}\right\ra_{[\rk,\infty]}-c^\lb_{i }=
\sum_{\ell=\rk}^\iy b_{i }^\lb(\ell)B_2(\ell)
  -c^\lb_{i }.
\el\ee
   Remembering the notation (\ref{hhat}) for ${\bf \hat h}_{\rk -i}$, but extended as an infinite vector by adding $0$'s, define the  $\al\times \beta$ matrix for $1\leq \al\leq \ka+1$ and $0\leq \beta\leq \rk$,
   \be \label{NN'}\bl \NR_{\al,\beta}&:=\left(\left\la  {\bf b}_{i }^{ \lb \KR_1 },{\bf \hat h} _{\rk -j-1}
 \right\ra\right)_{{1\leq i \leq \al }\atop{0\leq j \leq \beta-1}} 
%
= 
\left(\begin{array}{cc}
&({\bf b}_{ 1}^{\lb\KR_1})^\top\\
&\vdots
\\&({\bf b}_{\al }^{\lb\KR_1})^\top
\end{array}\right) 
 \left({\bf \hat h}_{\rk -1},\dots,{\bf \hat h}_{\rk-\beta}
\right)
  . \el \ee
The following square matrices of sizes $\ka$ and $\rk$ will also be useful: ($\ka+\sg=\rk$)
\be\bl  \label{NN''}
\NR_{\ka}:= \NR_{\ka,\ka},~~~~\NR'_\rk &:=\left(
\begin{array}{c}
\NR_{\ka,\rk}
\\   \hline 
O_{\sg,\rk}\end{array}\right) .
\el\ee
Also remember   
  the radii $a< \rho_1<\sg_1<\sg_2<R<R+<\rho_2<\rho_3<a^{-1}$.
 
 \begin{lemma} \label{EPK'}The following holds :
 \be\bl
 \left[\ER_\rk\PR_\rk(\Id-\KR_1^\top)\right]_{ i,j}
 & 
  =-\oint_{\ga_{ \sg_2 }-\Ga_0}\frac{d\widehat\beta_j(v)}{v^{i+1}}
 \mbox{~~for $0\leq i,j\leq \rk-1$}
 \\
  \left[\ER_\rk\PR_\rk(-B_1-\KR_2^\top B_2)\right]_i&=
 -\oint_{\ga_{ \sg_2 }-\Ga_0}\frac{d\widehat\beta_\rk(v)}{v^{i+1}}
\mbox{~~for $0\leq i \leq \rk-1$}
 \\
 (1\!-\!\dt_{i,\ka} )\! \left[\ER_\rk\PR_\rk(\Id \!-\!\KR_1^\top)\right]_{ i,j}+b^{(\lb)}_{i+1}(j)
 &=\oint_{\Ga^{(i)}_0}d\al_i(u)\oint_{\ga_{\sg_2}-\Ga_0^+}
\frac{d\beta_j (v)}{u-v}
~~\mbox{for $ 
 \left\{{0\leq i\leq \ka }\atop{0\leq  j\leq \rk-1}\right.$}
 \el
  \label{EPK}\ee
  \be \label{c'k}\bl
 c'_{i+1}- (1-\dt_{i,\ka})  \bigl({\mathcal E}_\rk &\PR_{\rk}(B_1+ \KR_2^\top B_2)\bigr)( i)~~~~~~~~~~~~~~~~\mbox{for}~~0\leq i\leq \ka
  \\&=
\oint_{\Ga^{(i)}_0}d \al_i(u )\left[\oint_{\ga_{\sg_2}-\Ga_0^+}\frac{d \beta_\rk (v )}{u -v }
+\dt_{i,\ka}\oint_{\ga^+_{R}} \frac{   d\theta_r (v)}{   u-v   
   }
\right],
 \el \ee
  where $\Ga^{(i)}_0$ stands for 
 \be\label{contour-i}\Ga^{(i)}_0=\Id_{0\leq i\leq \ka-1}\Ga_0  + \Id_{i=\ka} \ga_{\rho_2}
 .\ee
 The first and third equation in (\ref{EPK}) can be summarized by the following $\rk\times \rk$-matrix identity:
\be\label{NEPK}\bl
\left(I_\rk+\NR'_\rk
 \right)\ER_\rk\PR_\rk(I-\KR_1^\top)=
  \left(
\begin{array}{ll}\left(\dis\oint_{\Ga_0}d\widehat\al_i(u_i)\oint_{\ga_{\sg_2}-\Ga_0^+}
\frac{d\widehat\beta_j (v_j)}{u_i-v_j}
\frac{P_{\bm \lb}(u_i)}{P_{\bm \lb}(v_j)} \right)_{{0\leq i \leq \ka-1}\atop{0\leq j\leq \rk-1}}
\\ \\
\hline \\
~~~~~\left(-\dis\oint_{\ga_{ \sg_2 }-\Ga_0}\frac{d\widehat\beta_j(v_j)}{v_j^{i+1}}\right)_{{\ka\leq i \leq \rk-1}\atop{0\leq j\leq \rk-1}}
\end{array}\right).
 \el\ee

   \end{lemma}
   
 \noindent \underline  {\em Remark 2}: The entries of the matrix (\ref{NEPK}) involve 
  $d\widehat\beta_j(v_j)$ integrations about the annulus $\ga_{\sg_2}-\Ga_0^+$, which contains only one pole $v_j=1/t$, the pole of $h^{-1}(v)$; the $\lb_i\in \ga_R$ being outside $\ga_{\sg_2}-\Ga_0^+$. See Fig 7. The latter contour will have to be understood 
   as nested annuli: $\ga_{s_jr_j}\ni v_j$ of radii $s_{\ka-1}<\dots<s_0<\frac 1t <r_0<\dots<r_{\ka-1}<\sg_2$, depending on the index $0\leq j\leq \rk-1$. This will play a role in the proof of Proposition \ref{LemmaInt0}. The same remark holds for  the $d\widehat\al_i(u_i)$ integration about $\Ga_0$, although here it turns out not to play any role.
  
  \bigbreak



\proof
%
%
To establish the expression below for $\left(  {\mathcal K}_{1 }^{ } {\mathcal E}_\rk \right)_{k\ell}$ for $0\leq k,\ell\leq \rk-1$, one uses formula (\ref{Kmatrix}) for the kernel ${\mathcal K}_{ }^{\infty} $; one notices that in $\stackrel{*}{=}$ below we can replace  $\ga_{\sg_1}\to \Ga_0$, since $\sg_1<\sg_2$ and the integrand has a pole at $0$ only. One also uses:
\be \label{hh}
   \left[ \frac 1h\right]_{[0,k]}\!\!\!(\eta)
 = 
 \sum_{i=0}^k {e_{i}}\eta^i
 = 
  \eta^{k+1}\sum_{j=0}^k\frac{e_{k-j}}{\eta^{j+1}},
 \mbox{ and so }  h(\eta)\left[ \frac 1h\right]_{[0,k]}\!\!\!(\eta)=1+a_{k+1}\eta^{k+1}+\dots,\ee
the latter being a convergent series in $\Ga_0$. Altogether this gives us, using (\ref{hh}) above:
$$\footnotesize\bl
 \left(  {\mathcal K}_{1 }^{ } {\mathcal E}_\rk \right)_{k,\ell}&=\sum_{i=\ell}^{\rk-1}
( {\mathcal K}_1)_{k,i}e_{i-\ell} =\sum_{i=0 }^{\rk-1-\ell}
 ({\mathcal K} _1)_{k,\ell+i} e_i
  &\stackrel{*}{=}\oint_{\ga_{\sg_2}}\frac{d\ze~\ze^{\rk-k-1}}{(2\pi \I)^2h(\ze)}
 \oint_{\ga_{\sg_1}\to \Ga_0}\frac{d\eta~h({\eta})}{(\ze-\eta)\eta^{\rk-\ell}}
\sum_{i=0}^{ \rk-1-\ell }\eta^i e_i
 \el$$
 $$\footnotesize\bl  
 &= \oint_{\ga_{\sg_2}}\frac{d\ze~\ze^{\rk-k-1}}{(2\pi \I)^2h(\ze)}
  \oint_{\Ga_0}\frac{d\eta~}{(\ze-\eta)\eta^{\rk-\ell}}
  h({\eta})\left[\frac1{h(\eta)}\right]_{[0,\rk-\ell-1]}    =
 \oint_{\ga_{\sg_2}}\tfrac{d\ze~\ze^{\rk-k-1}}{ 2\pi \I  h(\ze)}
\zeta^{\ell-\rk}=E_{k-\ell }
 \el $$
and so, we have
\be\left[(\Id- {\mathcal K}_1) {\mathcal E}_\rk \right]_{i,j} =e_{i-j }-E_{i-j },\mbox{ and so }
\left[\ER_\rk\PR_\rk(\Id-\KR_1 ^\top)\right]_{ i,j}
   =e_{i+j-\rk+1}-E_{i+j-\rk+1}.
\label{KE} \ee
 The {\em first identity} in (\ref{EPK}) then follows from using the expressions (\ref{h/h}), (\ref{E}) and the definition (\ref{al-beta-hat}) of $d\widehat \beta_i$. The {\em third identity} in (\ref{EPK}) follows from (\ref{KE}) combined with the expressions (\ref{cb}) for $b_{i+1}^\lb(j)$ for $0\leq i\leq \ka-1$.
 %

  \medbreak
\noindent We now prove the {\em second identity} of (\ref{EPK}). To begin with, we compute $(B_1+ \KR_2^\top B_2)(\ell)$. For $0\leq \ell\leq r-1$, one has, using the fact that the $v$-integral in $\stackrel{*}{=}$ has a pole at $v=u$ only, and using the fact that passing the $\ze\in \ga_{\sg_2}$ contour inside $\ga_{\rho_1}$ goes through two contours, the $\ga_{\sg_1}$ and $\ga_{\rho_1}$-contours, yielding two residues  in $\stackrel{***}{=}$, one canceling with $B^{\infty}(\ell)$ and another being\footnote{Indeed, for $|u|<\sg_1$, we have for $\al \geq 1$,  $$\oint_{\ga_{\sg_1}}\frac{d\eta}{2\pi\I \eta^{\al}(\eta-u)}
   =0$$} $=0$, since $\ell \leq r-1$.
 So, we need   (remember $\rho_1<\sg_1<\sg_2$)
  $$ \bl (B_1&+ \KR_2^\top B_2)(\ell)
    = B^{\infty}(\ell) +\sum_{j=r}^\infty
  (\KR^{ }_2)_{ j,\ell}B^{\iy}(j)   
 \\&\stackrel{*}{=}B^{\infty}(\ell) + \oint_{\ga_{\sg_1}}\frac{h(\eta)d\eta}{(2\pi\I)^4 \eta^{r-\ell }}\oint_{\ga_{\sg_2}} \frac{ d\ze}{(\ze-\eta)h(\ze)}
  \oint_{\ga_{\rho_1}} \!\!  \frac{ du}
  {  \GR_{ }(u)}
 \underbrace{ \oint_{\ga_{\sg_1}}   {h(v) dv   \over  (  v\!-\!u) }\overbrace{ \frac{\ze^{r-1}}{v^r}\sum_{j=r}^\infty \left(\frac{v}{\ze}\right)^j}^{\mbox{\tiny$ (\ze-v)^{-1} $}}  
 }_{\mbox{\footnotesize$ \frac{h(u)}{\ze-u}$}}
 \\
  &   \stackrel{**}{=}B^{\infty}(\ell)
     + \oint_{\ga_{\sg_1}}\frac{h(\eta)d\eta}{(2\pi\I)^3\eta^{\rk-\ell}}\oint_{\ga_{\sg_2}}\frac{d\ze}{(\ze-\eta)h(\ze)}
  \oint_{\ga_{\rho_1}} \!\!  \frac{ h(u)du}
  { (\ze-u)\GR (u)}
   \el$$ $$\bl
   &\stackrel{***}{=}B^{\infty}(\ell)+ \oint_{\ga_{\sg_1}}\frac{h(\eta)d\eta}{(2\pi\I)^3 \eta^{\rk-\ell}}\oint_{\ga_{ \rho_1^-}}\frac{d\ze}{(\ze-\eta) h(\ze)}
  \oint_{\ga_{\rho_1}} \!\!  \frac{ h(u)du}
  {  (\ze-u)\GR_{}(u)}
  \\&~~~~~+
 \underbrace{ \oint_{\ga_{\sg_1}}\frac{h(\eta)d\eta}{(2\pi\I)^2 \eta^{\rk-\ell }}
  \oint_{\ga_{ \rho_1 }}\frac{  du}{(u-\eta)  \GR_{}(u)}
  }_{=-B^{\infty}(\ell)}
  +\underbrace{\oint_{\ga_{\sg_1}}\frac{ d\eta}{(2\pi\I)^2 \eta^{\rk-\ell }} 
  \oint_{\ga_{\rho_1}} \!\!  \frac{ h(u)du}
  {  (\eta-u)\GR_{}(u)}}_{=0}
\\%
   &\stackrel{****}{=}  \oint_{\ga_{\sg_1}}\frac{\eta^{\ell-\rk}h(\eta)d\eta}{(2\pi\I)^3}\oint_{\ga_{ \rho_1^-}}\frac{d\ze}{(\ze-\eta) h(\ze)}
  \oint_{\ga_{\rho_1}} \!\!  \frac{ h(u)du}
  { (\ze-u)\GR_{}(u)}.
   \el$$
  Next we have that for $0\leq \ell\leq \rk-1$, 
$$
\left(\PR_{\rk}(B_1+ \KR_2^\top B_2)\right)(\ell)
=(B_1+ \KR_2^\top B_2)(\rk-\ell-1);
$$
then using (\ref{hh}) and $(\ER_\rk)_{i\ell}$, we have (remember the contours below (\ref{radii}))
$$ \footnotesize\bl&\left({\mathcal E}_\rk \PR_{\rk}(B_1+ \KR_2^\top B_2)
\right)(i)
 \\&=
\oint_{\ga_{ \rho_1^-}}\frac{d\ze}{(2\pi\I)^3   h(\ze)} \oint_{\ga_{\rho_1}} \!\!  \frac{ h(u)du}
  {  (\ze-u)\GR_{ }(u)}
  \oint_{\ga_{\sg_1}}\frac{  d\eta}{  (\ze-\eta)}
 \left(h(\eta)\sum_{\ell=0}^k\frac{e_{i-\ell}}{\eta^{\ell+1}}\right)
  \\
  \\ & = \oint_{\ga_{ \rho_1^-}}\frac{d\ze}{(2\pi \I)^3 h(\ze)}
  \oint_{\ga_{\rho_1}} \!\!  \frac{ h(u)du}
  {  (\ze-u)\GR_{ }(u)}\oint_{\ga_{\sg_1}\to \Ga_0+\Ga_{\zeta}}      \frac{d\eta}{( \ze-\eta )\eta^{i+1} }h(\eta) \left[ \frac 1h\right]_{[0,i]}\!\!\!(\eta)
   \\
     \\ & = \oint_{\ga_{ \rho_1^-}}\frac{d\ze}{(2\pi \I)^3 h(\ze)}
  \oint_{\ga_{\rho_1}} \!\!  \frac{ h(u)du}
  {  (\ze-u)\GR_{}(u)}\oint_{\Ga_{0} }      \frac{d\eta}{( \ze-\eta ) \eta^{i+1}}
  h(\eta) \left[ \frac 1h\right]_{[0,i]}\!\!\!(\eta)
 \\&+ \oint_{\ga_{ \rho_1^-}}\frac{d\ze}{(2\pi \I)^3 h(\ze)}
  \oint_{\ga_{\rho_1}} \!\!  \frac{ h(u)du}
  {  (\ze-u)\GR_{}(u)}\oint_{ \Ga_{\zeta}}      \frac{d\eta}{( \ze-\eta )\eta^{i+1} }h(\eta) \left[ \frac 1h\right]_{[0,i]}\!\!\!(\eta)  
 \el$$
 \be\bl \footnotesize  \\ &\stackrel{ *}{=}  \underbrace{ \oint_{\ga_{ \rho_1^-}}\frac{d\ze}{(2\pi \I)^2 \zeta^{i+1}h(\ze)}
  \oint_{\ga_{\rho_1}\to\ga_{\sg_2}^+ } \!\!  \frac{ h(u)du}
  {  (\ze-u)\GR_{ }(u)}  }_{(1')}
 - \oint_{\ga_{\rho_1}} \!\!  \frac{ h(u)du}
  { (2\pi \I)^2 \GR_{ }(u)}\oint_{\ga_{ \rho_1^-}\to\Ga_0}\frac{\left[ \frac 1h\right]_{[0,i]}(\ze)d\ze}{  (\ze-u)\ze^{i+1}}
\\& =
  \oint_{\ga_{\sg_2}-\Ga_0}\frac{d\widehat\beta_\rk(\ze)}{\ze^{i+1}}\mbox{  (replacing $\ga_{\rho_1}$ in $(1')$ above by 
$\ga_{\sg_2+}$)},
\label{EPB}\el \ee
upon replacing in the first expression in $\stackrel{ *}{=}  $ above the contour $\ga_{\rho_1}$ by 
$\ga_{\sg_2^+}$ and next upon replacing in the second expression the contour $\ga_{\rho^-_1}$ by $\Ga_0$ and then the finite series $\left[ \frac 1h\right]_{[0,i]}$ by the convergent series $\frac 1h$. This ends the proof of the {\em second identity} of (\ref{EPB}).  

For future use, yet another way of writing the last expression in (\ref{EPB}), is to write the contribution from the $\Ga_0$-contour differently, namely multiply the $\ze$-integrand above and below by the polynomial $P_{\bm \lb}^\vr(\ze)$ for $\vr=0\mbox{ or }1$, and represent $\frac{\Phi(\ze)}{h(\ze)P_{\bm \lb}^\vr(\ze)}$ as a residue about $u=\ze$ and then replace $\Ga_\ze $ by $\Ga_0 +$, yielding another representation for (\ref{EPB}),
\be \bl   
(1')+& \oint_{\Ga_{0}}\frac{P_{\bm \lb}^\vr(\ze)d\ze}{ 2\pi \I   \ze^{i+1}} \oint_{\Ga_\ze \rightarrow \Ga_0+} \frac{ \Phi(u) du}{   h(u)P_{\bm \lb}^\vr(u)(\ze-u)}\mbox{~~}
  \\ &  \stackrel{ }{=}   {(1')}
 + \oint_{\Ga_0} d\widehat\al_i(\ze)  
\oint_{\Ga_{0+} }\frac{d\widehat\beta_{\rk}(u)}{\ze-u}
\stackrel{  }{=} {(1')}
 + \oint_{\Ga_0} d \al_i(\ze)  
\oint_{\Ga_{0+} }\frac{d \beta_{\rk}(u)}{\ze-u}.
 \el \label{EPB'}\ee

   We now prove {\em identity 
    (\ref{c'k})}. For $0\leq i\leq \ka-1$,  $c'_{i+1}$ is defined in (\ref{c'}), $B_2(\ell)$ in (\ref{RABK}) and (\ref{B1B2'}), $c_{i+1}$ in (\ref{cb}). The measures $d\al_i$ and $d\beta_j$ are defined in (\ref{al-beta}).  Then the fourth equality is obtained from the third by writing the triple integral as a difference of two such integrals, one works out the $\frac{-1}{w-u}$ part and one  replaces in the last integral $z\to u$ and $u\to \ze$,
$$\bl
 {c^{\lb'}_{i+1}} &=\sum_{\ell=\rk}^\iy b_{i+1}^\lb(\ell)B_2(\ell)-c_{i+1}^\lb  \hspace*{7cm}
\el$$ 
 \be\bl
&=\oint_{\Ga_0}\frac{dw}{2\pi \I w^{i+1}}
\oint_{\ga_{\sg_2}}\frac{du}{2\pi \I h(u)}
\tfrac{\frac{P_{\bm \lb}(w)}{P_{\bm \lb}(u)}\!-\!1}{w-u}
\oint_{\ga_{\rho_1}}\frac{dz}{(2\pi \I)^2 \GR(z)}
\overbrace{\oint_{\ga_{\sg_1}} \frac{  h(v)dv}{   (v\!-\!z)}
    \overbrace{\left(\tfrac{1}{u} \sum_{\ell=r}^\iy \tfrac{v^{\ell-r}}{u^{\ell-r}}\right)}^{\tiny{(u-v)^{-1} }}}^{\frac{h(z)}{ u-z}}
\\&\hspace*{11cm}-\underbrace{c^\lb_{i+1}}_*
%
 \\&=
\oint_{\Ga_0}\frac{dw}{2\pi \I w^{i+1}}
\oint_{\ga_{\sg_2}}\frac{du}{2\pi \I h(u)}
\frac{\frac{P_{\bm \lb}(w)}{P_{\bm \lb}(u)}-1}{w-u}
\oint_{ \begin{array}{c}\ga_{\rho_1}\to \ga_{\sg_2+}\\ \mbox{\tiny cancels}* \end{array}}\frac{h(z)dz}{2\pi \I (u-z)\GR(z)} 
\\&=
\oint_{\Ga_0} d\al_i(w)  
\oint_{\ga_{\sg_2} }\frac{d\beta_{\rk}(u)}{w-u}
 +
\underbrace{\oint_{\ga_{ \sg_2}}\frac{d\ze}{(2\pi \I)^2 \zeta^{i+1}h(\ze)}
  \oint_{\ga_{ \sg_2+}} \!\!  \frac{ h(u)du}
  {  (\ze-u)\GR_{ }(u)} 
}_{(1)}
\el\label{c'i}\ee
Then subtracting from (\ref{c'i}) the last alternative expression (\ref{EPB'}) for $\left({\mathcal E}_\rk \PR_{\rk}(B_1+ \KR_2^\top B_2)
\right)(i)$, one sees that the double integrals (1) and (1') in these expressions cancel and one obtains {\em identity (\ref{c'k})} for $0\leq i< \ka$ in Lemma \ref{EPK'}.
Finally, for $i=\ka$, we have from (\ref{RABK}), (\ref{cb}), (\ref{al-beta})
and from summing the geometric series, as in (\ref{c'i}):
  $$\bl
 &c^{\prime \lb}_{\ka+1}
  :=\sum_{\ell=\rk}^\iy b_{\ka+1}^\lb(\ell)B_2(\ell)-c^\lb_{\ka+1}
\\
&= \oint_{ \ga_{\sg_2}}\frac{    du}{  2\pi \I  h(u) P_{\bm\lb}(u)
 }
 \oint_{ \ga_{\rho_2}}  \frac{ \FR (v)dv}{2\pi \I( u-v)  }
 \oint_{\ga_{\rho_1}}\frac{dw}{(2\pi \I)^2  \GR (w)}\overbrace{\oint_{\ga_{\sg_1}}
    \frac{h(z) dz}{  (z-w)}\frac{1}{u-z} 
     }^{\frac{h(w)}{u-w}} 
      -c^\lb_{\ka+1}
  \\
    &= \oint_{ \ga_{\sg_2}}\frac{    du}{  2\pi \I  h(u) P_{\bm \lb}(u)}\oint_{ \ga_{\rho_2}}  \frac{ \FR (v)dv}{2\pi \I(u-v)  }
 \oint_{\ga_{\sg_2+}}\frac{ h(w) dw}{ 2\pi \I   (u-w)  \GR (w)} 
 \\
 &~~~ +\Bigl(\underbrace{\oint_{ \ga_{\sg_2}}+\oint_{\Ga_{\bm \lb} } }_{ \oint_{\ga_R+}}\Bigr)\frac{   du}{  2\pi \I  \GR(u)P_{\bm \lb}(u)
   }\oint_{ \ga_{\rho_2}}  \frac{ \FR (v)dv}{2\pi \I(u-v)  }
   \\&= 
   \oint_{\ga_{\rho_2}} d\al_\ka(v)  
\Bigl[\oint_{\ga_{\sg_2}-\Ga^+_0}\frac{d\beta_{\rk}(u)}{v-u}
  +%
 \oint_{\ga_{R+}} \frac{   d\theta_\rk (u)}{   v-u  
   }
      \Bigr],  \el
  $$
since $d\beta_{\rk}(u)$ is pole-free at $u=0$. 

To finally prove (\ref{NEPK}), using the definition (\ref{NN'}) of $\NR_{\al,\rk}$ for $0\leq \al\leq \ka+1$, we have: 
$$\bl
 \NR_{\al,\rk}\ER_\rk\PR_\rk 
&=
\left(\begin{array}{cc}
&({\bf b}_{ 1}^{\lb\KR_1})^\top\\
&\vdots
\\&({\bf b}_{\al  }^{\lb\KR_1})^\top
\end{array}\right) 
 \left({\bf \widehat h}_{\rk -1},\dots,{\bf \widehat h}_{0}
\right) \ER_\rk\PR_\rk
=
\left(\begin{array}{cc}
&({\bf b}_{ 1}^{\lb\KR_1})^\top\\
&\vdots
\\&({\bf b}_{\al  }^{\lb\KR_1})^\top
\end{array}\right) 
 I_{\rk}
\\&=\left(\begin{array}{cc}
&({\bf b}_{ 1}^{\lb\KR_1})^\top\\
&\vdots
\\&({\bf b}_{\al  }^{\lb\KR_1})^\top
\end{array}\right)
=({\bf b}^{(\lb)}_{i }(j))_{{1\leq i \leq \al }\atop{0\leq  j\leq \rk-1}}
(I-\KR_1^\top)^{-1},\el$$
and so, we deduce the following  identity, involving $\NR_\rk'$ as in (\ref{NN''}),
\be\label{NN'''}
\bl
  \left(I_\rk+\NR'_\rk
  \right)\ER_\rk\PR_\rk(I-\KR_1^\top) 
 =   \ER_\rk\PR_\rk(I-\KR_1^\top)+\left(\begin{array}{llll}( {\bf b}_{i+1}^{(\lb)  }(j)
 )_{{{0\leq i \leq \ka-1}\atop{0\leq j\leq \rk-1}}\atop{}}
\\   \hline
~~~~~~~~O_{\sg,\rk}\end{array}\right),
\el\ee 
which by the first and third identity in (\ref{EPK}) leads to the matrix identity (\ref{NEPK}). Fore future use, we replace the dummy integration variables $u$ and $v$ by $u_i$ and $v_j$ according to the rows and columns and one replaces the measures $d\al_i$ and $d\beta_j$ by 
$d\widehat\al_i$ and $d\widehat\beta_j$, at the expense of having the $\lb$-dependent polynomials $P_{\bm \lb}(u)$ and $P_{\bm \lb}(v)$ appear (see (\ref{al-beta-hat})).
This ends the proof of Lemma \ref{EPK'}.\qed 
 
%


   
   \section{An integral representation of the $\mathbb L$-kernel 
    }
  
   In Proposition \ref{newL} we found the expression for the kernel 
   $\mathbb L=\frac{\LR}{\LR(1)}(\mathbb L^\lb)$ in terms of a ${\bm\lb}$-dependent kernel $ \mathbb L^\lb  $, where the operator  $\LR$  essentially amounts to $d{\bm \mu}(\bm \lb)$-integration. The main difficulty stems from the presence of an innerproduct between two $\bm \lb$-dependent infinite vectors, which in addition contain a resolvent; see (\ref{Lixy}). So, the main goal of this section is to first express this innerproduct as a determinant of a matrix $ \MR_{\ka+1}$ of size $\ka+1$ (\ref{M'}), still containing innerproducts;  secondly to enlarge the matrix to size $\rk+1$ in order to remove the innerproducts by row and column operations. Finally one takes out the (${\bm \lb}$-free) $d\widehat\al_i(u_i)$ rowwise  and   $d\widehat\beta_j (v_j) $-integrations  columnwise from the matrix, ending up with being able to pull the $d{\bm \mu}(\bm \lb)$-integration all the way to the right; this leads to having to $d{\bm \mu}(\bm \lb)$-integrate pure rational functions (\ref{Vplus}) in ${\bm \lb}$, $u_i$ and $v_j$. So, the main point of this section is the integral representation (\ref{LRfull}) in Proposition \ref{Prop5.2}.

   
  Remembering the definition (\ref{NN'}) of $\NR_\ka=\left(\left\la {\bf b}_{i+1}^{(\lb)\KR },{\bf \hat h} _{\rk -j-1}
\right\ra\right)_{0\leq i,j\leq \ka-1}$, we define an additional matrix of size $\ka+1$,
   \be \label{M'}
 \MR_{\ka+1}=
\left(\begin{array}{llllll}
   & &\vline \hspace*{.2cm}\star_0
\\
& I_\ka+\NR_\ka& \vline  \\
 &  &\vline\hspace*{.2cm}\vdots& &&
\\ 
&  &\vline \hspace*{.2cm}\star_{\ka-1}
\\ \hline \\
    \left\la {\bf b}_{\ka+1}^{\lb\KR_1},{\bf \hat h} _{\rk -1} 
\right\ra &
\dots
 \left\la {\bf b}_{\ka+1}^{\lb\KR_1},{\bf \hat h} _{\rk -\ka}\right\ra
&\vline \hspace*{.2cm}\star_{\ka}
\\
\end{array}\!\!\!\!\right),
\ee
with
   $$\star_i:=\left\la {\bf b}_{i+1}^\lb,{\bf   B}^{\infty \KR^\top}_{ }\right\ra_{[0,\rk-1]}+c'_{i+1}
~~\mbox{for $0\leq i\leq \ka$}.$$
%
Remember the kernel ${\mathbb L}$ in (\ref{Lsca}) involves the two expressions (\ref{Kratio}) below:

 \begin{lemma} \label{PropM}
 Each of the following expressions have a determinantal representation, being aware that $\KR^{(\lb)}=\KR^{(\lb)}(0,0)$ and $\det (I-\KR ^{ \iy })=\det(I-\KR_1)$, (see (\ref{RABK}) and (\ref{K1K2})):
\be\label{Kratio} \bl & \frac{\det (I-\KR^{(\lb)})}{\det (I-\KR ^{ \iy })}
=\det (I+\NR_\ka)
\\
&\frac{\det(I-\KR^\lb)}{\det(I-\KR ^\iy)} \left( \langle(I-\KR^\lb)^{-1} {\bf A}^\lb,  {\bf B}^\lb\rangle_{\geq 0}-c_{\ka+1}^{(\lb)} \right)  
 =   \det (\MR_{\ka+1}).
\el 
\ee


\end{lemma}

\proof 
Using the identity for vectors $e_i,f_j \in \BC^m$ (in the application, $m=\iy$) 
$$\det(I_\rk+\sum_{i=1}^{\al} e_i \otimes f_i)=\det \left(I_\al+(\la e_i,f_j\ra)_{1\leq i,j\leq \al}\right),
 $$
one checks 
$$
\bl
\det&( I  -\KR^{(\lb)})
  =
\det\left(\begin{array}{ccccc}
I_{{ \rk }}-\KR^{(\lb)}_1& \vline &O
\\ \hline 
-\KR^{(\lb)}_2 & \vline & I_{ \iy }
\end{array}
\right)
  =\det(I_{ \rk }-\KR^{(\lb)}_1)
\\&=\det(
I-\KR_1^{ }+\sum^{\ka }_{i =1}  {{\bf b}_{i  }^\lb  \otimes  {\bf \hat h}_{r -i}})
 \\&=
\det(I-\KR_1^{ })\det(I_\rk+\sum^{\ka }_{i =1}  {{\bf b}_{i  }^{(\lb) \KR_1} \otimes  {\bf \hat h}_{r -i}}
) =\det(I-\KR_1^{ })\det(I_\ka+\NR_\ka
)
\el
$$
establishing the first identity of (\ref{Kratio}). Next,
using the identity 
\be\bl
%
\langle (I-K)^{-1}A,B  \rangle =\frac{\det(I-K+  A\otimes B)-\det(I-K )}{\det(I-K )},
\el\ee
 it further follows that from (\ref{Kmatrix}) and (\ref{Bvector})\footnote{Set $v^\KR=(\Id-\KR^\iy)^{-1}v$ for $v\in \BC^\iy$.}, 
$$\bl
&\frac{\det(I-\KR^\lb)}{\det(I-\KR^\iy)} \left( \langle(I-\KR^\lb)^{-1} {\bf A}^\lb,  {\bf B}^\lb\rangle_{\geq 0} -c_{\ka+1}^{(\lb)} \right) 
\\&=\frac1{\det(I-\KR^\iy )}\left(
\det({\bf I}-\KR^{(\lb)}  +{\bf    {\bf A}}^{\lb}\otimes {\bf   {\bf B}}^{\lb})-\det(I-\KR^\lb )
(1+ c_{\ka+1}^{(\lb)}  )\right)
\\&=\frac 1 {\det(I-\KR^\iy )}
\footnotesize\left[
\bl &\det\left(I-\KR^{(\infty)}+\sum^{\ka }_{i =1}  {\widetilde{\bf b}_{i  }^\lb  \otimes  \widetilde{\bf \hat h}_{r -i}} 
+ { {\bf A}^{(\lb)}\otimes  \left[{\bf    B^{(\infty)}}+ \sum^{\ka }_{i =1}c^\lb_{i } \widetilde{\bf \hat h}_{r -i}\right] }\right)
\\
&-\det\left(I-  \KR^\iy+\sum^{\ka }_{i =1} \widetilde{\bf b}_{i  }^\lb  \otimes  \widetilde{\bf \hat h}_{r -i}
\right)(1+ c_{\ka+1}^{(\lb)}  ) \el \right]
 \\&={\footnotesize\det\left(\Id +\sum^{\ka }_{i =1} \widetilde{\bf b}_{i  }^{\lb\KR }  \otimes \underbrace{ \widetilde{\bf \hat h} _{r -i}}_{=: \widetilde f_i}
+\widetilde{\bf b}_{\ka+1}^{\lb\KR}
\otimes  \underbrace{\left[{\bf    B^{(\infty  )}}+ \sum^{\ka }_{i =1}c^\lb_{i } \widetilde{\bf \hat h} _{r -i}\right] }_{ =:\widetilde f_{\ka+1}}\right)
 }
\\
&-\det\left(\Id +\sum^{\ka }_{i =1} \widetilde{\bf b}_{i  }^{\lb\KR }  \otimes  \underbrace{ \widetilde{\bf \hat h} _{r -i}}_{=: \widetilde f_i}
\right) (1+c_{\ka+1}^{(\lb)}   )   
\el$$
%
%
\be\bl  \stackrel{ }{=}&\det \bigl(\Id+\la \widetilde{\bf b}_{i  }^{\lb\KR } ,\widetilde f_j\ra_{1\leq i,j\leq \ka+1}\bigr)-\det \bigl(\Id+\la \widetilde{\bf b}_{i  }^{\lb\KR } ,\widetilde f_j\ra _{1\leq i,j\leq \ka }\bigr) (1+c_{\ka+1}^{(\lb)} )
\\ \stackrel{*}{=}& \det \left(\left(\begin{array}{cccccccccc}
 &&&  0\\
& \Id_{\ka}&& \vdots &&& 
\\0& \dots&& -c_{\ka+1}^{(\lb)} 
\end{array}\!\!\!\!\!\!\!\!\!\right)+\la \widetilde{\bf b}_{i  }^{\lb\KR }  ,\widetilde f_j\ra_{1\leq i,j\leq \ka+1} \right)
\\
  \stackrel{**}{=}&\det(\MR_{\ka+1})
\el\label{matrixM}\ee
Equality $\stackrel{**}{=}$ above is obtained from the matrix in $\stackrel{* }{=}$, by subtracting from the last column $\la \widetilde {\bf b}_i^{\lb \KR }, \widetilde f_{\ka+1}\ra _{1\leq i \leq \ka+1}$   a linear combination of the first $\ka$ columns, after using the definition (\ref{c'}) of $c^{\prime \lb}_{i }$, and by moving $(I-\KR )^{-1}$ to the other side of the inner-product, by using the definition  (\ref{B1B2}) of $ {\bf B}^{\infty \KR^\top} $ and finally by replacing the infinite vectors $\widetilde{\bf b}_i^{\lb  }, 
{\bf B}^{\infty \KR^\top}$ by finite $\rk$-vectors $ {\bf b}_i^{\lb  }, ({\bf B}^{\infty \KR^\top})_1={\bf B}^{\infty \KR^\top}_1,\dots
$, since the inner-product is taken in the range $[0,\rk-1]$; to be precise, for $1\leq i\leq \ka+1$:
$$\footnotesize\bl
& \left\{\bl
&\left\la \widetilde{\bf b}_i^{\lb \KR},\widetilde f_{\ka+1} \right\ra -\sum_{j=1}^{\ka}c_j^{(\lb)}\left(\left\la \widetilde{\bf b}_i^{\lb \KR},\widetilde{\widehat {\bf h}}_{\rk-j}
\right\ra+\dt_{ij} \right)
 \\
&~~~~~~~~~~~~~~~~~~~~~~~~~\mbox{       for $1\leq i\leq \ka$}
\\
&\left\la \widetilde{\bf b}_{\ka+1}^{\lb \KR},\widetilde f_{\ka+1} \right\ra
-c^\lb_{\ka+1 } -\sum_{j=1}^{\ka}c_{j}^{(\lb)} \left\la \widetilde{\bf b}_{\ka+1}^{\lb \KR},\widetilde{\widehat {\bf h}}_{\rk-j}
\right\ra  
\el \right\}
\\
 &~~~~~~~~~~~=\left\la \widetilde{\bf b}_i^{\lb \KR},  {\bf  B}^{\infty 
 }
\right\ra-c^\lb_{i }
  =\left\la {\bf b}_i^{\lb  }, {\bf   B}_1^{\infty  \KR^\top
 }
\right\ra_{[0,\rk-1]}+
c^{\prime \lb}_{i },
\el
$$ 
which is the last column of $\MR_{\ka+1}$ in (\ref{M'}), after using $\la \widetilde{\bf b}_{i  }^{\lb\KR } ,\widetilde f_j\ra=\la  {\bf b}_{i  }^{\lb\KR_1 } ,\hat h_{\rk-j}\ra$ for $1\leq i \leq \ka+1$ and $1\leq j\leq \ka\leq \rk$. 
This ends the proof of Lemma \ref{PropM}.\qed


  
 \noindent We now set:
    $$\bl 
    V_{\rk}^{(\bm \lb)}  (u;v)&:=V_{\rk}^{(\bm \lb)}(u_0,\dots,u_{\ka-1};v_0,\dots,v_{\rk-1})\\&:=\left(\prod_{i=0}^{\ka-1} P_{\bm \lb}(u_i)\right)
    \det\left( \begin{array}{lllll}
 \left(\frac{1}{u_i-v_j}\frac{1}{P_{\bm \lb}(v_j)}
 \right)_{{0\leq i\leq \ka-1}\atop{0\leq j\leq \rk-1}}
 \\ \\
 \hline
 \hspace{.4cm}\left(\frac{1}{v_j^{\rk-i}} \right)_{{0\leq i\leq \sg-1}\atop{0\leq j\leq \rk-1}}
 \end{array}\right)
 \el $$
 \newpage
 \vspace*{-1cm}
 \be\label{Vplus}  \bl
 V^{(\bm \lb)}_{\rk+1}(u;v)&=
 V^{(\bm \lb)}_{\rk+1}(u_0,\dots,u_{\ka };v_0,\dots,v_{\rk }) \\&:=\left(\prod_{i=0}^{\ka-1}  P_{\bm \lb} (u_i)\right)
   \det\left( \begin{array}{lllll}
\left(\frac{1}{u_i-v_j}\frac{1}{P_{\bm \lb}(v_j)}
\right)_{{0\leq i\leq \ka }\atop{0\leq j\leq \rk }}
\\ \\
\hline
\hspace{.4cm}\left(\frac{1}{v_j^{\rk-i}} \right)_{{0\leq i\leq \sg-1}\atop{0\leq j\leq \rk }}
\end{array}\right).
 \el\ee
Lemma \ref{PropM} tells us that using the definition (\ref{Ltil}) of $\LR$, the ratios (\ref{Kratio}) and the matrices $\NR_\ka$ and $\MR_{\ka+1}$ defined  in (\ref{M'}),    
 \be \label{LR3'}\bl
    \LR(1) 
  &= \det(\Id-\KR_1^{ }) \oint_{ \ga_R ^\ka}
d  \bm\mu(\bm\lb)
\det\left(\Id+\NR_\ka\right)
 \\
   \LR (-\SR(\lb))  
  & =  \det(\Id-\KR_1^{ })\oint_{ \ga_R ^\ka} d  \bm\mu(\bm\lb) 
\det\left(\Id+\NR_\ka\right)( -\SR(\lb))
\\
&
 \hspace{-2.8cm}   \LR    \left( \langle(I-\KR^\lb)^{-1}{\bf A}^\lb, {\bf B}^\lb\rangle_{\geq 0}-c_{\ka+1}^{(\lb)}  \right)  
 =
 \det(\Id-\KR_1^{ }) \oint_{\ga_R^\ka}d  \bm\mu(\bm\lb )
   \det (\MR_{\ka+1}).
\el \ee


 \begin{proposition} \label{Prop5.2} The first expression (\ref{LR3'}) and the sum of the second and third expression (\ref{LR3'}) turn into multiple integrals, involving the measures (\ref{al-beta}) and the determinants (\ref{Vplus}), setting $\ga:=\tfrac{\rk(\rk-1)}{2}+\tfrac{\sg(\sg+1)}{2}$, remembering the notation (\ref{contour-i}) for $\Ga_0^{(i)}$, and the contours with radii $a< \rho_1<\sg_1<\sg_2<R<R+<\rho_2<\rho_3<a^{-1}$  and the nesting of $v_j$-contours (Remark 2 after Lemma \ref{EPK'}),
\be\label{LR1}\bl
   \LR(1) 
  &= (-1)^{\ga }   
  \left(\prod_{ {0\leq i\leq \ka-1}\atop{0\leq j\leq \rk-1}}\oint_{\Ga_0}d\widehat\al_i(u_i) 
  \oint_{\ga_{\sg_2}-\Ga^+_0 }  {d\widehat\beta_j (v_j)} 
   \right)
  \oint_{ \ga_R ^\ka}
d  \bm\mu(\bm\lb)   V_{\rk}^{(\bm \lb)}  (u;v)
,\el \ee 
\be\label{LRfull}\bl
    \LR &    \Bigl(-\SR(\lb)+ \langle(I-  K^\lb)^{-1}A^\lb, B^\lb\rangle_{\geq 0}-c_{\ka+1}^\lb
     \Bigr) ={\frak L}_1+ {\frak L}_2:=(-1)^{ \ga  }  \\
&
\times
\footnotesize \left(\bl
      &  -   \!\oint_{\ga_{\rho_2}}d \al_\ka(u_\ka )
     \oint_{\ga_{\rho_3}-\ga_{R+}}   \frac{d\widehat\theta_\rk(v_\rk)}{u_\ka\!-\!v_\rk}
 \left(\prod_{ {0\leq i\leq \ka-1}\atop{0\leq j\leq \rk-1}}\oint_{\Ga_0}d\widehat\al_i(u_i) 
  \oint_{\ga_{\sg_2}\!-\!\Ga_0^+ }    d\widehat\beta_j (v_j) \right)    
   \\&\qquad\qquad\qquad\qquad\qquad\qquad\qquad\qquad\times\oint_{ \ga_R ^\ka}  d \bm \mu(\bm\lb) \frac{V_{\rk}^{(\bm \lb)}  (u;v)}{P_{\bm \lb}(v_\rk)}
  \\&+(-1)^\sg \left(\prod_{ {0\leq i\leq \ka }\atop{0\leq j\leq \rk }}
 \oint_{\Ga^{(i)}_0}d\widehat\al_i(u_i) 
  \oint_{\ga_{\sg_2}-\Ga_0^+ }  {d\widehat\beta_j (v_j)} \right) 
  \oint_{ \ga_R ^\ka}d\bm\mu(\bm\lb )  V_{\rk+1}^{(\bm \lb)}  (u;v)
     \el\right)
 .\el \ee
 So, referring to (\ref{Lsca}), we have in terms of (\ref{LR1}) and (\ref{LRfull}),
\be   \label{LR1LR2}
  {\mathbb L}
=-  {\mathbb L}_0 
+(-1)^\ka   \left(   {\mathbb L} _1
+  {\mathbb L}_2\right)=-  {\mathbb L}_0 +
(-1)^\ka \frac{ {\frak L}_1+ {\frak L}_2 }{\LR(1)}  .\ee
The multiple integrals above will have to be understood as in Remark 2, following expression (\ref{NEPK}).

\end{proposition}


   \proof  
\noindent {\bf  Computation of $\det (I_\ka+\NR_\ka)$:}

 \noindent {\bf (i)} $\ka\leq \rk$: One enlarges the matrix $I_\ka+\NR_\ka$ (see (\ref{NN''})) to a matrix of size $\rk$ with same determinant, one multiplies and divides by the determinants $\det {\mathcal E}_\rk \PR_\rk=(-1)^{ \frac{\rk(\rk-1)}2}$ and $\det(I_{\rk}-\KR^\top_1)$ and one uses the matrix identity (\ref{NEPK}). Then one pulls out the $d\widehat \al_i$ and $d\widehat\beta_j$-integrations from the rows and columns of the determinant of the matrix (\ref{NEPK}) and one permutes the rows of the lower part of the matrix (\ref{NEPK}) (producing $\tfrac{\sg(\sg-1)}{2}+\sg$ sign changes), yielding a representation as a multiple integral of the matrix $V^{(\bm \lb)}_{\rk }(u;v)$ as in (\ref{Vplus}):
 %
%
\be\bl\label{I+N}
\det (I_\ka+\NR_\ka)
 &=\det\left(\begin{array}{llll}I_\ka+\NR_\ka  
 & \vline &
\left\la {\bf b}_{i+1}^{(\lb)\KR_1 },{\bf \hat h} _{\rk -j-1}
\right\ra_{{0\leq i \leq \ka-1}\atop{\ka\leq j\leq \rk-1}}
\\
\hline
O_{\sg,\ka} &\vline & I_{\sg}\end{array}
\right)=\det\left(I_\rk+\NR'_\rk 
\right)
%
\\&= \frac{\det(I-\KR_1)^{-1}}{\det \ER_\rk\PR_\rk}\det\left[\left(I_\rk+\NR'_\rk
  \right)\ER_\rk\PR_\rk(I-\KR_1^\top)\right]
%
%
\\
&=\frac{(-1)^{\ga }}{\det(I-\KR_1)} 
  \left(\prod_{ {0\leq i\leq \ka-1}\atop{0\leq j\leq \rk-1}}\oint_{\Ga_0}d\widehat\al_i(u_i) 
  \oint_{\ga_{\sg_2}-\Ga_0 }  {d\widehat\beta_j (v_j)} 
   \right)
   V_{\rk}^{(\bm \lb)}  (u;v).
   \el \ee
 
 
%
  %

  \noindent {\bf (ii)} $\ka>\rk$ : Setting $s=-\sg=\ka-\rk>0$, the matrix $(I_\ka+\NR_\ka)$ has then the following form, since ${\bf \hat h} _{\rk -j-1}=0$ for $j\geq \rk$,
  $$
  I_\ka+\NR_\ka=\left(\begin{array}{lllll}
  ~~~~~~I_\rk+\NR_\rk&\vline&O_{\rk,s}
  \\ \hline
  \left\la {\bf b}_{i+1}^{(\lb)\KR_1 },{\bf \hat h} _{\rk -j-1}
\right\ra_{{\rk\leq i \leq \ka-1}\atop{0\leq j\leq \rk-1}}&\vline&I_s
  \end{array}\right)
  $$
 and so 
 $$\det(I_\ka+\NR_\ka)=\det(  I_\rk+\NR_\rk)
 ,$$
 which reduces the problem to formula (\ref{I+N}) evaluated at $\ka=\rk$. 
 So, multiplying this expression (\ref{I+N}) 
  with $\det(\Id-\KR_1^{ })$, using (\ref{LR3'})  and integrating with regard  to $d\bm \mu(\bm\lb)$ gives the first formula (\ref{LR1}), using (\ref{RABK}) and (\ref{al-beta}) for $\SR(\lb)$,
and
 \be\label{LR2}\bl
  \LR (-\SR(\lb))  
  & =  (-1)^{ \ga-1}  
 \oint_{\ga_{\rho_2} }d \al_\ka(u_\ka ) \oint_{\ga_{\rho_3}}  \frac{   d\widehat\theta_r (v_\rk)}{   u_\ka-v_\rk   }
 \\&\times \left(\prod_{ {0\leq i\leq \ka-1}\atop{0\leq j\leq \rk-1}}\oint_{\Ga_0}d\widehat\al_i(u_i) 
  \oint_{\ga_{\sg_2}-\Ga_0 }  {d\widehat\beta_j (v_j)} 
   \right)
  \oint_{\ga_R^\ka}  d \bm \mu(\bm\lb) \frac{ V_{\rk}^{(\bm \lb)}  (u;v)   }{P(v_\rk)}
\el\ee

\medbreak
  \noindent {\bf  Computation of $\det (\MR_{\ka+1})$:}
   
  \noindent {\bf (i)} $\ka\leq \rk$: We define two matrices $\AR $ and $\AR'$ of size $\rk+1$:
$$\AR \AR' :=\left(\begin{array}{lllll}
\ER \PR (I-\KR_1^\top)&\vline&O_{\rk,1}
\\ \hline
~~~~~O_{1,\rk}&\vline& 1\end{array}
\right)
\left(\begin{array}{lllll}
~~~~~~I_\rk &\vline & \begin{array}{lll}
-{\bf \widetilde B}^{\infty \KR^\top}_{0 }\\ ~~~~\vdots \\ -{\bf \widetilde B}^{\infty \KR_1^\top}_{ \rk-1}
\end{array}
\\ \hline ~~~O_{1,\rk}&\vline& 1
\end{array}
\right)
$$
with
 $$\det \AR \det \AR' =
(-1)^{ \frac{\rk(\rk-1)}2}  \det(I-\KR_1)^{ }
$$
 One embeds the matrix $\MR_{\ka+1}$ (see (\ref{M'})), in a larger matrix of size $\rk+1$, as before, one moves the $\ka+1$th row and column to last position in the matrix $\stackrel{*}{=}$ below, one multiplies the matrix to the right by $\AR $, which acts on the $I_\rk+\NR'_\rk$-part and on the last row, giving the matrix to the right of $\stackrel{\star\star}{=}$, using  the representation (\ref{NN'''}). Then one multiplies by the matrix $\AR' $, which has the effect of removing the inner-product in the terms $\star_i$'s (as in (\ref{M'})), at the expense of introducing $-\ER_\rk\PR_\rk 
  (B_1+\KR_1^\top B_2)$ in the right hand column. In $\stackrel{\star\star\star}{=}$ one goes back to the expression (\ref{NN'''}). So, one finds
  $$\bl
    \det &~\MR_{\ka+1}
   \stackrel{*}{=}\det\left(\begin{array}{llll} \MR_{\ka+1}  
 & \vline &
\left\la {\bf b}_{i+1}^{(\lb)\KR_1 },{\bf \hat h} _{\rk -j-1}
\right\ra_{{0\leq i \leq \ka }\atop{\ka\leq j\leq \rk-1}}
\\ &\vline \\
\hline  
O_{\sg,\ka} &\vline & I_{\sg}\end{array}
\right)
\el $$
$$  \bl&=\frac{(-1)^{ \frac{\rk(\rk-1)}2}  }{\det(I-\KR_1) }
 \det\left[
\left(
\begin{array}{lllll}
~~~~~~~~~~~~I_\rk+\NR'_\rk & \vline&\begin{array}{lll}
\star_0\\ \vdots \\ \star_{\ka-1}\\0\\
\vdots\\ 0
\end{array}
\\ \hline 
 \left\la {\bf b}_{\ka+1}^{(\lb)\KR_1 },{\bf \hat h} _{\rk -j-1}
\right\ra_{ {0\leq j\leq \rk-1}} 
&\vline& \star_\ka
\end{array}
\right)\AR \AR'\right]
 \el$$ 
$$ \bl& \stackrel{\star\star}{=}\frac{(-1)^{ \frac{\rk(\rk-1)}2}  }{\det(I\!-\!\KR_1) }
 \det \!\!
\left(
\begin{array}{lllll}
 \ER_\rk\PR_\rk(I-\KR_1^\top)+\left(\begin{array}{llll}( {\bf b}_{i+1}^{(\lb)  }(j)
 )_{{{0\leq i \leq \ka-1}\atop{0\leq j\leq \rk-1}}\atop{}}
\\   \hline
~~~~~~~~O_{\sg,\rk}\end{array}\right)
 & \vline&\begin{array}{lll}
\star_0\\ \vdots \\ \star_{\ka-1}\\0\\
\vdots\\ 0
\end{array}
\\ \hline 
  {\bf b}_{\ka+1}^{(\lb)  }(0)~~~~\dots  \dots  ~~~~
{\bf b}_{\ka+1}^{(\lb)  }(\rk\!-\!1)
&\vline& \star_\ka
\end{array}
\right)  \AR'
\\&\stackrel{\star\star\star}{=}\frac{(-1)^{ \frac{\rk(\rk-1)}2}  }{\det(I\!-\!\KR_1) }\det\!\!
\left(
\begin{array}{lllll}\left(I_\rk+\NR'_\rk
  \right)\ER_\rk\PR_\rk(I\!-\!\KR_1^\top)
&\vline  \ER_\rk\PR_\rk 
  (-B_1\!-\!\KR_1^\top B_2)+\left(
  \begin{array}{lll}
c'_1
   \\ \vdots \\ c'_{\ka}
   \\ \hline O_{\sg,1}%
\end{array}
\right) 
  \\ \hline
    {\bf b}_{\ka+1}^{(\lb)  }(0) \dots  
{\bf b}_{\ka+1}^{(\lb)  }(\rk\!-\!1)&\vline
~~~~~~~~~~~~~~~~c'_{\ka+1} 
\end{array}
\right)
  \el$$
One then uses again identity (\ref{c'k}), (\ref{NEPK}) and (\ref{cb}), one permutes as before the rows $\ka+1,\dots,\rk$ and one takes out the $-$sign. Then the new expression obtained has a last column which can be written as a sum of two columns:
%
%
$$ \left(\begin{array}{ll}\left(\oint_{\Ga_0^{(i)}}d\al_i(u_i)\oint_{\ga_{\sg_2}-\Ga_0^+}
\frac{d\beta_\rk (v_\rk)}{u_i-v_\rk}\right)_{{0\leq i \leq \ka-1}}
 \\ \\
\hline  
~~~~~\left( \oint_{\ga_{ \sg_2 }-\Ga_{0}^+}\frac{d\widehat\beta_\rk(v_\rk)}{v_\rk^{\rk-i}}\right)_{{0\leq i \leq \sg-1}}
\\ \\ \hline
\oint_{\Ga^{(\ka)}_0}d \al_\ka(u_\ka ) \oint_{\ga_{\sg_2}-\Ga_0^+}\frac{d \beta_\rk (v_\rk )}{u_\ka -v_\rk }
\end{array}
\right)
+
\left(
\begin{array}{ccc}
0\\\vdots \\ 0 \\ \\ \hline \oint_{\Ga^{(\ka)}_0}d \al_\ka(u_\ka ) \oint_{\ga^+_{R }} \frac{   d\theta_r (v_\rk)}{   u_\ka-v_\rk   }
\end{array}\right)
$$
and so the determinant above can be written as the sum of two determinants, with $\ga$ as in (\ref{I+N}),
$$   
 =\frac{(-1)^{\ga} }{\det(I-\KR_1)} \left(
 \bl &\det\left[
\begin{array}{ll}\left(\oint_{\Ga_0^{(i)}}d\al_i(u_i)\oint_{\ga_{\sg_2}-\Ga_0^+}
\frac{d\beta_j (v_j)}{u_i-v_j}\right)_{{0\leq i \leq \ka-1}\atop{0\leq j\leq \rk}}
 \\ \\
\hline \\
~~~~~\left( \oint_{\ga_{ \sg_2 }-\Ga_{0}^+}\frac{d\widehat\beta_j(v_j)}{v_j^{\rk-i}}\right)_{{0\leq i \leq \sg-1}\atop{0\leq j\leq \rk }}
\\ \\ \hline
\left(\oint_{\Ga_0^{(\ka)} }d\al_\ka(u_\ka)\oint_{\ga_{\sg_2}-\Ga_0^+}
\frac{d\beta_j (v_j)}{u_\ka-v_j}\right)_{ {0\leq j\leq \rk}}
 \end{array}\right] 
\\ \\&
+\oint_{\Ga^{(\ka)}_0}d \al_\ka(u_\ka ) 
\oint_{\ga^+_{R}}   \frac{d\theta_\rk(v_\rk)}{u_\ka-v_\rk}
\\&~~~\times \det\left[
\begin{array}{ll}\left(\oint_{\Ga_0}d\al_i(u_i)\oint_{\ga_{\sg_2}-\Ga_0^+}
\frac{   d \beta_j (v_j)}{u_i-v_j}\right)_{{0\leq i \leq \ka-1}\atop{0\leq j\leq \rk-1}}
\\ \\
\hline\\
~~~~~\left( \oint_{\ga_{ \sg_2 }-\Ga_0}\frac{d\widehat  \beta_j(v_j)}{v_j^{\rk-i}}\right)_{{0\leq i \leq \sg-1}\atop{0\leq j\leq \rk-1}}
\end{array}\right]
\el\right)$$
One then moves the last row of the first determinant above $\sg$ steps up, at the expense of $(-1)^\sg$. One then finds:
 \be \footnotesize\label{Mk+1} \bl   
 =& \frac{(-1)^{ \ga-\sg   }}{\det(I-\KR_1)} 
 \\& 
 \times\left( \bl &\left(\prod_{ {0\leq i\leq \ka }\atop{0\leq j\leq \rk }}\oint_{\Ga^{(i)}_0}d\widehat\al_i(u_i) 
  \oint_{\ga_{\sg_2}-\Ga_0^+ }  {d\widehat\beta_j (v_j)} \right)
    V_{\rk+1}^{(\bm \lb)}  (u;v)
\\&+ (-1)^\sg \!\oint_{\Ga^{(\ka)}_0}d \al_\ka(u_\ka )\oint_{\ga^+_{R }}   \frac{d\widehat\theta_\rk(v_\rk)}{u_\ka\!-\!v_\rk}
\\
&~~~\times\left(\prod_{ {0\leq i\leq \ka-1}\atop{0\leq j\leq \rk-1}}\oint_{\Ga_0}d\widehat\al_i(u_i) 
  \oint_{\ga_{\sg_2}-\Ga_0^+ }  {d\widehat\beta_j (v_j)}\right)  \frac{V_{\rk}^{(\bm \lb)}  (u;v)}{P_{\bm\lb }(v_\rk)} 
  \el \right).
  \el\ee
Multiplying the expression 
 (\ref{Mk+1}) with $\det(\Id-\KR_1^{ })$, using 
%
%
 (\ref{LR3'})  and integrating with regard  to $d\bm \mu(\bm\lb)$ gives (\ref{LR3}) below:
\be\footnotesize\label{LR3}\bl
    \LR &    \Bigl( \langle(I-\widetilde K^\lb)^{-1}A^\lb, B^\lb\rangle_{\geq 0}-c_{\ka+1}^\lb
     \Bigr) =(-1)^{ \ga-\sg  }  \\
&
\times
\left(\bl
&\left(\prod_{ {0\leq i\leq \ka }\atop{0\leq j\leq \rk }}
 \oint_{\Ga^{(i)}_0}d\widehat\al_i(u_i) 
  \oint_{\ga_{\sg_2}-\Ga_0^+ }  {d\widehat\beta_j (v_j)} \right) 
  \oint_{\ga_R^\ka}d\bm\mu(\bm\lb )  V_{\rk+1}^{(\bm \lb)}  (u;v)
     \\& + (-1)^\sg \!\oint_{\Ga^{(\ka)}_0}d \al_\ka(u_\ka )
     \oint_{\ga_{R}^+}   \frac{d\widehat\theta_\rk(v_\rk)}{u_\ka\!-\!v_\rk}
\\&\times\left(\prod_{ {0\leq i\leq \ka-1}\atop{0\leq j\leq \rk-1}}\oint_{\Ga_0}d\widehat\al_i(u_i) 
  \oint_{\ga_{\sg_2}\!-\!\Ga_0^+ }    d\widehat\beta_j (v_j) \right)    \oint_{\ga_R^\ka}  d \bm \mu(\bm\lb) \frac{V_{\rk}^{(\bm \lb)}  (u;v)}{P_{\bm\lb}(v_\rk)}
     \el\right)
.\el \ee
Finally, adding formulas (\ref{LR2}) and (\ref{LR3}) and noticing that the integral $\LR (-\SR(\lb)) $ as in (\ref{LR2}) appears in (\ref{LR3}), except for a different contour in the $d\widehat\theta_\rk(v_\rk)$-integral, we obtain expression (\ref{LRfull}).

 \noindent {\bf (ii)} $\ka> \rk$: One proceeds like case {\bf (ii)} in the computation of $\det(I_\ka+\NR_\ka)$, after formula (\ref{I+N}). One shifts both the $\rk$th column and row to the end and goes on as above. One then finds $s=-\sg=\ka-\rk>0$ zero-columns, except for $I_s$, as before. Then take the determinant {\em about $I_s$}, yielding (\ref{LRfull}) for $\ka=\rk$, establishing Proposition \ref{Prop5.2}. \qed


  \section{The $\prod_{i=0}^{\ka-1}d\widehat \al_i$ and 
   $\prod_{j=0}^{\ka-1}d\widehat \beta_j$ integrations
   }

 The first point of this section (Proposition \ref{LemmaIntW}) is to show that the $\prod_{i=0}^{\ka-1}d\widehat \al_i(u_i)$ integration about $\Ga_0$ of a Vandermonde $\Dt_\ka(u)$ times a function amounts to evaluating that function at $u=0$. 
  The second similar point (Proposition \ref{LemmaInt0}) is to show that  the $\prod_{i=0}^{\ka-1}d\widehat \beta_j(v_j)$ integration about $\ga_{\sg_2}-\Ga_0 \ni 1/t$ of a Vandermonde times a product of appropriate powers of $\frac{h_0(v_j)}{\prod_{i\neq j}(v_j-v_i)}$ vanishes.
 \begin{proposition}\label{LemmaIntW}
Given a function $F(u_0,\dots,u_{m-1}; v)$ holomorphic in the $u_i$'s at the origin and depending on another set of variables $v$, then the following identity holds for $1\leq \ell\leq m$:
\be\bl\label{Id0}
\prod_{i=0}^{\ell-1}&\oint_{\Ga_0} d\widehat \al_i(u_i)
\Dt_m(u_0,\dots,u_{m-1})F(u_0,\dots,u_{m-1}; v)
\\&=(-1)^{\frac{\ell(\ell-1)}2}\left(\prod_{i=\ell}^{m-1}(-u_i)^\ell\right)
\Dt_{m-\ell}(u_\ell,\dots,u_{m-1})
F(\overbrace{0,\dots,0}^{\ell},u_\ell,\dots,u_{m-1};v)
\\&=(-1)^{\frac{m(m-1)}2} F(0,\dots,0;v)\mbox{,    if~~}\ell=m
\el\ee

\end{proposition}
\proof The identity (\ref{Id0}) holds for $\ell=1$. Assuming it holds for $\ell\geq 1$, we now prove the identity for $\ell+1$. If one denotes the right hand side of (\ref{Id0}) by $I_\ell (u_\ell,\dots,u_{m-1})F( {0,\dots,0} ,u_\ell,\dots,u_{m-1};v)$, one   computes the following, noticing that the expression in $\stackrel{*}{=}$ has a simple pole at $u_\ell=0$:
$$\bl\oint_{\Ga_0}&\frac{du_\ell}{2\pi \I u_\ell^{\ell+1}}I_\ell (u_\ell,\dots,u_{m-1}) F(0,\dots,0,u_\ell,\dots,u_{m-1};v)
\\&\stackrel{*}{=}\oint_{\Ga_0}\frac{du_\ell}{2\pi \I u_\ell^{\ell+1}}(-1)^{\frac{\ell(\ell-1)}2} 
(-u_\ell)^\ell \left(\prod_{i=\ell+1}^{m-1} (-u_i)^\ell (u_\ell-u_i)\right)\Dt(u_{\ell+1},\dots,u_{m-1})
\\&~~~~~~~~~~~~~~~~~~~~~~~~~~\times F(0,\dots,0,u_\ell,\dots,u_{\rk-1};v)
\\&=I_{\ell+1} F(\underbrace{0,\dots,0}_{\ell+1},u_{\ell+1},\dots,u_{\rk-1};v),
\el$$
%
ending the proof of Proposition \ref{LemmaIntW}. \qed

Remembering 
 the  notation $[\sum_{i\geq 0}a_iz^i]_{[0,\ell]}:=
 \sum_{i= 0}^{\ell}a_iz^i
 $ and $[\sum_{i\geq 0}a_iz^i]_{ \ell }:=
  a_\ell  
 $, we state and prove the following Lemma : 
 \begin{lemma} \label{LemmaRes}For $f(\lb)=\sum_{i\geq 0} a_i \lb^i$
, we have    
\newline \underline{for $|\ze|<|\lb|$ }
$$
\oint_{\ga_R}\frac{d\lb  f(\lb)\ze^{r+1}
  }{2\pi \I(\lb-\ze )\lb^{r+1}}
  = 
  \oint_{\ga_R}\frac{d\lb  
  }{2\pi \I  }
\sum_{k\geq 0}a_k\lb^k \sum_{i\geq r+1}\frac{\ze^i}{\lb^{i+1}}
=\sum_{i\geq r+1}a_i\zeta^i
=f(\ze)-[f(\ze)]_{[0,r]},$$
\newline \underline{for $|\ze|>|\lb|$ } 
$$
\oint_{\ga_R}\frac{d\lb  f(\lb)\ze^{r+1}
  }{2\pi \I(\lb-\ze )\lb^{r+1}}
  = 
 -\zeta^{r+1}\sum_{i=0 }^{r }\frac{a_{r-i}}{\zeta^{i+1}}
    = - {[f(\ze)]_{[0,r]} }.$$
\end{lemma}

Remember $P_{\bm \lb}(u)$ as in (\ref{P}), the measure $d\mu(\lb)=\frac{h_0(\lb)d\lb}{2\pi \I \lb^{\rk+1}}$ as in (\ref{mu}), with $\oint_{\Ga_0} d\mu(\lb)=1$, and with $h_0(z)=\frac1\psi (1-tz)^{n+1}=\frac 1\psi+\al_1 z+\dots+z^\rk+\dots$. Setting for $\vr=0$ or $1$, and arbitrary variables $v_0,\dots,v_{\ka }$, 
\be
 Q_{\ka +\vr}(v):=
  \prod_0^{\ka-1+\vr}(v-v_j)
\label{Q}\ee
 we have:

\begin{lemma}\label{LemmaRes'}
 For $|v_i|<R$, with $0\leq i\leq \ka-1$ and $|v_\ka|>R$
, we have for $u=(u_0,\dots,u_{\ka-1})$ and $\vr=0$ or $1$:
\be\bl\label{LemmaRes1}
\Lb^{(\vr)}&:=\oint_{ \Ga_R }
\frac{d \mu(\lb)}{(\lb-v_\ka)^\vr}
\prod_{k=0}^{\ka-1}\frac{\lb- u_k }{\lb-v_k }\Bigr|_{u =0}
\\&=   
\frac{1}{\sg !  }\left(\frac{d}{dz}\right)^{\sg}\frac{[h_0(z)]_{[0,\sg]}}{Q_{\ka+\vr} (z)}\Bigr|_{z=0} 
+
\sum_{i=0}^{\ka-1}\frac{  h_0(v_i)   }{v_i^{ \sg+1}Q_{\ka+\vr}'(v_i)} 
 =:(\Lb^{(\vr)}_0+\Lb^{(\vr)}_1)(v_0,\dots,v_{\ka-1+\vr}).
\el\ee
If $|v_\ka|<R$ and $\vr=1$, then the same formula holds, except that the sum in $\Lb^{(\vr)}_1$ ranges from $0$ to $\ka$.


\end{lemma}
\proof We integrate identity  
\be\bl\label{lb-dep}
  \frac1{(\lb-v_\rk)^\vr} \prod_{k=0}^{\ka-1}\frac{\lb -u_k}{\lb -v_k}
&=  \sum_{i=0}^{\ka-1+\vr}\frac{\prod_{j=0}^{\ka-1} (v_i-u_j)}{(\lb -v_i) Q_{\ka+\vr}'(v_i)}+1-\vr .
\el\ee
with respect to $d\mu(\lb)$ and set $u=0$, taking into account the notation before formula (\ref{Q}). One uses the first formula of Lemma \ref{LemmaRes} for the $|v_i|<R$ and the second one for $|v_\ka| >R$ and also the fact that $\rk=\ka+\sg$ (\ref{rho,r}), yielding:
$$\bl
 & \oint_{ \Ga_R } 
   \frac{d \mu(\lb)}{(\lb-v_\ka)^\vr}
\prod_{i=0}^{\ka-1}\frac{\lb- u_i }{\lb-v_i }\Bigr|_{u =0}
\\ & =\oint_{ \Ga_R }
 \frac{h_0(\lb)d\lb}{2\pi \I \lb^{\rk+1}}\left(\sum_{i=0}^{\ka-1+\vr}\frac{\prod_{j=0}^{\ka-1}(v_i-u_j)}{(\lb-v_i)Q_{\ka+\vr}'(v_i)} +1-\vr\right)\Bigr|_{u =0}
\el$$
$$\bl \\&=
\sum_{i=0}^{\ka-1 }\frac{\left(h_0(v_i)-[h_0(v_i)]_{[0,\rk]}\right)\prod_{j=0}^{\ka-1} (v_i\!-\!u_j)}{v_i^{\rk+1}Q_{\ka+\vr}'(v_i)}
 - \frac{\vr\left([h_0(v_\ka)]_{[0,\rk]}\right)\prod_{j=0}^{\ka-1} (v_\ka\!-\!u_j)}{v_\ka^{\rk+1}Q_{\ka+\vr}'(v_\ka)}\Bigr|_{u=0}\!+\!1\!-\!\vr
\\&=\sum_{i=0}^{\ka-1}\frac{ h_0(v_i)   }{v_i^{ \sg+1}Q_{\ka+\vr}'(v_i)}-\oint_{\Ga(v_0,\dots,v_{\ka-1})+\vr\{  v_\ka\}+(1-\vr)\Ga_\iy  }
\frac{[h_0(z)]_{[0,\rk]}dz}{2\pi \I z^{\sg+1}Q_{\ka+\vr}(z)
}
\\&=\sum_{i=0}^{\ka-1}\frac{ h_0(v_i)   }{v_i^{ \sg+1}Q_{\ka+\vr}'(v_i)}+\oint_{\Ga_0}
\frac{[h_0(z)]_{[0,\rk]}dz}{2\pi \I z^{\sg+1}Q_{\ka+\vr}(z)
 }
\\&=  
\sum_{i=0}^{\ka-1}\frac{  h_0(v_i)   }{v_i^{ \sg+1}Q_{\ka+\vr}'(v_i)}+
\frac{1}{\sg !  }(\frac{d}{dz})^{\sg}\frac{[h_0(z)]_{[0,\sg]}}{Q_{\ka+\vr} (z)}\Bigr|_{z=0} .
\el$$
establishing (\ref{LemmaRes1}). The case $|v_\ka|<R$ and $\vr=1$ follows readily, except that the first sum ranges from $0$ to $\ka$.\qed

  
  We will need the following Proposition, remembering from (\ref{LemmaRes1}) that for $k=0,1$, we have
$\Lb^{(\vr)}_k   := \Lb^{(\vr)}_k (v_0,\dots,v_{\ka-1+\vr})$ and set $\Lb^{(\vr)}_{1}=:\sum_{i=0}^{\ka-1}\Lb^{(\vr)}_{1,v_i}$.

\begin{proposition}\label{LemmaInt0} Given $\ka$ arbitrary distinct integers $0\leq m_\ell\leq \rk-1$, the following holds:
%
\be \label{LemmaInt0'}\bl
 \left(\prod_{\ell=0}^{\ka-1 }\oint_{\ga_{\sg_2}-\Ga_0 } \!\!\frac{d\widehat\beta_{m_\ell} (v_\ell)} {v_\ell^\rk}\right) 
   \Dt_\rk( v ) \left[ (\Lb^{(\vr)}_0+\sum_{i=0}^{\ka-1}\Lb^{(\vr)}_{1,v_i})^\ka -(\Lb^{(\vr)}_0)^\ka\right]
=0.
 \el
 \ee
\end{proposition}

\proof We give the proof for $\vr=0$. At first remember Remark 2 and the comment, following (\ref{NEPK}) and (\ref{LR1LR2}). 
 It is advantageous to make a coordinate change $v_j\to u_j =1/t-v_j$, turning the nested annuli $ \ga_{s_j,r_j}$, mentioned in Remark 2, after deformations  to circles $C_i$, with increasing radii, having the property that $u_i\in C_i,~~u_{i+1}\notin C_i$.

Notice that the two expressions below have a piece in common for any $0\leq \ell\leq\ka-1 $:
\be\bl
\Dt(u_0,\dots,u_{\rk-1})
&=
\Dt_\ell(u_0,\dots,u_{\ell-1})\prod_{0\leq i\leq \ell -1}(u_i-u_{\ell })
\prod_{{0\leq i\leq \ell-1}\atop{\ell +1\leq k\leq \rk-1}}(u_i-u_k)\Dt_{\rk-\ell}(u_\ell,\dots,u_{\rk-1})
\\
Q_\ka'(u_{\ell })&=\prod_{0\leq i \leq\ell -1}(u_{\ell }-u_i)
\prod_{\ell +1\leq i \leq\ka-1}(u_{\ell }-u_i)
.\el \label{DelQ}\ee

 One expands the power in the square bracket of (\ref{LemmaInt0'}), using the partition $\sum_{i=1}^{\al+1} k_i=\ka$, with $1\leq k_i\leq \ka$ and $k_{\al+1}<\ka$, and using numbers $0\leq \ell_1<\ell_2<\dots \ell_{ \al}\leq \ka-1 $. 
 Then removing all integrations, but the first $\ell_1+1$ ones, we find, omitting the subscript $1$ in $\ell_1$,
\be \label{Int1}
\bl
\left(\prod_{j=0}^{\ell}\oint_{ \ga_{s_j,r_j}}
\frac{dv_j}{2\pi \I v_j^{m_j+1}h(v_j)}\right)
\Dt_\rk(v)\Lb_0^{k_{\al+1}}\prod_{i=1}^{\al}(\Lb_{1,v_{\ell_i}})^{k_i}
\el 
\ee
%



Noticing that in $u$-coordinates, $h_0(v)= (tu)^{n+1}$ and $h(v)= u^{n+1}(1+a^2-a^2tu)^n$,
 the expression (\ref{Int1}) above reads  in the new $u_i$-coordinates for $\ell \leq \ka-1$: (leaving out multiplicative constants) 
\be \label{Int2}
\bl
\prod_{j=0}^{\ell }\oint_{C_j}&
\frac{g_j(u_j)du_j}{2\pi \I u_j^{n+1} }
  \left( \frac{  u_{\ell }^{n+1}  }{ \prod_{0\leq i <\ell }(u_{\ell }-u_i)}\right)^{k_1}
 \prod_{0\leq i \leq\ell -1}(u_{\ell }-u_i) \\
 &\times \Dt_\ell(u_0,\dots,u_{\ell -1})
F(u_{\ell +1},\dots,u_{\rk-1}).
\el 
\ee
From the integration point of view the variables $u_{\ell +1},\dots,u_{\rk-1}$ can be viewed as constants and will also be dropped. Expanding $\prod_{j=0}^{\ell } g_j(u_j)$ as a power series in the monomials $\prod_{j=0}^{\ell_1}  u_j^{\al_j}$, inserting this series in the expression, one needs to consider each term in this sum; so a typical term, after moving the $u_\ell$-integration up-front,  reads as follows (set $u_{\ell_1}=u$ and $k_1=k$):
\be\footnotesize\label{Int3}
\bl
\oint_{C_{\ell }}&\frac{u_{\ell }^{(n+1)(k -1) +\al_{\ell }} du_{\ell }}{2\pi \I }
\prod_{j=0}^{\ell -1}
\oint_{C_j}\frac{du_j}{2\pi \I  u_j^{n+1-\al_j}(u_{\ell }-u_j)^{k -1}}
\det\left(u_j^i \right)_{0\leq j,i\leq \ell -1}\\
&\stackrel{*}{=}\oint_{C_{\ell}}\frac{u_{ }^{(n+1)(k -1) +\al_{\ell}} du_{ }}{2\pi \I }
 \det\left(\oint_{C_j}\frac{z^{i+\al_j-n-1}dz}{2\pi \I   (u_{  }-z)^{k-1}}\right)_{0\leq j,i\leq \ell -1}\\&=\oint_{C_{\ell_1}}\frac{u_{ }^{(n+1)(k-1) +\al_{\ell}} du_{ }}{2\pi \I }\det\left( 
 \frac{
 \left( {k+n-\al_j-i-2}\atop{k-2} \right)
  }{u^{k+n-\al_j-i-1}}
  \right)_{0\leq j,i\leq \ell -1}
\\
&=
\frac1{\ell (k-2)!}\det\left( (n-i-\al_j+k-2)_{ k-2}  
\right)_{0\leq j,i\leq \ell -1}
\oint_{\Ga_0}\frac{u^{ \sum_0^{\ell } \al_i -\frac{\ell(\ell-1)}{2} }
du}
{2\pi \I u^{(n-\ell+1)(\ell-k+1 )    }  }\el\ee
Equality $\stackrel{*}{=}$ latter is obtained by distributing the $u_j$-integration over the rows and after replacing the dummy $u_j\in C_j$'s (now about unrelated circles) by variables $z\in \Ga_0$. Then one works out the integral\footnote{Using the integration 
$ 
\oint_{\Ga_0}\frac{dz}{2\pi \I z^{m+1}(u-z)^{k-1}}
=\frac{\left( {m+k-2}\atop{k-2} \right)}{u^{m+k-1}}.
$ }
inside the determinant. 
%
Notice that each entry in the matrix above, (\ref{Int3}), can be written as a polynomial $\sum_{\beta=0}^{k-2}(-i)^\beta c_{\beta}(\al_j)$ of degree $k-2$ in $-i$ ; therefore each column is a linear combination of $k-1$ columns, implying that the matrix has $\det=0$, if $k-1<\ell$. So, it remains to consider the case where $k\geq \ell+1$, which splits into two cases: either $\ell-k+1< 0$ or $\ell-k+1=0$.

\noindent {\em Case $\ell-k+1< 0$}. The integral in (\ref{Int3}) has zero residue for $n$ large enough, since $\al_i\geq 0$ for $0\leq \ell, k \leq \rk$.

\noindent {\em Case $\ell-k+1=0$}. Then  the integral in (\ref{Int3}) would be $\neq 0$ if the exponent  
\be \sum_0^{\ell } \al_i -\frac{\ell(\ell-1)}{2} =-1.
\label{Int5}\ee
This equality implies that for some $0\leq i\neq i'<\ell$ we must have $\al_i=\al_{i'}$ and thus two identical columns imply that the  determinant above $=0$. Indeed, after putting the integers $\al_i$ in increasing order, if no two one would be identical, one would have 
$$
\al_{i }\geq \al_{i-1}+1 , \mbox{  for  } 1\leq i\leq \ell-1,
$$
implying that $\al_{i }\geq \al_0+i$ for $1\leq i\leq \ell-1$. Therefore (\ref{Int5}) can be written:
$$
-1=\al_\ell+\sum_0^{\ell-1} \al_i -\frac{\ell(\ell-1)}{2} \geq
 \al_\ell+ \sum_0^{\ell-1} i  + \ell \al_0-\frac{\ell(\ell-1)}{2} =\al_\ell+\ell\al_0  \geq 0,
  $$
a contradiction! So if (\ref{Int5}) holds, two $\al_i$'s must be equal (none being $\al_\ell$), implying the vanishing  of the determinant in (\ref{Int3}). This ends the proof of Proposition \ref{LemmaInt0}. \qed

  
  \section{The ${\mathbb L}_{n,\rk,\rho}$-kernel in its simple form}
  
  The purpose of this section is to perform  the $d\bm\mu(\bm\lb ) $ integrations in the kernel $\mathbb L$, as in (\ref{LR1LR2}), containing the multiple integrals (\ref{LR1}) and (\ref{LRfull}). They involve the determinants $V_\rk^{(\bm \lb)}(u,v)$ and $V_{\rk+1}^{(\bm \lb)}(u,v)$, which will be expanded according to the $\sg$ last rows. The $\kappa$-fold $d\bm\mu(\bm\lb ) $ integration turns into the $\ka$th power of a single $d\mu(\lb)$-integration about $\ga_R$ (done in Lemmas \ref{LemmaRes} and \ref{LemmaRes'}), for which only the ``{\em leading term}" survives, using Schur expansion arguments for symmetric functions.  
   Then applying Propositions \ref{LemmaIntW} and \ref{LemmaInt0} to the  $\prod_{i=0}^{\ka-1}d\widehat \al_i$-integration about $\Ga_0$ and the $\prod_{i=0}^{\ka-1}d\widehat \beta_i$-integration about $\ga_{\sg_2}-\Ga_0  $  leads to an enormous collapse. It finally gives a simple expression (\ref{K1}) for the kernel, which unfortunately is still not in the appropriate form to do the asymptotics. Putting this kernel in an appropriate form will be done in the next section.
 
  Define
\be\bl   \Om_1(u,v  ) &:= \left(
\prod_{\ell=0}^{\rk-1}\oint_{\ga_{\sg_2}-\Ga_0}\frac{du_\ell  }{2\pi \I u_\ell^{\ka+\ell+1}  h(u_\ell) }\frac{ u_\ell-v}{u_\ell-u}
\right)\Dt_\rk(u_0,\dots,u_{\rk-1})
,\el\ee
and set $\Om_1=\Om_1(0,0)$.  Also, remember  the kernel ${\mathbb L}={\mathbb L}_{n,\tau,\rho}(x_1,y_1;x_2,y_2)$ from (\ref{LR1LR2}), with  ${\frak L}_i$ defined in (\ref{LRfull}):
 \be  \label{Lsca'}
  {\mathbb L}
=-  {\mathbb L}_0 
+(-1)^\ka \left(   {\mathbb L} _1
+  {\mathbb L}_2\right) =-  {\mathbb L}_0 +
(-1)^\ka \frac{ {\frak L}_1+ {\frak L}_2 }{\LR(1)}  ,\ee
  \begin{proposition}\label{Prop7.1}
 The ${\mathbb L}_i$-parts of the kernel read as follows:
where (remember $\Phi(v) =\oint_{\ga_{\sg_2+}}\frac{h(u)du}{2\pi \I \GR(u)(v-u)}$ from (\ref{al-beta}) )
\be \label{K1}
\bl
-  {\mathbb L}_0 
  &=\Id_{x_1>x_2 }\oint_{\ga_{\rho_1}}\frac{\FR(u) du}{2\pi \I u^\ka \GR(u)}
\\
 (-1)^\ka\!({\mathbb L} _1\!+\!{\mathbb L} _2)
&= -\!\oint_{\ga_{\rho_2}} \frac{u^\rk \FR(u)du}{2\pi \I} \oint_{\ga_{\sg_2}-\Ga_0} \frac{ \Phi(v)dv }{2\pi \I h(v) v^{\rho}}\frac{\Om_1(u,v)}{(u\!-\!v)\Om_1 }
 -\oint_{\ga_{\rho_2}} \frac{\FR(u)du}{2\pi \I u^\ka \GR(u)}
\el \ee 
  \end{proposition}
  
  The two expressions (\ref{LR1}) and (\ref{LRfull}) in Proposition (\ref{Prop5.2}) contain prominently the two integrals appearing in the Lemma below, with usual contours of radii $a< \rho_1<\sg_1<\sg_2<R<R+<\rho_2<\rho_3<a^{-1}$:  
  \begin{lemma} The following multiple integrals can be evaluated in terms of the functions $\FR$, $\Phi$ and $\Om_1(u,v)$. Set $\vr=0$ or $=1$. So we have :
   \be\label{Int1'}\bl
 \left(\prod_{ {0\leq i\leq \ka-1}\atop{0\leq j\leq \rk-1}}\oint_{\Ga_0}d\widehat\al_i(u_i) 
  \oint_{\ga_{\sg_2}\!-\!\Ga^+_0 }  {d\widehat\beta_j (v_j)} 
 \!\!  \right)&
    \oint_{ \ga_R ^\ka} d\bm \mu(\bm\lb)
\frac 
{V_\rk^{(\bm \lb)} (u;v)}{(P_{\bm\lb}(v_\rk))^\vr}
 \\&\hspace*{-2cm}= \frac{(-1)^{ \frac    {\ka(\ka-1)}2}}{ (-v_\rk)^{\vr\ka }\psi^\ka}   \Om_1(0,0),~ \mbox{for $v_\rk\in \ga_{\rho_3}-\ga_{R}^+$}.
  \el\ee
and so, from (\ref{LR1}) and (\ref{LRfull}),  
\be\label{LLfrak}\bl\LR(1)&=\frac{(-1)^{\ga+\frac{\ka(\ka-1)}2}}{\psi^\ka}\Om_1(0,0) \\
(-1)^\ka {\frak L}_1&=-\frac{(-1)^{\ga+\frac{\ka(\ka-1)}{2}}}{\psi^\ka} 
\oint_{\ga_{\rho_2}}d \al_\ka(u_\ka )
     \oint_{\ga_{\rho_3}-\ga_{R+}}   \frac{d\widehat\theta_\rk(v_\rk)}{v_\rk^\ka (u_\ka\!-\!v_\rk)}\Om_1(0,0).
 \el\ee 
Also \footnote{\label{foot10}$(-1)^\star=(-1)^{\ka +\frac{(\ka+1)(\ka+2)}2 +(\rk+1)\ka+\ka\sg+\rk+\sg}$ and $(-1)^{(\rk+1)\ka+\ka\sg+\rk+\sg}=(-1)^\ka$ and $\frac{(\ka+1)(\ka+2)}2 -\frac{ \ka (\ka-1)}2=2\ka+1$}  (here $v_\rk$ is integrated about $\ga_{\sg_2}-\Ga_0^+$)
\be\label{Int1''} \bl (-1)^\ka {\frak L}_2=&(-1)^{\ga+\ka +\sg}\left(\prod_{ {0\leq i\leq \ka }\atop{0\leq j\leq \rk }}\oint_{\Ga^{(i)}_0}d\widehat\al_i(u_i) 
  \oint_{\ga_{\sg_2}\!-\!\Ga_0^+ }    d\widehat\beta_j (v_j) \right)    
    \oint_{ \Ga_R ^\ka}  d \bm \mu(\bm\lb)  V^{(\bm \lb)}_{\rk+1}   (u;v) 
  \\&     =\frac{(-1)^{ \ga+\star} }{\psi^\ka}\oint_{\ga_{\rho_2}} \frac{u^\rk \FR(u)du}{2\pi \I} \oint_{\ga_{\sg_2}-\Ga_0} \frac{ \Phi(v)dv }{2\pi \I h(v) v^{\rho}}\frac{\Om_1(u,v)}{(u-v)  } .\el\ee  
     \end{lemma}
  
  \proof {\bf Step 1.  Proof of identities (\ref{Int1'}) and (\ref{LLfrak}).} Expanding the determinant $V_\rk^{(\lb)} (u;v) $ in the expression below as a sum of products of determinants of sizes $\ka$ and $\sg$, upon using the partitions of the indices $J:=(0,\dots,\rk-1)=J_\sg\cup J_\ka$, with the corresponding variables $v_J,~v_{J_\ka},~v_{J_\sg}$, and noticing that the integrand of the integral $\oint_{   \ga_R ^\ka  } $ decouples, we find (after dividing and multiplying by $\Dt_\ka(u)$)
 \be\footnotesize\bl
 \left(\prod_{ {0\leq i\leq \ka-1}\atop{0\leq j\leq \rk-1}}\oint_{\Ga_0}d\widehat\al_i(u_i) 
  \oint_{\ga_{\sg_2}\!-\!\Ga_0 }  {d\widehat\beta_j (v_j)} 
   \right)
    \oint_{ \ga_R ^\ka} d\bm \mu(\bm\lb)
\frac 
{V_\rk^{(\bm \lb)} (u;v)}{(P_{\bm \lb}(v_\rk))^\vr}
 \label{intint} \el\ee
 $$\footnotesize\bl
 =& \left(\prod_{i=0}^{\ka-1}\oint_{\Ga_0}d\widehat\al_i(u_i)
 \right)\Dt_\ka(u)\left(\prod_{j=0}^{\rk-1 }\oint_{\ga_{\sg_2}-\Ga_0 }  {d\widehat\beta_j (v_j)} 
  \right)
  \sum_{J_\ka }(-1)^{\vr(J_\ka)}
 \\&\times\underbrace{ \frac{\det\left(\frac{1}{u_i-v_j}\right)_{{0\leq i\leq \ka-1}\atop{j\in J_\ka}} }{\Dt_\ka(u)}}_{\ast }
   \underbrace{\det (v_j^{i-\rk})_{{0\leq i\leq \sg-1}\atop{j\in J_\sg}} }_{\ast\ast}
\left( \underbrace{  
 \oint_{  \ga_R  } \frac{d \mu(\lb)}{ ( \lb-v_\rk)^\vr } \frac{\prod_{i=0}^{\ka-1}(\lb-u_i)}{\prod_{j\in J_\ka}(\lb-v_j)}
 }_{(\ast\ast\ast)} \right) ^\ka
 \el$$
At first, we use a standard identity for $(\ast)$ and the fact that $\prod_{j\in J}v_j =\prod_{j\in J_\sg} v_j
 \prod_{j\in J_\ka}v_j$ for $(\ast\ast)$. Moreover, the integral $(\ast\ast\ast)$ above is exactly like (\ref{LemmaRes1}), but applied to to the variables $(v_0,\dots,v_{\ka-1})\to v_{J_\ka}$ and $v_\ka\to v_{\rk}$; we will need to evaluate this integral at $u=0$. This gives us:
%
 %
\be\bl\label{vdm}
(\ast)&=\frac1{\Dt_\ka(u)}\det\left(\frac{1}{u_i-v_j}\right)_{{0\leq i\leq \ka-1}\atop{j\in J_\ka}} =
(-1)^{\frac 12 \ka(\ka+1)}
\frac{\Dt_\ka(v_{J_\ka})}{\prod_{{0\leq i\leq \ka-1}\atop{j\in J_\ka}} (v_j-u_i)},
\\
&\qquad\qquad\qquad\qquad\qquad\qquad~~~~ ~=(-1)^{\frac 12 \ka(\ka+1)}
\frac{\Dt_\ka(v_{J_\ka})}{\prod_{j\in J_\ka}  v^\ka_j },\mbox{  for $u=0$}
\\
(\ast\ast)&=
  \frac{\Dt_\sg(v_{J_\sg})\prod_{j\in J_\ka} v_j^\sg}{\prod_{j\in J_\sg}   v_j^\ka   \prod_{j=0}^{\rk-1}v_j^\sg}, ~\mbox{and}~~~(\ast\ast\ast)\bigr|_{u=0}=(\Lb_0^{(\vr)}+\Lb_1^{(\vr)})(v_{J_\ka},\vr v_\rk).
 \el\ee
 Therefore the right hand side of the expression (\ref{intint}) has the form $\left(\prod_{i=0}^{\ka-1}\oint_{\Ga_0}d\widehat\al_i(u_i)
 \right)\Dt_\ka(u)F(u_0,\dots,u_{\ka-1};v)$, with $d\widehat\al_i(u_i)=\frac{du_i}{2\pi \I u_i^{i+1}}$, exactly as in the last line of (\ref{Id0}) in Proposition \ref{LemmaIntW}, setting $m=\ka$. So, the expression (\ref{intint}) simplifies to, using (\ref{vdm}), (remember $\ka+\sg=\rk$)
 \be\label{intint1}\bl
(-1)^{\frac 12 \ka(\ka+1)} &\left(\prod_{j=0}^{\rk-1 }\oint_{\ga_{\sg_2}-\Ga_0 }  \frac{d\widehat\beta_j (v_j)} {v_j^{ \ka+\sg}}
  \right)
\\& \times  { \sum_{J_\ka }(-1)^{\vr(J_\ka)}\Dt_\ka(v_{J_\ka}) \Dt_\sg(v_{J_\sg})\left(\prod_{j\in J_\ka} v_j^\sg\right)}
 \left[ (\Lb_0^{(\vr)}+\Lb_1^{(\vr)})(v_{J_\ka},\vr v_\rk)\right]^\ka
  \el\ee
  The claim is that all terms in the expansion of the $\ka$th power of the square bracket have integral$=0$, except for the $(\Lb_0^{(\vr)})^\ka$-term. For convenience replace the $v_{J_\ka}$-variables by $(v_1,\dots,v_\ka)$, having for effect to change the $j$-label in the $d\widehat\beta_j (v_j)$-integrals to some $0\leq m_j\leq \rk-1$. Then Proposition \ref{LemmaInt0} shows that the $\ka$th power of the square bracket in (\ref{intint1}) can simply be replaced by $\left[ (\Lb_0^{(\vr)}(v_{J_\ka},\vr v_\rk)\right]^\ka$.
  
 %
 %
 Consider now the corresponding polynomials $Q_J(z)=\prod_{i\in J}(z-v_i)$, with $Q_J=Q_{J_\ka}Q_{J_\sg}$ to the partitions    $J  =J_\ka\cup J_\sg$ mentioned before. The $\vr_j(v_J) $, $c_j(v_J)$, and the $a_j(v_J)$ below are symmetric polynomials in the respective variables. So we have, using (\ref{h'}):  
%
%
$$\bl
 \frac1{Q_{J\cup\vr v_\rk} (z)}&=\sum_{j=0}^\infty \tfrac {c_j(v_J,\vr v_\rk)}{j!}z^j 
  \mbox{, with }\left\{ \begin{array}{lll}(\tfrac {d}{dz})^j \frac1{Q_{J\cup\vr v_\rk} (z)}\big|_{z=0}=c_j(v_J,\vr  \rk )  \\
     c_0(v_{J \cup \vr \rk} )=\frac{(-1)^{\rk+\vr}}{\prod_0^{\rk-1+\vr}v_i} \end{array}\right.
\\Q_{J_\sg}&=\prod_{i\in J_\sg}  (z\!-\!v_i)\!=\!\sum_{i=0}^\sg 
  z^{\sg-i}\vr_i(-v_{J_\sg}),\mbox{with } \left\{\begin{array}{llll}(\tfrac{d }{dz })^k Q_{J_\sg}\big|_{z=0}=
   k!\vr_{\sg-k} (-v_{J_\sg}),
   \\ \vr_0=1\end{array}\right.
\\ h_0(z)&=\sum_0^{n+1} \al_i z^i, \mbox{  with  }\left(\tfrac{d}{dz}\right)^{i} h_0\big|_{z=0} =i!\al_i
\mbox{ and }\al_0=\tfrac 1\psi
\el$$
and so\footnote{It is understood that $\vr v_\rk$ does not appear when $\vr=0$.}
%
\be\footnotesize\bl \label{Lambda}
\Lb_0^{(\vr)}&=\frac{1}{\sg !  }\left(\frac{d}{dz}\right)^{\sg} \frac{[h_0(z)]_{[0,\sg]}}{Q_{J_\ka\cup \vr v_\rk} (z)}\Bigr|_{z=0}
  =\frac{1}{\sg !  }\left(\frac{d }{dz }\right)^{\sg}\frac{ h_0(z) }{Q_{J_\ka \cup \vr v_{\rk}} (z)}\Bigr|_{z=0}
\\&=
\frac{1}{\sg !  }\sum_{i+j+k=\sg} \left(  {\sg}\atop{i~j~k}\right)
\left(\frac{d}{dz}\right)^{i} h_0 \left(\frac{d}{dz}\right)^{j}\frac1{Q_{J_\ka\cup \vr v_\rk }(z)}\left(\frac{d}{dz}\right)^{k}
 Q_{J_\sg} \Bigr|_{z=0}
 \\&=
\frac{c^{(\vr)}_0}{\sg !  } \sum_{k=0}^\sg   k!\vr_{\sg-k}(-v_{J_\sg})\left[\sum_{i+j =\sg-k} \left(\!  {\sg}\atop{i~j~k}\!\right)
\frac{i!\al_i c_j(v_{J_\ka\cup \vr v_{\rk}})}{c^{(\vr)}_0}\right]
\\&= c^\vr_0 \left(\al_0   
+\sum_{i=1}^{ \sg }a^\vr_i(v_{J_\ka}) \vr_i(v_{J_\sg})\right)
 =\tfrac{(-1)^{\rk+\vr}}{\prod_0^{\rk-1+\vr}v_i}\left(\frac1\psi+\sum_{i=1}^{ \sg }a^\vr_i(v_{J_\ka}) \vr_i(v_{J_\sg})\right),
 \el\ee 
 where the first term is obtained by setting $k=\sg$.
  
  Based on the Pieri formula, we will be using the following expansion of products of $\ga_i$th powers of symmetric functions in Schur polynomials $S_\lb(v_{J_\sg})$, where $\sum_{i=0}^\sg \ga_i=\ka$,~$\ga_i\geq0$:
  \be\bl
 \prod_{1}^\sg  \vr_i(v_{J_\sg})^{\ga_i}
 =
 \sum_{{0\leq \lb_1\leq \sum_1^\sg \ga_i=\ka-\ga_0}\atop{0<\lb_1^{\top}\leq \sg},~ {|\lb|\leq \sum_1^\sg i\ga_i}}
 C_\lb(\ka-\ga_0) S_\lb(v_{J_\sg}),
 \el\label{sym}\ee
 and by the Jacobi-Trudy formula,
 \be\bl S_\lb(v_{J_\sg}) =\frac{\det\left(v_{j}^{\lb_i+\sg-i} \right)_{{1\leq i\leq \sg}\atop {j\in J_\sg}}}{\det\left(v_{j}^{\sg-i} \right)_{{1\leq i\leq \sg}\atop {j\in J_\sg}}} 
 & =(-1)^{\frac 12 \sg(\sg-1)}
 \left(\prod_{j\in J_\sg} v_j^\sg \right)
 \frac{\det\left(v_{j}^{\lb_i -i} \right)_{{1\leq i\leq \sg}\atop {j\in J_\sg}}}{ \Dt_\sg(v_{J_\sg})} 
 .\el\label{sym1}\ee  
%
 %
 So, expanding the $\ka$th power of  $\Lb_0^{(\vr)}$ and using (\ref{sym}) and (\ref{sym1}), one finds a leading term corresponding to $\ga_0=\ka$ and further terms for $0\leq \ga_0<\ka$:%
 \be\label{Lambdak}\bl
 \Lb_0^{(\vr)\ka}  
&=\frac{(-1)^{(\rk+\vr)\ka }}{\prod_{j=0}^{\rk-1+\vr}v_j^\ka}
\left(\al_0^\ka   
+\sum_{{\sum_{0}^{\sg } \ga_i=\ka-1}\atop{0\leq \ga_0<\ka}}\left({\ka}\atop{\ga_0\dots \ga_\sg}\right)
    \prod_{0}^\sg  \left(a^\vr_i\vr_i(v_{J_\sg})\right)^{\ga_i}%
  \right) 
  \\
&=\frac{(-1)^{(\rk+\vr)\ka }}{\prod_{j=0}^{\rk-1+\vr}v_j^\ka}
\left(\al_0^\ka   
+\frac{(-1)^{\frac 12 \sg(\sg-1)}(\prod_{j\in J_\sg}v^\sg_{ j})} 
{ \Dt_\sg(v_{J_\sg})}\sum_{{\sum_{0}^{\sg } \ga_i=\ka}\atop{0\leq \ga_0<\ka}}\left({\ka}\atop{\ga_0\dots \ga_\sg}\right)
    \prod_{0}^\sg  ( a^\vr_i) ^{\ga_i}%
  \right.
  \\
  &\times \left.\sum_{{0\leq \lb_1\leq  \ka-\ga_0>0}\atop{0<\lb_1^{\top}\leq \sg},~ {
   }}
   C_\lb (\ka-\ga_0)  
 {\det\left(v_{j}^{\lb_i -i} \right)_{{1\leq i\leq \sg}\atop {j\in J_\sg}}}\right) 
=:\frac{(-1)^{(\rk+\vr)\ka }}{v_\rk^{\vr\ka}\prod_{j=0}^{\rk-1 }v_j^\ka}(\al_0^\ka+R)\el\ee
Substituting $ \Lb_0^{(\vr)\ka}$ of (\ref{Lambdak}) in (\ref{intint1}), 
 one looks separately at the two pieces, first $\al_0^\ka$ and then $R$. The first one reads, using the underbracket in (\ref{intint1}),
$$\bl
(-1)^{(\rk+\vr)\ka+\frac 12 \ka(\ka+1)}&\left(\frac{\al_0}{v_\rk^{\vr }}\right)^\ka  \left(\prod_{j=0}^{\rk-1 }\oint_{\ga_{\sg_2}-\Ga_0 }  \frac{d\widehat\beta_j (v_j)}  {v_j^{2\ka+\sg}}
  \right)\sum_{J_\ka }(-1)^{\vr(J_\ka)}
\left( \prod_{j\in J_\ka} v_j^\sg  \right)\Dt_\ka(v_{J_\ka})\Dt_\sg(v_{J_\sg})
 \\& =(-1)^{(\sg+\rk+\vr) \ka+\frac 12 \ka(\ka+1)}\left(\frac{\al_0}{v_\rk^{\vr }}\right)^\ka \left(\prod_{j=0}^{\rk-1 }\oint_{\ga_{\sg_2}-\Ga_0 }  \frac{d\widehat\beta_j (v_j)}  {v_j^{2\ka+\sg}}
  \right)\Dt_\rk(v)
  \\& =(-1)^{ \frac { \ka(\ka-1)}2}\left(\frac{1}{(-v_\rk)^{\vr }\psi}\right)^\ka \Om_1(0,0),
  \el$$
 using (\ref{al-beta}), with $(-1)^{(\sg+\rk+\vr) \ka+\frac 12 \ka(\ka+1)}=(-1)^{\frac 12 \ka(\ka-1)+\ka \vr}$. 
 %
  The second term of (\ref{intint1}) containing $R$ reads as follows, up to signs and factors containing $v_\rk$,

 $$\bl
\frac{1}{v_\rk^{\vr \ka}} \sum_{{\sum_{0}^{\sg } \ga_i=\ka}\atop{0\leq \ga_0<\ka}}\left({\ka}\atop{\ga_0\dots \ga_\sg}\right)
 &
   \al_0^{\ga_0} 
 \left(\prod_{j=0}^{\rk-1 }\oint_{\ga_{\sg_2}-\Ga_0 } \frac {d\widehat\beta_j (v_j)}  {v_j^{2\ka}}
  \right)
  \prod_{1}^\sg   a_i ^{\ga_i}
  \sum_{{0\leq \lb_1\leq  \ka-\ga_0>0}\atop{0<\lb_1^{\top}\leq \sg} }
   C_\lb
 \\
 &\times  \sum_{J_\ka }(-1)^{\vr(J_\ka)}
 \Dt_\ka(v_{J_\ka})
 \det (v_j^{\lb_i-i})_{{1\leq i\leq \sg}\atop{j\in J_\sg}};
    \el   $$
  the sum on the second line is a determinant of a matrix with two equal rows, since $\lb_1\leq \ka-\ga_0\leq \ka$  and  therefore it vanishes:
    $$\bl
    \sum_{J_\ka }(-1)^{\vr(J_\ka)}
 \Dt_\ka(v_{J_\ka})
 \det (v_j^{\lb_i-i})_{{1\leq i\leq \sg}\atop{j\in J_\sg}}
 =\det\left(
 \begin{array}{llll}
 v_j^0\\
 \vdots\\
 v_j^{\ka-1}
 \\ \\
 v_j^{\lb_1-1}
 \\ \vdots \\
 v_j^{\lb_\sg-\sg}
 \end{array}
 \right)_{j\in J_\sg} =0
. \el$$
This ends the proof of identity (\ref{Int1'}). Identities (\ref{LLfrak}) follow at once from the expressions (\ref{LR1}) and (\ref{LRfull}) for $\LR(1)$ and ${\frak L}_1$.

{\bf Step 2.  Proof of identity (\ref{Int1''}).} Indeed, using now the partitions $J_\sg\cup J_{\ka+1} $ of the indices $(0,\dots,\rk)$ and setting $J_{\sg}^c=J_{\ka+1}$, the determinant $V_{\rk+1}^{ }  (u;v)$, as in (\ref{Vplus}), can be expanded in a similar way as before and thus the integral expression in (\ref{Int1''}) can be written as follows:
 \be\label{Int1'''}\bl
 &(-1)^{\ga+\sg}{\frak L}_2 
\\&=\oint_{\ga_{\rho_2}}d \al_\ka(u_\ka)
 \left(\prod_{ {0\leq i\leq \ka-1}\atop{0\leq j\leq \rk }}\oint_{\Ga_0}d\widehat\al_i(u_i)   \oint_{\ga_{\sg_2}\!-\!\Ga_0^+ }  {d\widehat\beta_j (v_j)} 
   \right)
    \oint_{ \ga_R ^\ka} 
 d\bm \mu(\bm\lb)  V_{\rk+1}^{(\bm \lb)} (u;v) 
 \\
 &= \oint_{\ga_{\rho_2}}d \al_\ka(u_\ka)
 \left(\prod_{i=0}^{\ka-1}\oint_{\Ga_0}d\widehat\al_i(u_i)
 \right)\Dt_\ka(u)\left(\prod_{j=0}^{\rk  }\oint_{\ga_{\sg_2}-\Ga_0 }  {d\widehat\beta_j (v_j)} 
  \right)
  \sum_{J_\ka }(-1)^{\vr(J_\ka)}
 \\&\times\underbrace{ \frac{\det\left(\frac{1}{u_i-v_j}\right)_{{0\leq i\leq \ka }\atop{j\in J_{\ka+1 } } } }{\Dt_\ka(u)}}_{\ast }
   \underbrace{\det (v_j^{i-\rk})_{{0\leq i\leq \sg-1}\atop{j\in J_\sg}} }_{\ast\ast}
 \underbrace{ \left( \oint_{  \ga_{_R}  } 
 d \mu(\lb) 
 \frac{\prod_0^{\ka-1}(\lb-u_i)}{\prod_{j\in J_{\ka+1}}(\lb-v_j)}\right) ^\ka
 }_{\ast\ast\ast}.
 \el\ee
In the same way as before, using (\ref{Lambdak}) and using the same sequence of arguments, one computes the following expressions evaluated at $u:=(u_0=\dots=u_{\ka-1})=0$:
$$ \bl
(*)\Bigr|_{u=0}&=(-1)^{\frac 12 (\ka+1)(\ka+2)}
\tfrac{\Dt_{\ka+1}(v_{J_{\ka+1}})\Dt_{\ka+1}(u)}{\prod_{{0\leq i\leq \ka}\atop{j\in J_{\ka+1}}}{ (v_j-u_i) \Dt_\ka(u)} }\Bigr|_{u=0}~,~ (**) =\tfrac{\Dt_\sg(v_{J_\sg})}{\prod_{j\in J_\sg}(v_j^\ka v_j^\sg)}
\\&=
(-1)^{\frac 12 (\ka+1)(\ka+2)}\tfrac{\Dt_{\ka+1}(v_{J_{\ka+1}}) u_\ka^\ka \prod_{j\in J_{\sg}}(v_j^\ka(v_j-u_\ka))    }{\prod_{j=0}^{\rk}(v_j^\ka(v_j-u_\ka))}
\\ 
(***)\Bigr|_{u=0}&=(c'_0)^\ka 
 \left(\al_0
+\sum_{i=1}^{ \sg }a'_i \vr_i(v_{J_\sg})\right)^\ka  \mbox{,~~with~~} \left\{\begin{array}{lll}a'_i=a'_i(v_0,\dots,v_{\rk })
\\
\al_0=h(0)=\frac 1\psi, ~\vr_0=1
\\
c'_0=\frac{(-1)^{(\rk+1)  }}{\prod_{j=0}^{\rk }v_j}
\end{array}\right.
\\
&=\frac{(-1)^{(\rk+1)\ka }}{\prod_{j=0}^{\rk }v_j^\ka}
\left(\al_0^\ka    
+\sum_{{\sum_{0}^{\sg } \ga_i=\ka}\atop{0\leq \ga_0<\ka}}\left({\ka}\atop{\ga_0\dots \ga_\sg}\right)
    \prod_{0}^\sg  \left(a'_i\vr_i(v_{J_\sg})\right)^{\ga_i}%
  \right) 
\el $$   
 $$\bl
&=\frac{(-1)^{(\rk+1)\ka }}{\prod_{j=0}^{\rk }v_j^\ka}
\left(\al_0^\ka
+\frac{(-1)^{\frac 12 \sg(\sg-1)}(\prod_{j\in J_\sg}v^\sg_{ j})} 
{ \Dt_\sg(v_{J_\sg})}\sum_{{\sum_{0}^{\sg } \ga_i=\ka}\atop{0\leq \ga_0<\ka}}\left({\ka}\atop{\ga_0\dots \ga_\sg}\right)
    \prod_{0}^\sg  ( a'_i) ^{\ga_i}%
  \right.
  \\
  &\times \left.\sum_{{0\leq \lb_1\leq  \ka-\ga_0>0}\atop{0<\lb_1^{\top}\leq \sg},~ {
   }}
   C_\lb (\ka-\ga_0)  
 {\det\left(v_{j}^{\lb_i -i} \right)_{{1\leq i\leq \sg}\atop {j\in J_\sg}}}\right) 
=:\frac{(-1)^{(\rk+1)\ka }} {\prod_{j=0}^{\rk  }v_j^\ka}(\al_0^\ka+R').\el$$
As before, using Propositions  \ref{LemmaIntW}  and the same argument as before concerning the $d\widehat \beta_j$-integrations, 
 and inserting the product of these three formulas into the previous formula, the latter expression is a sum of expressions depending on $\ga_i$ and partitions $\lb_j$, with $0\leq\lb_1\leq \sum_1^\sg \ga_i=\ka-\ga_0$ for $0\leq \ga_0\leq \ka$; the first term corresponds to $\ga_0=\ka$, implying all $\lb_i=0$. This gives us in either case, up to a multiplicative constant, $=1$ when $\ga_0=\ka$:
$$\bl \tfrac{ (-1)^{\frac {(\ka+1)(\ka+2)}2+(\rk+1)\ka }}{\psi^\ka}
\oint_{\ga_{\rho_2}}
& u_\ka^\ka d\al_\ka(u_\ka) 
 \left(\prod_{j=0}^{\rk   }
 \oint_{\ga_{\sg_2}-\Ga_0 }  \frac{d\widehat\beta_j (v_j)}{v_j^{2\ka}(v_j-u_\ka)} 
  \right)
  \\&~~~~~\times\sum_{J_\sg}(-1)^{\vr(J_\sg)}
  \Dt_{\ka+1}(v_{J_{\ka+1}})\det\left[v_{j}^{\lb_i-i}(v_{j}-u_\ka)\right]_{{i=\sg,\dots,1}\atop{j\in J_\sg}}
\el$$
$$\bl
&=\tfrac{(-1)^{\frac{(\ka+1)(\ka+2)}2 +(\rk+1)\ka}}{\psi^\ka}
\oint_{\ga_{\rho_2}}
u_\ka^\ka d\al_\ka(u_\ka)
 \left(\prod_{j=0}^{\rk  }\oint_{\ga_{\sg_2}-\Ga_0 }  \frac{d\widehat\beta_j (v_j)}{v_j^{2\ka}(v_j-u_\ka)} 
  \right)\det\left[\begin{array}{cccc}
  v_j^0 \\ \vdots \\ v_j^\ka \\ 
  v_j^{\lb_\sg-\sg}(v_j-u_\ka)
  \\\vdots\\ 
  v_j^{\lb_1-1}(v_j-u_\ka)
  \end{array}\right]_{0\leq j\leq \rk},
\\
&=\left\{\bl
&\tfrac{(-1)^{\frac{(\ka+1)(\ka+2)}2 +(\rk+1)\ka+\ka\sg+\rk}}{\psi^\ka}\oint_{\ga_{\rho_2}} \frac{u^\rk \FR(u)du}{2\pi \I} \oint_{\ga_{\sg_2}-\Ga_0} \frac{ \Phi(v)dv }{2\pi \I h(v) v^{\rho}}\frac{\Om_1(u,v)}{(u-v)  }
,\mbox{  if $\ga_0=\ka$ 
 }
\\
&0, \mbox{  if $0\leq\ga_0<\ka$}
&\el
\right.\el$$
The last equality follows from noticing that the determinant above vanishes, unless all $\lb_j=0$ (since $\lb_1\leq \ka-\ga_0$). In the latter case, the determinant above equals, after row operations : $$\bl
 &   \frac{(-1)^{\ka\sg}  u_\ka ^\sg}{\prod_{j=0}^{\rk}v_j^\sg}\Dt_{\rk+1}(v)
 =\frac{(-1)^{\ka\sg}  u_\ka ^\sg}{\prod_{j=0}^{\rk}v_j^\sg}\Dt_{\rk }(v)
 \prod_{j=0}^{\rk-1}(v_\rk-v_j)
  =\frac{(-1)^{\ka\sg+\rk}  u_\ka ^\sg}{\prod_{j=0}^{\rk}v_j^\sg}\Dt_{\rk }(v)
 \prod_{j=0}^{\rk-1}(v_j-v_\rk). \el$$
  Then one replaces the integration variables $u_\ka$ and $v_\rk$ by $u$ and $v$ in the last equality. This yields the desired expression  for (\ref{Int1'''}) and thus (\ref{Int1''}) for ${\frak L}_2$.\qed

  \noindent {\em Proof of Proposition \ref{Prop7.1}}: The ratios of the  integrals $(-1)^\ka{\frak L}_1$ and $\LR(1)$ in (\ref{LLfrak}) and the ratios of $(-1)^\ka{\frak L}_2$ in (\ref{Int1''}) and $\LR(1)$ in (\ref{LLfrak}), one finds
   $$\bl  (-1)^{\ka} \frac{{\frak L}_1}{\LR(1)}
 &=-\oint_{\ga_{\rho_2}} d\al_\ka (u)
  \oint_{\ga_{\rho_3}-\ga_{R}^+}
  \frac{ d\widehat \theta_\rk(v)}{v^\ka(u-v)}
  \\&=\oint_{\ga_{\rho_2}} du \FR(u) 
  \oint_{\ga_{\rho_3}-\ga_{R}^+}
  \frac{ dv}{2\pi \I \GR(v)v^\ka(u-v)}=-\oint_{\ga_{\rho_2}} 
  \frac{\FR(u)du}{2\pi \I u^\ka\GR(u)}
  \\ \\
  (-1)^\ka\frac{{\frak L}_2}{\LR(1)}&=
\frac{ (-1)^{\star\star} }{\Om_1(0,0)} \oint_{\ga_{\rho_2}} \frac{u^\rk \FR(u)du}{2\pi \I} \oint_{\ga_{\sg_2}-\Ga_0} \frac{ \Phi(v)dv }{2\pi \I h(v) v^{\rho}}\frac{\Om_1(u,v)}{(u-v)  },\el$$ 
%
 with $(-1)^{\star\star}=(-1)^{\ka+\frac{(\ka+1)(\ka+2)}{2}-\frac{\ka(\ka-1)}{2}+(\rk+1)\ka+\ka\sg+\rk+\sg}=-1$ using footnote \ref{foot10}. The single integral for $\frac{{\frak L}_1}{\LR(1)}$ comes from working out the integral about $\ga_{\rho_3}-\ga_{R}^+$. Summing up these two expressions gives formula (\ref{K1}) and thus ends the proof of Proposition \ref{Prop7.1}.\qed

\section{Preparing the ${\mathbb L}_{n,\rk,\rho}$- kernel for asymptotics}

In order to perform the asymptotics, we must express the  contour integrals in ${\mathbb L}$ (as in (\ref{K1})) in terms of only two contour integrals, namely $\Ga_0\ni 0$ and $\ga_{\sg_2}\!-\!\Ga_0\ni 1/t$.  Indeed, $0$ and $1/t$ are the only poles of the integrands of kernel (\ref{K1}). Doing so transforms the simple kernel (\ref{K1}) into a more complicated one ${\mathbb L}= -
{\mathbb L}_0'+\sum_{i=1}^4
{\mathbb L}'_i $ with the ${\mathbb L}'_i$'s as in (\ref{Lfinal'}). 
 
To do so, we define the following functions:
\be\label{Om}\bl
\Om(u,v)&=\frac{(-1)^{\frac{\rk(\rk-1)}{2}}}{ \rk  !}
  \left(\prod_{\ell=0}^{\rk-1}\oint_{\ga_{\sg_2}-\Ga_0}\frac{dv_\ell}{2\pi \I v_\ell^{\rho}h(v_\ell)} \frac{v-v_\ell}{u-v_\ell} \right)
 \Dt_{\rk  }^2(v_0,\dots,v_{\rk-1  })
\\ 
\Om^\pm_{\rk \mp 1}(u,v)&=
\frac{(-1)^{\frac{\rk(\rk-1)}{2}}}{(\rk \mp 1)!}
  \left(\prod_{\ell=0}^{\rk-1 \mp 1}\oint_{\ga_{\sg_2}-\Ga_0}\frac{((u-v_\ell)(v-v_\ell))^{\pm 1}dv_\ell}{2\pi \I v_\ell^{\rho}h(v_\ell)}  \right)
 \Dt_{\rk \mp 1}^2(v_0,\dots,v_{\rk-1 \mp 1})
, \el\ee
 \noindent and the following auxiliary functions:
 \be 
  \label{Omtil}\bl 
 \Om_1(u,v  ) &:= \left(
\prod_{\ell=0}^{\rk-1}\oint_{\ga_{\sg_2}-\Ga_0}\frac{du_\ell  }{2\pi \I u_\ell^{\ka+\ell+1}  h(u_\ell) }\frac{ u_\ell-v}{u_\ell-u}
\right)\Dt_\rk(u_0,\dots,u_{\rk-1})
\\
  \Om^{(k)}_2(u)  &:=
  \left(
\prod_{{\ell=0}\atop{\ell\neq k}}^{\rk-1}\oint_{\ga_{\sg_2}-\Ga_0}\frac{du_\ell}{2\pi \I u_\ell^{\ka+\ell+1}  h(u_\ell)}
\right)     \Dt_{\rk}(u_0,\dots,u_{k-1},u, u_{k+1},\dots,u_{\rk-1}).
\el \ee 
The kernel ${\mathbb L}$ as in (\ref{Lsca'}) contains a nested triple integral depending on $\Om_1(u,v)$, itself a $\rk$-uple integral, with $h,\FR,\GR$ behaving for $t\to 0$ as indicated in (\ref{series}). From Fig. the following contours can be replaced by new ones (with $\widetilde\Ga^+$ slightly larger than $\widetilde\Ga $):
\be \label{contours}\ga_{\sg_2}-\Ga_0 \to\widetilde \Ga \mbox{ and }
\ga_{\rho_2}\mbox{ or } \ga_{\sg_2^+}
\to\widetilde \Ga^+ +\Ga_0^-.
\ee

 \newpage

\setlength{\unitlength}{0.015in}\begin{picture}(0,0)

\put(155,  50 ){\makebox(0,0){\includegraphics[width=130mm,height=150mm]{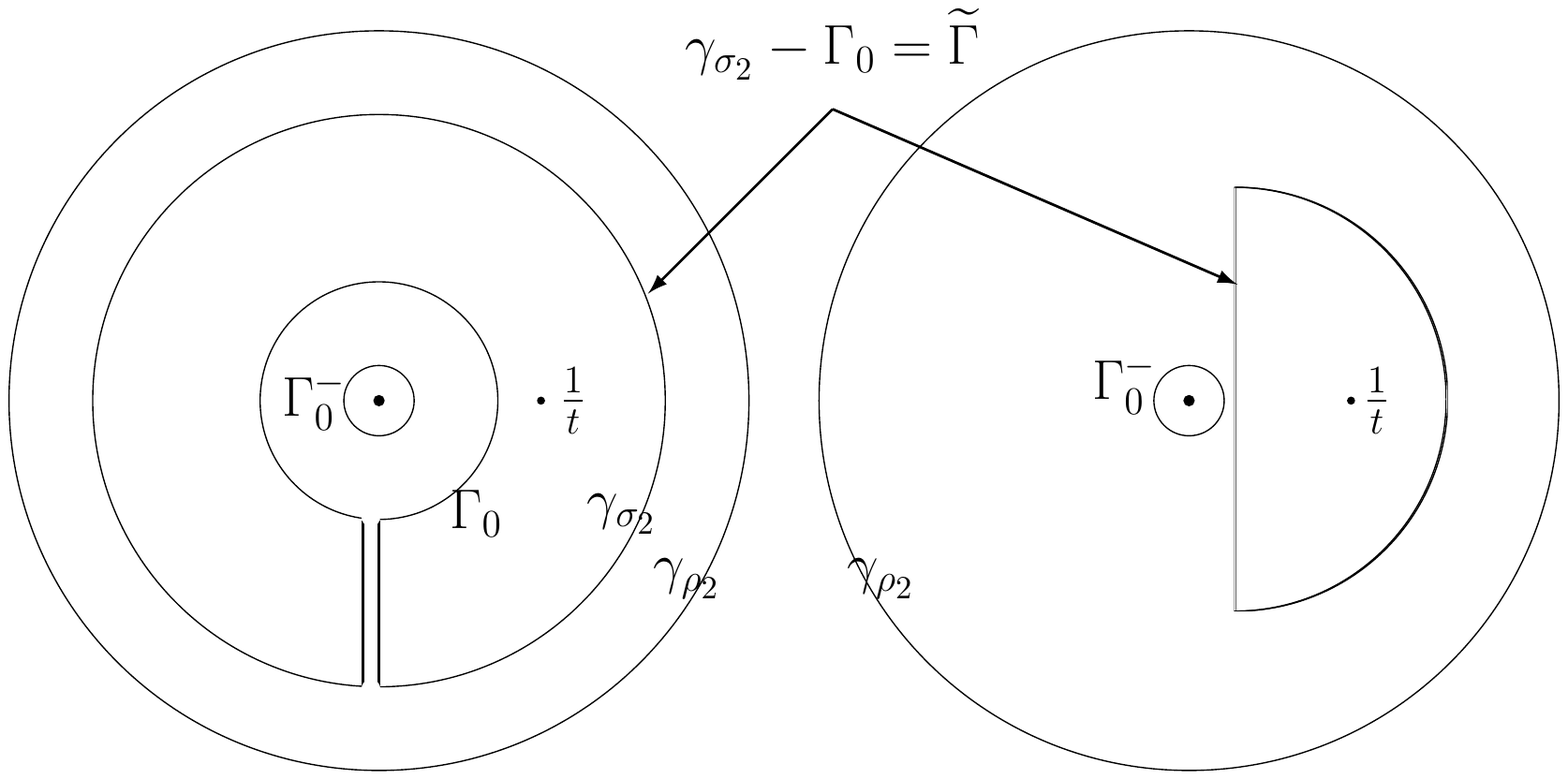}}}

\end{picture}

\vspace*{1cm}
 
Fig. 8: Turning the contour $\ga_{\sg_2}-\Ga_0$ into a semicircle $\widetilde \Ga$ of radius $1/t^2$ and containing $1/t$, to the right of $\Ga_0^-$. 

\vspace*{.6cm}

\begin{proposition}\label{Prop-L'} The kernel  ${\mathbb L}={\mathbb L}_{n,\rk,\rho} (x_1,y_1;x_2,y_2)$ has the following form:
\be \label{Lfinal}
{\mathbb L}= -
{\mathbb L}_0'+\sum_{i=1}^4
{\mathbb L}'_i ,
\ee
where ($h,~\FR\mbox{ and } \GR$ are defined in (\ref{series}))
 \be\bl\label{Lfinal'}
-{\mathbb L}'_0&:
=-\Id_{x_1>x_2}\Id_{y_1\geq y_2}\oint_{ \Ga_0} 
\frac{   \FR(w) dw}{2\pi \I w^\ka \GR(w)  }-\Id_{x_1>x_2}\Id_{y_1= y_2} \tfrac{(-a^2t)^{x_1-x_2-1}}{ 1+a^{-2} }
\\
{\mathbb L}'_1+{\mathbb L}'_2&:=\oint_{\Ga_0^-}\frac{du}{(2\pi \I)^2}
\oint_{\widetilde \Ga}\frac{dv}{u-v}\left(\frac{u^\rk \FR(u)}{v^\rho \GR(v)}+\frac{\frac{h}{\GR}(u)}{v^\ka \frac{h}{\FR}(v)}\right)
\frac{\Om(u,v)}{\Om(0,0)}
\\
{\mathbb L}'_3&:=-\oint_{\widetilde \Ga}\frac{du}{(2\pi \I)^2}
\oint_{\widetilde \Ga}dv\frac{u^{-\ka}\frac{\FR}{h}(u)} 
 {v^\rho \GR(v)}
\frac{\Om^+_{\rk-1}(u,v)}{\Om(0,0)}
 \\%
 {\mathbb L}'_4 &:=\oint_{\Ga_0^-}\frac{du}{(2\pi \I)^2}
\oint_{\Ga_0^-}dv \frac{u^{ \rk}\FR(u) } 
 {\frac{ \GR }{h}(v )}
\frac{\Om^-_{\rk+1}(u,v )}{\Om(0,0)}
.\el\ee
\end{proposition}
Before proving this proposition, we need the following Lemma:
\begin{lemma} \label{Lemma Id1}The following identities hold:
\be\bl
(i)~&\Om_1(u,v) =\Om(u,v)
\\
(ii)~& \sum_{k=0}^{\rk-1}
 \frac{\Om_2^{(k)}(v)}{u^{k+1-\rk}}
    = \Om^+_{\rk-1}(u,v)
  \\
 (iii)~&\oint_{\ga_{\sg_2}-\Ga_0}
 \frac{ dw}{2\pi \I w^\rho h(w)}
 \frac{\Om(u,w)}{(w-u)(w-v)}
  =\Om_{\rk+1}^-(u,v)
  \\
 (iv)~ &\oint_{\ga_{\sg_2}-\Ga_0}
 \frac{ dw}{2\pi \I w^\rho h(w)}
 \frac{\Om^+_{\rk-1}(v,w)}{(w-u) }
  =\frac{\Om (u,v)-\Om(0,0)}{ u-v}.
 \el\label{Id1}\ee

\end{lemma}
\proof At first we prove (\ref{Id1}){\it(i)}; indeed, remembering $\rho=\rk+\ka$,
$$\footnotesize\bl
  \Om_1(u,v)
  &= \left(\prod_{\ell=0}^{\rk-1}\oint_{\ga_{\sg_2}-\Ga_0}\frac{du_\ell}{2\pi \I u_\ell^{\ka+\ell+1}h(u_\ell)}\frac{v-u_\ell}{u-u_\ell}\right)\Dt_\rk(u_0,\dots,u_{\rk-1})
 \\&=\left(\prod_{\ell=0}^{\rk-1}\oint_{\ga_{\sg_2}-\Ga_0}\frac{du_\ell}{2\pi \I u_\ell^{\ka +1}h(u_\ell)}\frac{v-u_\ell}{u-u_\ell}\right)\underbrace{\left(\prod_{\ell=0}^{\rk-1} u_\ell^{-\ell}\right)}_{*}\Dt_\rk(u_0,\dots,u_{\rk-1})
  \\&=\frac{(-1)^{\frac{\rk(\rk-1)}{2}}}{r!}\left(\prod_{\ell=0}^{\rk-1}\oint_{\ga_{\sg_2}-\Ga_0}\frac{du_\ell}{2\pi \I u_\ell^{\ka+\rk}h(u_\ell)}\frac{v-u_\ell}{u-u_\ell}\right)
\Dt_\rk(u_0,\dots,u_{\rk-1})^2
\\&=\Om(u,v);
  \el$$
 the presence of the Vandermonde $\Dt_\rk$ following
 the expression $(*)$ above enables us to replace this product by
 $$\bl
  \frac1{r!}\det(u_i^{-j})_{0\leq i,j\leq \rk-1}=
  \frac{(-1)^{\frac{\rk(\rk-1)}{2}}}{r!} 
  \prod_{\ell=0}^{\rk-1}\left(\frac1{u_\ell^{\rk-1}}\right)\Dt_\rk(u_0,\dots,u_{\rk-1})
 .\el$$
 Proof of (\ref{Id1}){\em(ii)}:
Upon relabeling,
 $$v_\ell=\left\{  \bl u_\ell,~~0\leq \ell\leq k-1
                            \\ u_{\ell+1},~~k\leq \ell <\rk-1 \el \right.
                            \mbox{  and  }
     \lb_\ell(k)=\left\{ \bl \ell,~~~0\leq \ell\leq k-1
     \\
     \ell+1,~~~k\leq \ell\leq \rk-1  \el\right.                       
 , $$ 
  the Vandermonde below reads                      
 $$
 \Dt_\rk(u_0,\dots,u_{k-1},u,u_{k+1},\dots,u_{\rk-1})
 =(-1)^{\rk-k-1}\prod_{\ell=0}^{\rk-2}(u-v_\ell)\Dt_{\rk-1}(v_0,\dots,v_{\rk-2})
 ,$$
 and $\Om_2^{(k)}$ becomes
 \be\Om_2^{(k)}(u)=
 (-1)^{\rk-k-1}\left(\prod_{\ell=0}^{\rk-2}\oint_{\ga_{\sg_2}-\Ga_0}\frac{(u-v_\ell)dv_\ell}{2\pi \I v_\ell^{\ka+1}h(v_\ell)}\underbrace{\frac1{v_\ell^{\lb_\ell(k)}}}_{*}\right)
 \Dt_{\rk-1}(v_0,\dots,v_{\rk-2})
 .\label{Id1'}\ee
 Notice that in the integral above the ratio $(*)$ can be replaced, as before, by a Vandermonde; one then uses the representation of Schur  polynomials $S_\lb$ as a ratio of Vandermonde-like determinants:   
 \be\footnotesize\label{Id1''}\bl\frac1{(\rk-1)!} \det &\left(\frac{1}{v_i^{\lb_j(k)}}\right)_{0\leq i,j\leq \rk-1}   =\frac1{(\rk-1)!} 
 \Dt_{\rk-1}( \frac{1}{v_0},\dots,\frac{1}{v_{\rk-2}})
  S_{1^{\rk-k-1}}(\frac1{v_0},\dots,
 \frac1{v_{\rk-2}})
 \\&=\frac{(-1)^{\frac{(\rk-1)(\rk-2)}{2}}}{(\rk-1)!} 
\left(\prod_{\ell=0}^{\rk-2}\frac1{v_\ell^{\rk-2}}\right)\Dt_{\rk-1}(  {v_0},\dots, {v_{\rk-2}}) S_{1^{\rk-k-1}}(\frac1{v_0},\dots,\frac1{v_{\rk-2}}),
 \el\ee
 where the $S_{1^{\rk-k-1}}=e_{\rk-1-k}$ are elementary symmetric polynomials: 
 \be\label{Id1'''}\sum_{k=0}^{\rk-1}(-w)^{\rk-k-1}e_{\rk-1-k}(\frac1{v_0},\dots,
 \frac1{v_{\rk-2}})=\prod_{\ell=0}^{\rk-1}\left(1-\frac w{v_\ell}\right)=(-1)^{\rk-1}\left(\prod_{\ell=0}^{\rk-1}\frac1{v_\ell}(w-v_\ell)\right)
 .\ee
 Then the generating function of $\Om_2^{(k)}$, as in (\ref{Id1'}), with (\ref{Id1''}) and (\ref{Id1'''}) inserted, reads
 $$\footnotesize\bl\sum_{k=0}^{\rk-1}
 \frac{\Om_2^{(k)}(u)}{w^{k+1-\rk}}
 &=\frac{(-1)^{\frac{\rk(\rk-1}{2}}}{(\rk-1)!}
  \left(\prod_{\ell=0}^{\rk-2}\oint\frac{(u-v_\ell)(w-v_\ell)dv_\ell}{2\pi \I v_\ell^{\ka+\rk}h(v_\ell)}  \right)
 \Dt_{\rk-1}^2(v_0,\dots,v_{\rk-2})
  =\Om^+_{\rk-1}(u,w),
 \el$$
remembering from (\ref{sg'}) that $\ka+\rk=\rho$; this establishes (\ref{Id1}){\em (ii)}. 
\medbreak
\noindent Proof of  (\ref{Id1}){\em (iii)}: the expression on the left hand side of {\em (iii)}, upon inserting $\Om(u,w)$ as in (\ref{Om}), upon calling $w=v_\rk$, and upon using $\Dt^2_\rk(v_0,\dots,v_{\rk-1})\prod_{\ell=0}^{\rk-1}(v_\rk-v_\ell)=\Dt^2_{\rk+1}(v_0,\dots,v_{\rk })\prod_{\ell=0}^{\rk-1}(v_\rk-v_\ell)^{-1}
$ can be written as
$$\bl
\frac{(-1)^{\frac{\rk(\rk-1)}{2}}}{ \rk  !}
\left(\prod_{\ell=0}^{\rk }\oint_{\ga_{\sg_2}-\Ga_0}\frac{dv_\ell}{2\pi \I v_\ell^{\rho}h(v_\ell)(u-v_\ell)} \right)\underbrace{\frac{(v-v_\rk)^{-1}}{\prod_{{i=0}\atop{i\neq \rk}}^{\rk }(v_\rk-v_i)}}_{*} 
 \Dt_{\rk+1  }^2(v_0,\dots,v_{\rk-1  },v_\rk)
.\el$$
In the expression (*), the variable $v_\rk$ is singled out; by symmetry of the variables $v_1,\dots,v_\rk$, one may as well single out any $v_k$ and in particular replace the expression (*) by 
$$
\frac1{\rk+1}\sum _{k=0}^\rk\frac1{v-v_k}\prod_{{i=0}\atop{i\neq k}}^{\rk }\frac{1}{ v_k-v_i }=
\frac{ \frac1{\rk+1}}{\prod_{{i=0} }^{\rk }(v -v_i)}.
$$
Doing so establishes at once formula  (\ref{Id1}){\em (iii)}.
\medbreak
\noindent Proof of  (\ref{Id1}){\em (iv)}: similarly the expression on the left hand side, with $\Om_{\rk-1}^+$ inserted read as follows, upon renaming $w=v_{\rk-1}$:
\be\footnotesize\bl\frac{(-1)^{\frac{\rk(\rk-1)}{2}}}{(\rk-1)!}
    \left(\prod_{\ell=0}^{\rk-1 }\oint_{\ga_{\sg_2}-\Ga_0}\frac{dv_\ell}{2\pi \I v_\ell^{\rho}h(v_\ell)} \underbrace{\frac1{(v_{\rk-1}-u)}\prod_{{i=0}\atop{i\neq \rk-1}}^{\rk-1}\frac{v-v_i }{v_{\rk-1}-v_i}}_{*} \right)
 \Dt_{\rk   }^2(v_0,\dots,v_{\rk-1  } )
.\el\label{Id1''''}\ee
In the expression (*), the variable $v_{\rk-1}$ is singled out and can by symmetry of the remaining integrand, be replaced by the following sum:
$$
\frac1{\rk}\sum _{{i=0}\atop{i\neq k}}^{\rk-1}\frac1{  v_k-u }  \prod_{{i=0}\atop{i\neq k}}^{\rk }
 \frac{v-v_i }{v_{k}-v_i}=
 \frac{\prod_{i=0}^{\rk-1 }\frac{v-v_i}{u-v_i}-1}{\rk(u-v)}.
$$
This insertion in expression (\ref{Id1''''}) establishes identity ( \ref{Id1})({\em (iv)}) and finishes the proof of Lemma \ref{Lemma Id1}.\qed 

 \noindent{\em Proof of Proposition \ref{Prop-L'}:}
Remember from (\ref{Lsca}),  (\ref{Lsca'}) and Proposition \ref{Prop7.1}, the following formula
\be  \label{Lsca''}
  {\mathbb L}
=-  {\mathbb L}_0 
+(-1)^\ka \left(   {\mathbb L} _1
+  {\mathbb L}_2\right),\ee
with (where $\ga_{\sg_2}-\Ga_0\simeq \widetilde\Ga$ and so $\ga_{\rho_2}\simeq \widetilde \Ga^+ +\Ga^-_0$, with $\widetilde\Ga^+$ is slightly larger than $\widetilde\Ga $ from Fig.) 
\be \label{K1'}
\bl
 (-1)^\ka({\mathbb L} _1\!+\!{\mathbb L} _2)
=& -\oint_{\ga_{\rho_2}\simeq \widetilde \Ga ^+ +\Ga^-_0} \frac{u^\rk \FR(u)du}{2\pi \I} \oint_{\ga_{\sg_2}-\Ga_0\simeq \widetilde \Ga} \frac{ \Phi(v)dv }{2\pi \I h(v) v^{\rho}}\frac{\Om_1(u,v)}{(u\!-\!v)\Om_1(0,0) }\\&\footnotesize -\oint_{\ga_{\rho_2}} \tfrac{\FR(u)du}{2\pi \I u^\ka \GR(u)}.
\el \ee 
The first line in (\ref{K1'}) equals the first line in the formula (\ref{K2}) below. Indeed, the $u$-integration on the first line of (\ref{K1}) involves an integration about $\Ga_0^-$ and one about $\widetilde\Ga^+$, the latter  containing $u=v$ and $u=v_j$ for $0\leq j\leq \rk-1$, coming from $\Om_1(u,v)$. So the $u\in \widetilde\Ga^+$-integration decomposes into an integration about $u=v$ and a contour $\Ga ({\bf v}):=\Ga(v_0,\dots,v_{\rk-1}) =\sum_{\ell=0}^{\rk-1} \Ga_{v_\ell}$, with $v\notin  \Ga ({\bf v})$ and $\Ga_{v_\ell}$ a small contour about $v_\ell$. This yields the following expression for (\ref{K1}), after interchanging the $u$ and $v$-integrations and using $\ka=\rho-\rk$:
\vspace*{-.4cm}\be \footnotesize \label{K2}\bl %
  =&
 \overbrace{\frac 1{\Om_1(0,0)} \oint_{\ga_{\sg_2}-\Ga_0} \frac{ \Phi(v)dv }{2\pi \I v^{\rho} h(v) }
\left(-\ER_1(v) 
-\ER_2(v)
\right)}^{(*)}
 - \overbrace{\oint_{\ga_{\sg_2}-\Ga_0} \frac{ \Phi(v)dv }{2\pi \I v^\ka h(v)  }  \FR(v)}^{(**)}
-\overbrace{\oint_{\ga_{\rho_2}} \tfrac{\FR(u)du}{2\pi \I u^\ka \GR(u)}
}^{***}
\el
\ee
where
\be\label{E12} \bl
\ER_1(v)&:=\oint_{\Ga^{-}_0} \frac{ \FR(u)u^\rk du}{2\pi \I (u-v)} 
\Om_1(u,v)
\\
\ER_2(v)&:=\oint_{\Ga ({\bf v})}
\frac{\FR(u)u^\rk du}{2\pi \I (u-v)}
\left(\prod_{j=0}^{\rk-1} \oint_{\widetilde\Ga}
\frac{dv_j}{2\pi \I v_j^{j+\ka+1}h(v_j)}\frac{v-v_j}{u-v_j}\right)\Dt_\rk(v_0,\dots,v_{\rk-1})
 ,\el\ee
 with  
$$\footnotesize\bl
\ER_2(v)&=\sum_{\ell=0}^{\rk-1}
\oint_{ \widetilde\Ga}
\frac{dv_\ell \FR(v_\ell)v_\ell^\rk}{2\pi \I v_\ell^{\ell+\ka+1}h(v_\ell)}\frac{v-v_\ell}{v_\ell-v } 
\left(\prod_{{j=0}\atop{j\neq \ell}}^{\rk-1}
\oint_{\widetilde\Ga}
\frac{dv_j}{2\pi \I v_j^{j+\ka+1}h(v_j)}\right)
 \left(\prod_{{j=0}\atop{j\neq \ell}}^{\rk-1}\frac{v-v_j}{v_\ell-v_j}\right) \Dt_\rk(v_0,\dots,v_{\rk-1})
\el$$
$$\footnotesize\bl &\stackrel{*}{=}
-
\oint_{\widetilde\Ga}
\frac{dw \FR(w ) }{2\pi \I w^{ \ka }h(w)}
\sum_{\ell=0}^{\rk-1}  w^{\rk-\ell-1}
\left(\prod_{{j=0}\atop{j\neq \ell}}^{\rk-1}
\oint_{\widetilde\Ga}
\frac{dv_j}{2\pi \I v_j^{j+\ka+1}h(v_j)}\right)
  \\&~~~~~\times
 \Dt_\rk(v_0,\dots,v_{\ell-1},v,v_{\ell+1},\dots,v_{\rk-1})
 \\&= 
-
\oint_{\widetilde\Ga}
\frac{dw \FR(w ) }{2\pi \I w^{ \ka }h(w)}\sum_{\ell=0}^{\rk-1} \frac{\Om_2^{(\ell)}(v)}{w^{\ell+1-\rk}}  
=-
\oint_{\widetilde\Ga}
\frac{dw \FR(w ) }{2\pi \I w^{ \ka }h(w)}\Om^+_{\rk-1}(w,v)
 ,\el $$
where $v_\ell$ in the expression $\stackrel{*}{=}$ can be replaced by $w$, because $v_\ell$ in the integrand is a dummy and thus the summand is really independent of $v_\ell$, although it does depend on the index $\ell$. 

Moreover the integrals $(*)$ and $(**)$  in (\ref{K2}), with the integral $\Phi(v)$ as in (\ref{al-beta}) substituted,  have the form (for an integer $\al\geq 0$ and $\JR(v)$ standing for $\ER_1(v) 
+\ER_2(v)$ or $\FR(v)$), up to a constant $\Om_1(0,0)$,
 \be\label{J-int}\bl
\oint_{\ga_{\sg_2}-\Ga_0}& \frac{  dv  \JR(v)}{2\pi \I v^\al h(v)  } 
\oint_{\ga_{\sg_2+}}\frac{h(u)du}{2\pi \I \GR(u)(v-u)}
\\&=
\left[\oint_{\widetilde \Ga} \frac{  dv   }{2\pi \I v^\al h(v)  } 
\oint_{\Ga_{0}^-}\frac{h(u)du}{2\pi \I \GR(u)(v-u)}
- \oint_{\widetilde \Ga}
\frac{  dv  }{2\pi \I v^\al \GR(v)  } 
\right] \JR(v)
, \el \ee
since the $u$-integral can be decomposed into $\ga_{\sg_2}-\Ga_0\simeq \widetilde \Ga$ and $\ga_{\sg_2^+}\simeq \widetilde \Ga^+ +\Ga_0^+$; one carries out the integral about $u=v$ and and one keeps the integral about $u=0$.  

One now applies (\ref{J-int}) to $\JR(v)=\ER_1(v) 
+\ER_2(v)$, as in (\ref{E12}). Doing so, the expression $(*)$ in (\ref{K2}) reads, upon systematically replacing $\Om_1(u,v)$ by $\Om(u,v)$ by virtue of identity (\ref{Id1}{\em (i)}, and using all the remaining identities in Lemma \ref{Lemma Id1}, %
$$\bl
 (\ref{K2})(*) &= \frac1{\Om (0,0)}
  \left[
 \oint_{\widetilde \Ga} \frac{  dv   }{2\pi \I v^\rho h(v)  } 
\oint_{\Ga_{0}^-}\frac{h(u)du}{2\pi \I \GR(u)(v-u)}
- \oint_{\widetilde \Ga}
\frac{  dv   }{2\pi \I v^\rho \GR(v)  } 
 \right]
 \\&\times \left(\oint_{\Ga^{-}_0} \frac{ \FR(w)w^\rk dw}{2\pi \I (v-w)} 
\Om(w,v)+
\oint_{\widetilde\Ga}
\frac{  \FR(w ) dw}{2\pi \I w^{ \ka }h(w)}\Om^+_{\rk-1}(w,v)\right)
\el$$
\be\footnotesize \label{contr1}\bl&=\oint_{\Ga_{0}^-}\frac{h(u)du}{2\pi \I \GR(u) }
 \left[
\bl&\oint_{\Ga^{-}_0} \frac{ \FR(w)w^\rk dw}{2\pi \I  } 
 \underbrace{\oint_{\widetilde \Ga} \frac{  dv   }{2\pi \I v^\rho h(v)  }\frac{ \Om(w,v)}{(v\!-\!u)(v\!-\!w)\Om(0,0)}}_{\stackrel{(4)}{=}\frac{\Om_{\rk+1}^-(w,u)}{\Om(0,0)} }
\\&+\oint_{\widetilde\Ga}
\frac{  \FR(w )dw }{2\pi \I w^{ \ka }h(w)}
\underbrace{\oint_{\widetilde \Ga} \frac{  dv   }{2\pi \I v^\rho h(v)  } \frac{\Om^+_{\rk-1}(w,v)}{(v-u)\Om(0,0)} }_{\stackrel{(2)}{=}\frac{1}{u-w}(\frac{\Om(u,w)}{\Om(0,0)}-1) }
\el\right]
\\&+\oint_{\widetilde \Ga}
\frac{  dv   }{2\pi \I v^\rho \GR(v)  }
\left(\underbrace{\oint_{\Ga^{-}_0} \frac{ \FR(w)w^\rk dw}{2\pi \I (w-v)} 
\frac{\Om(w,v)}{\Om}}_{(1)}\!-\!
\underbrace{\oint_{\widetilde\Ga}
\frac{  \FR(w ) dw}{2\pi \I w^{ \ka }h(w)}\frac{\Om^+_{\rk-1}(w,v)}{\Om}}_{(3)}\right).
.\el\ee
To deal with the remaining expressions $-(**)-(***)$ in (\ref{K2}), one again applies  (\ref{J-int}) to $\JR(v)=\FR(v)$, and one remembers that $\ga_{\rho_2}\simeq \widetilde\Ga^+ +
\Ga_0^-$ which gives after replacing $v\to w$ and in the last integral $u\to w$,
\be\label{contr2}\bl
-(**)-(***)&=-\oint_{\widetilde \Ga} \frac{   \FR(w) dw }{2\pi \I w^\ka h(w)  } 
\oint_{\Ga_{0}^-}\frac{h(u)du}{2\pi \I \GR(u)(w-u)}
+ \left(\oint_{\widetilde \Ga}-\oint_{\ga_{\rho_2}}\right)
\frac{   \FR(w) dw}{2\pi \I w^\ka \GR(w)  }
\\&=
- \oint_{\Ga_0^-}
\frac{   \FR(w) dw}{2\pi \I w^\ka \GR(w)  }+\oint_{\widetilde \Ga} \frac{   \FR(w) dw }{2\pi \I w^\ka h(w)  } 
\oint_{\Ga_{0}^-}\frac{h(u)du}{2\pi \I \GR(u)(u-w)}
.\el \ee
Adding the two contributions (\ref{contr1}) and (\ref{contr2}), renaming integration variables and noticing the cancellation of the term containing $-\frac{1}{u-w}$ in $\stackrel{(2)}{=}$ of (\ref{contr1}) with  the second term of  (\ref{contr2}) gives formula (\ref{Lfinal}), except for the ${\mathbb L}'_0$-term:
\be\bl
{\mathbb L}={\mathbb L}=-
{\mathbb L}_0'+\sum_{i=1}^4
{\mathbb L}'_i
=&\left(-{\mathbb L}_0- \oint_{\Ga_0^-}
\frac{   \FR(w) dw}{2\pi \I w^\ka \GR(w)  }\right)+\sum_1^4{\mathbb L}'_i
,\el\ee
%
To deal with the ${\mathbb L}'_0$-term above, one notices that, from (\ref{Lixy}), the integral about $\Ga_0^-$ can be preceded by $\Id_{x_1>x_2}$, because the integrand has no residue for $x_1\leq x_2$, justifying $\stackrel{*}{=}$. Also $\stackrel{**}{=}$ is valid, because for $y_1<y_2$ the integrand has no residue at $u=t^{-1}$. Next one moves the contour $\widetilde \Ga$ to $-\Ga_0-\Ga(-\frac{1}{a^2 t}) $, since the integrand has no pole at $\iy$. Given that $y_1\geq y_2$, the integrand has a residue at $u= {a^{-2}}{ t^{-1}}$, only when $y_1=y_2$, with value given in $\stackrel{***}{=}$ below:
$$\bl-{\mathbb L}'_0 &:=\Id_{x_1>x_2 }\oint_{\ga_{\rho_1}}\frac{\FR(u) du}{2\pi \I u^\ka \GR(u)}
 - \oint_{\Ga_0^-}
\frac{   \FR(w) dw}{2\pi \I w^\ka \GR(w)  }
 \stackrel{*}{=}\Id_{x_1>x_2}\oint_{\widetilde \Ga}
\frac{   \FR(w) dw}{2\pi \I w^\ka \GR(w)  } 
\\&\stackrel{**}{=}\Id_{x_1>x_2}\Id_{y_1\geq y_2}\oint_{\widetilde \Ga\to -\Ga_0-\Ga(-\frac{1}{a^2 t})} 
\frac{   \FR(w) dw}{2\pi \I w^\ka \GR(w)  }
\\&\stackrel{***}{=}-\Id_{x_1>x_2}\Id_{y_1\geq y_2}\oint_{  \Ga_0} 
\frac{   \FR(w) dw}{2\pi \I w^\ka \GR(w)  }-\Id_{x_1>x_2}\Id_{y_1= y_2} \tfrac{(-a^2t)^{x_1-x_2-1}}{ 1+a^{-2} },
\el$$
ending the proof of Proposition \ref{Prop-L'}.\qed
 
\section{The limit of ${\mathbb L}_{n,\rk,\rho} $ to the discrete tacnode kernel ${\mathbb L}^{\mbox{\tiny dTac}}_{\rk,\rho,\beta}$}

\begin{proposition} \label{mainTh'}The following limit holds:
 \be\label{Final0}\lim_{n\to \iy} {\mathbb L}_{n,\rk,\rho}(x_1,y_1;x_2,y_2)\Bigr|_{{x_i=\tau_i-\ka}\atop{y_i= \frac{y'_i}{\sqrt{2}}}}
={\mathbb L}^{\mbox{\tiny dTac}} _{\rk,\rho,\beta}(\tau_1 ,y'_1;\tau_2 ,y'_2) . \ee
  \end{proposition}
 \noindent{\em Proof of Proposition  \ref{mainTh'}:}~One takes the limit of the kernel ${\mathbb L}_{n,\tau,\rho}(x_1,y_1;x_2,y_2)=-
{\mathbb L}_0'+\sum_{i=1}^4
{\mathbb L}'_i$ as in (\ref{Lfinal}), with ${\mathbb L}'_i$ as in (\ref{Lfinal'}) and the $\Om$'s as in (\ref{Om}) and (\ref{Omtil}). Then one reads off  from (\ref{hFG'}) below the limits of the corresponding untilded functions  (\ref{series}) and the limit below to the Heaviside function (\ref{Heaviside}),
\be\label{hFG'}\bl
\widetilde h(u ) &=e^{-u ^2+2\beta u } ,~~
u^{\rk}{\widetilde \FR}_{ }(u )  =u ^{\rk-x_1}e^{-\frac 12 u ^2+(\beta+y_1\sqrt{2})u }
 ,~~
v^\rho {\widetilde \GR}_{ }(v) = v^{ \rk-x_2}e^{-\frac 12 v^2+(\beta+y_2\sqrt{2})v} 
\\
v^{\ka}\tfrac{\widetilde h }{\widetilde \FR} (v)&=
v^{x_1+\ka}e^{-\frac{v^2}2 +v(\beta-y_1\sqrt{2})},~\tfrac{\widetilde h }{\widetilde \GR} (u)=
u^{x_2+\ka}e^{-\frac{u^2}2 +u(\beta-y_2\sqrt{2})},
  \tfrac{\widetilde \FR(v)}{v^\ka\widetilde \GR(v)} =\tfrac{e^{v(y_1-y_2)\sqrt{2}}}{v^{x_1-x_2}},~
  \\  \footnotesize \Id_{x_1> x_2 }&\Id_{y_1\geq y_2} \oint_{\Ga_0} \tfrac{\widetilde \FR(v)}{2\pi \I v^\ka\widetilde \GR(v)} = {\mathbb H}^{x_1-x_2}(\sqrt{2}(y_1-y_2))
  .\el \ee
Also notice that the distribution $\Id_{y_1=y_2}$ in the second term of $-{\mathbb L}'_0$  tends to $0$. One then uses, besides $\rho=\ka+\rk$, the changes of variables $y_i\to y'_i$ and $x_i\to \tau_i$, given by  $y_i= \frac{y'_i}{\sqrt{2}}$  and $x_i+\ka=\tau_i$;  and so $\rk-x_i=\rho-\tau_i$. The kernel $\mathbb L_{n,\tau,\rho}$ contains several integrals about the contour $\widetilde \Ga$, as in Fig. 6. It consists on the one hand of a half circle, which we choose to have radius $1/t^2$, so that the integral about the half circle  tends to $0$. On the other hand the integral over the vertical part of $\widetilde \Ga$ tends to $\downarrow L_{0+} $, a  downgoing vertical line in $\BC$ to the right of $\Ga_0$. This at once establishes the limit statement (\ref{Final0}).\qed
%
\medbreak
\noindent{\em Proof of Theorem \ref{mainTh}:}
The limit statement of Theorem \ref{mainTh} follows immediately from Proposition \ref{mainTh'}, combined with the relation (\ref{Lquadr}) between the two kernels $\BK_{n,\rk,\rho}^{^{\textrm{\tiny red}}}  (\xi_1,\eta_1 ; \xi_2,\eta_2) $ and  
$ {\mathbb L}_{n,\rk,\rho}
(x_1,y_1;x_2,y_2)$, and omitting the primes. \qed 

 \end{document}